\documentclass[apj]{emulateapj}
\usepackage{natbib}
\usepackage{textcomp}
\newcommand{\nod}{\nodata}
\newcommand{\mpm}{$\,\,\pm\,\,$}
\newcommand{\magsqarc}{\ifmmode {{{\rm mag}/$\sq^{\arcsec}$}}
            \else {mag/$\sq^{\arcsec}$}
             \fi}
\newcommand{\kms}{\ifmmode{\,\mathrm{km~s^{-1}}} 
          \else km~s$^{-1}$\fi}

\newcommand{\sersic}{S\'{e}rsic}
\newcommand{\chisqrin}{$\chi^{2}_{in}$}
\newcommand{\chisqrgl}{$\chi^{2}_{gl}$}
\newcommand{\chisqr}{$\chi^2$}
\newcommand{\magarc}{\ifmmode {{{{\rm mag}~{\rm arcsec}}^{-2}}}
             \else {{{mag}$~${arcsec}$^{-2}$}}
             \fi}
\newcommand{\hunit}{km~s$^{-1}$~Mpc$^{-1}$}
\newcommand{\hub}{H$_{\hbox{\scriptsize 0}}$}
\newcommand{\etal}{et~al.\@}
\newcommand{\eg}{e.g.\@}
\newcommand{\ie}{i.e.\@}
\newcommand{\ionyou[2]}{#1$\;${\small\rmfamily{#2}\relax}}
\newcommand{\oiii}{[\ionyou{O}{III}]}

\shorttitle{Evolutionary History of Galactic Bulges}

\begin{document}

\title{The Evolutionary History of Galactic Bulges: Photometric and 
  Spectroscopic Studies of Distant Spheroids in the GOODS Fields}


\author {Lauren A. MacArthur\altaffilmark{1}, 
Richard S. Ellis\altaffilmark{1},
Tommaso Treu\altaffilmark{2},
Vivian U\altaffilmark{3}, 
Kevin Bundy\altaffilmark{4},
Sean Moran\altaffilmark{5}
} 

\altaffiltext{1}{Department of Astrophysics, California Institute of 
Technology, MS 105-24, Pasadena, CA 91125; lam@astro.caltech.edu, 
rse@astro.caltech.edu}
\altaffiltext{2}{Department of Physics, University of California,
Santa Barbara, CA 93106-9530; Sloan Fellow; Packard Fellow;
tt@physics.ucsb.edu} 
\altaffiltext{3}{Institute for Astronomy, University of Hawaii, 
2680 Woodlawn Drive, Honolulu, HI 96822; vivian@ifa.hawaii.edu} 
\altaffiltext{4}{Department of Astronomy \& Astrophysics, University of 
 Toronto, Toronto, ON M5S 3H4, Canada; Reinhardt Fellow; 
 bundy@astro.utoronto.ca}
\altaffiltext{5}{Department of Physics \& Astronomy, The Johns Hopkins 
University, Baltimore, MD 21218; moran@pha.jhu.edu}

%\date{\today}
  
\begin{abstract}

We report on the first results of a new study aimed at understanding
the diversity and evolutionary history of distant galactic bulges in
the context of now well-established trends for pure spheroidal
galaxies.  To this end, bulges have been isolated for a sample of 137
spiral galaxies within the redshift range 0.1\,$<$\,z$\,<$\,1.2 in the 
GOODS fields. Using proven photometric techniques we determine the 
characteristic parameters (size, surface brightness,
profile shape) of both the disk and bulge components in our sample.
In agreement with earlier work which utilized aperture colors, distant
bulges show a broader range of optical colors than would be the case
for passively-evolving populations.  To quantify the amount of recent 
star formation necessary to explain this result, we used the DEIMOS
spectrograph to secure stellar velocity dispersions for a sizeable
fraction of our sample. This has enabled us to compare the Fundamental
Plane of our distant bulges with that for spheroidal galaxies in a
similar redshift range.  Bulges of spiral galaxies with a
bulge-to-total luminosity ratio ($B/T$)\,$>$\,0.2 show very similar
patterns of evolution to those seen for pure spheroidals such that 
the stellar populations of all spheroids with 
$M$\,$>$\,$10^{11}$\,$M_{\odot}$ are homogeneously old, consistent 
with a single major burst of star formation at high redshift 
($z_f$\,$\gtrsim$\,2), while bulges with $M$\,$<$\,$10^{11}$\,$M_{\odot}$ 
must have had more recent stellar mass growth ($\sim$\,10\% in mass 
since $z$\,$\sim$1).  Although further
data spanning a wider range of redshift and mass is desirable, the
striking similarity between the assembly histories of bulges and low
mass spheroidals is difficult to reconcile with the picture whereby
the majority of large bulges form primarily via secular processes
within spiral galaxies.
\end{abstract}

\keywords{cosmology: observations --- galaxies: bulges --- 
galaxies: evolution --- galaxies: formation --- galaxies: high-redshift}

\section{Introduction}
  
The history of galactic bulges remains a key issue in studies of the origin of
the Hubble sequence. Originally thought to form at high redshift through 
dissipationless collapse (Eggen, Lynden-Bell \& Sandage 1962), their continued 
growth, as predicted in hierarchical models (Baugh \etal\ 1998), is consistent
with the diversity observed in their present-day stellar populations 
(Wyse, Gilmore \& Franx 1997). Local data alone, however, cannot distinguish 
between quite different hypotheses for bulge formation, including secular 
processes triggered by interactions and the evolution of bars (Kormendy \&
Kennicutt 2004; Combes 2006).

The assembly history of bulges is also central to understanding the strong 
correlations observed between the nuclear black hole mass and bulge properties 
(Magorrian \etal\ 1998; Gebhardt \etal\ 2000). The different physical scales 
involved in these local scaling relations represent a major theoretical 
challenge (Miralda-Escud\'e \& Kollmeier 2005). As ambiguities remain even 
if fairly precise observations are available for local samples, an important 
route to understanding this puzzle lies with undertaking observations at 
intermediate redshift. This is especially challenging for measurement of 
black hole masses where the sphere of influence remains unresolved 
(c.f.\@ Woo \etal\ 2006). However, as an alternative approach, it may be
more practical to attempt to measure the growth history of the bulges 
in which they reside. 

The arrival of deep multi-color imaging data from the {\it Hubble
Space Telescope} ({\it HST}) provided the first glimpse of the
photometric properties of bulges at intermediate redshift. In an early
paper, Ellis, Abraham \& Dickinson (2001, hereafter EAD), examined
aperture colors of bulges in 68 suitably oriented isolated spirals
with $I_{AB}$\,$<$\,24 in the northern and southern Hubble Deep Fields
(HDF). The authors found a remarkable diversity in bulge colors over
the redshift range 0.3\,$<$\,$z$\,$<$\,1 (using both spectroscopic and
photometrically determined redshifts), with few as red as a
passively-evolving track which matches the integrated colors of
luminous spheroidal galaxies observed in the HDFs \footnote{We will
adopt the term `spheroidal galaxy' to denote pure elliptical
%both elliptical and S0
systems, reserving `spheroidal component' where necessary to refer to
bulges within galaxies also harboring a disk.}. EAD concluded that bulges 
have suffered recent periodic episodes of rejuvenation consistent with
15--30\% growth in stellar mass since $z$\,$\simeq$\,1. These conclusions
were challenged by Koo \etal\ (2005, hereafter Koo05) who located 52 
luminous ($I_{AB}$\,$<$\,24) bulges in the shallower but wider field Groth 
Strip Survey for which a more elaborate photometric decomposition was
undertaken. They found that 85\% of their field sample had uniformly red
colors at $z$\,$\simeq$\,0.8 ($\delta (U-B)$\,$\simeq$\,$\pm$\,0.03), as 
red as present-day and distant cluster E/S0s.  Only a minority (8\%) showed
blue rest-frame colors, most of which occurred in interacting or
merging systems.

In the interim, much has been learned about the luminosity dependence
of evolution in the field spheroidal galaxy population. Treu \etal\
(2002), van~Dokkum \& Ellis (2003), Treu \etal\ (2005a\&b, hereafter
T05), van~der~Wel \etal\ (2005), and di Serego Alighieri \etal\
(2005) have undertaken comprehensive Fundamental Plane (FP) analyses
of several hundred field galaxies to $z$\,$\simeq$\,1.  Whereas the most
luminous spheroidal galaxies studied support the long-held view of
early collapse and subsequent passive evolution (\eg\ Bower, Lucey,
\& Ellis 1992), a surprising amount of recent star formation is
necessary to explain the scatter and FP offsets for lower-luminosity
galaxies. T05 find that as much as 20--40\% of the present dynamical
mass in systems with $M<10^{11}$\,$M_{\odot}$ formed since 
$z$\,$\simeq$\,1.2.  For these systems, spectroscopic signatures of 
recent star formation are visible and several show resolved blue cores,
consistent with recently-accreted gas-rich dwarfs.
 
The present paper aims to clarify the relationship between bulges and
spheroidals in the light of the above work. The first part is
motivated by the arrival of superior resolution multi-color ACS
imaging data in the Great Observatories Origins Deep Survey (GOODS)
fields; this represents a significant improvement over the WFPC2 data
used by EAD and Koo05. The GOODS ACS dataset offers the potential of
securing improved photometric parameters.  In particular, EAD chose
not to undertake bulge-to-disk (B/D) decomposition, arguing by example
with local data (de~Jong 1996) that aperture colors were adequate.  In
the present paper, we revisit this discussion using a larger sample
with equivalently-deep but higher spatial resolution, broader
wavelength coverage, spectroscopic redshifts, and employing
photometric decompositions of bulge and disk parameters.

The second component is concerned with securing dynamical estimates of 
the bulge stellar masses from resolved spectroscopy.  At intermediate
redshifts, resolved spectroscopy of high quality is now possible thanks to
instruments such as the Deep Imaging and Multi-Object Spectrograph (DEIMOS;
Faber \etal\ 2003) 
on the Keck II telescope. Following recent progress in interpreting the mass 
assembly history of field spheroidals (\eg\ T05), mass-to-light ratios ($M/L$)
determined from FP analyses are much superior to interpretations based 
solely on optical colors. Our goal is to secure the mass assembly 
history of a representative sample of intermediate redshift bulges and to 
compare trends found with the integrated properties of field spheroidals.

A plan of the paper follows. In \S\ref{sec:sample}, we discuss the
selection criteria we adopt for both the photometric and spectroscopic
sample of bulges in the GOODS fields. In \S\ref{sec:photom} we
describe the photometric techniques used for bulge/disk decomposition,
taking into account the effects of the ACS point spread function, and
derive rest-frame properties (k-corrections).  In
\S\ref{sec:colors} we compare our photometric results with
earlier measures of the diversity and star formation history of
intermediate redshift bulges. In \S\ref{sec:spec} we describe the
spectroscopic measurements undertaken with DEIMOS and their reduction
to the central stellar velocity dispersions necessary to construct the
FP.  Key issues of disk contamination, systematic rotation, and
relative aperture size are discussed.  In \S\ref{sec:spec_results} we
compare the FP of bulges with the trends now well-established for more
massive E/S0 galaxies.  We discuss the implications of both our
photometric and spectroscopic results in the context of various
formation hypotheses in \S\ref{sec:discuss}. We summarize our overall
findings in \S\ref{sec:summary}.

Throughout this paper, for all distance dependent quantities we adopt
a flat cosmological model with $\Omega_M$\,=\,0.3, $\Omega_{\Lambda}$\,=\,0.7,
and \hub\,=\,65~\hunit. We note that both diagnostics considered in
this study -- colors and evolution of mass-to-light ratios -- are
independent of the Hubble constant. All magnitudes are in the AB
system (Oke 1974) unless otherwise noted.

\section{Data}\label{sec:sample}
 
The primary dataset for this study is the GOODS public v1.0 data 
release (Giavalisco \etal\ 2004) which provides deep imaging in four
ACS passbands: F435W ($B$), F606W ($V$), F775W ($i$), and F850LP ($z$).
The depth and high resolution (0\farcs03/pix sampling) of the GOODS 
ACS photometry, allow for an examination of the structure and luminosity 
parameters of spheroidal galaxies and, in particular, spiral galaxy bulges 
can be isolated from their surrounding disk.  Spectroscopic data is essential
both for the photometric analysis where redshifts are required to create
rest-frame properties, and for the FP analysis where precise
bulge stellar velocity dispersions are required.

Spectroscopic data is drawn from two independent campaigns.  The first
is that of T05, who secured high signal/noise spectra of a magnitude-limited 
sample of isolated spheroid-dominated galaxies in the northern
GOODS field with $z_{AB}$\,$<$\,22.5 spanning the redshift range 
0.1\,$<$\,$z$\,$<$1.2 (see Figs.\@~1 \& 2 of T05 for a mosaic of this 
sample).  Precision central stellar velocity dispersions were secured 
for 181 of these galaxies via 1--2 night exposures undertaken with DEIMOS 
during 2003 April 1--5 at the Keck observatory.

The second component of our bulge sample arises from later Keck campaigns
during 2004 and 2005 dedicated to increasing the spiral sample for this 
project. Here, we selected isolated spirals from the northern and southern 
GOODS fields according to visual classifications presented by 
Bundy \etal\ (2005). This sample was also limited at $z_{AB}$\,$<$\,22.5 
with the additional criterion of a known spectroscopic redshift from the 
Keck Team Redshift Survey (Wirth \etal\ 2004). This yielded a further target 
sample of 45 spirals within 0.1\,$<$\,$z$\,$<$\,0.7.  A mosaic of 3-color 
ACS images for this second subset is shown in Figure~\ref{fig:mosaic}.

\begin{figure*}
    \centering
    \includegraphics[width=0.9\textwidth]{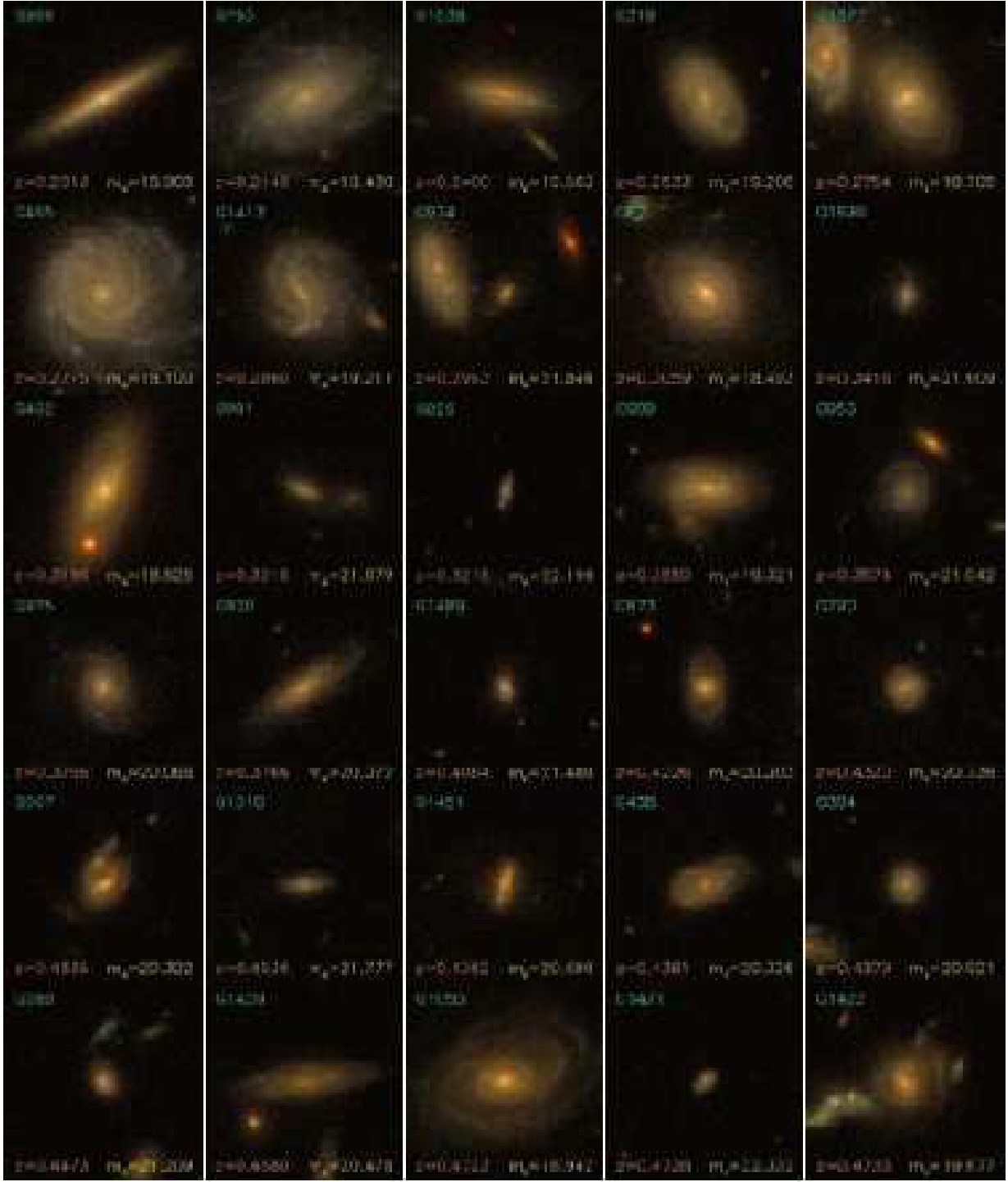} 
    \caption{Mosaic of 3-color composite ACS images (R=$z$, G=$i$, B=$V$), 
             7\arcsec\ on a side, of the new GOODS bulge sample, sorted by 
             redshift (lowest at {\it top left}, highest at 
             {\it bottom right}).
             For each galaxy the internal ID [{\it top left corner}], redshift 
             [{\it bottom left corner}], and total apparent $z$-band (F850LP) 
             magnitude [{\it bottom right corner}] are indicated.}
    \label{fig:mosaic}
\end{figure*}
\begin{figure*}[hp]
    \centering
    \includegraphics[width=0.9\textwidth]{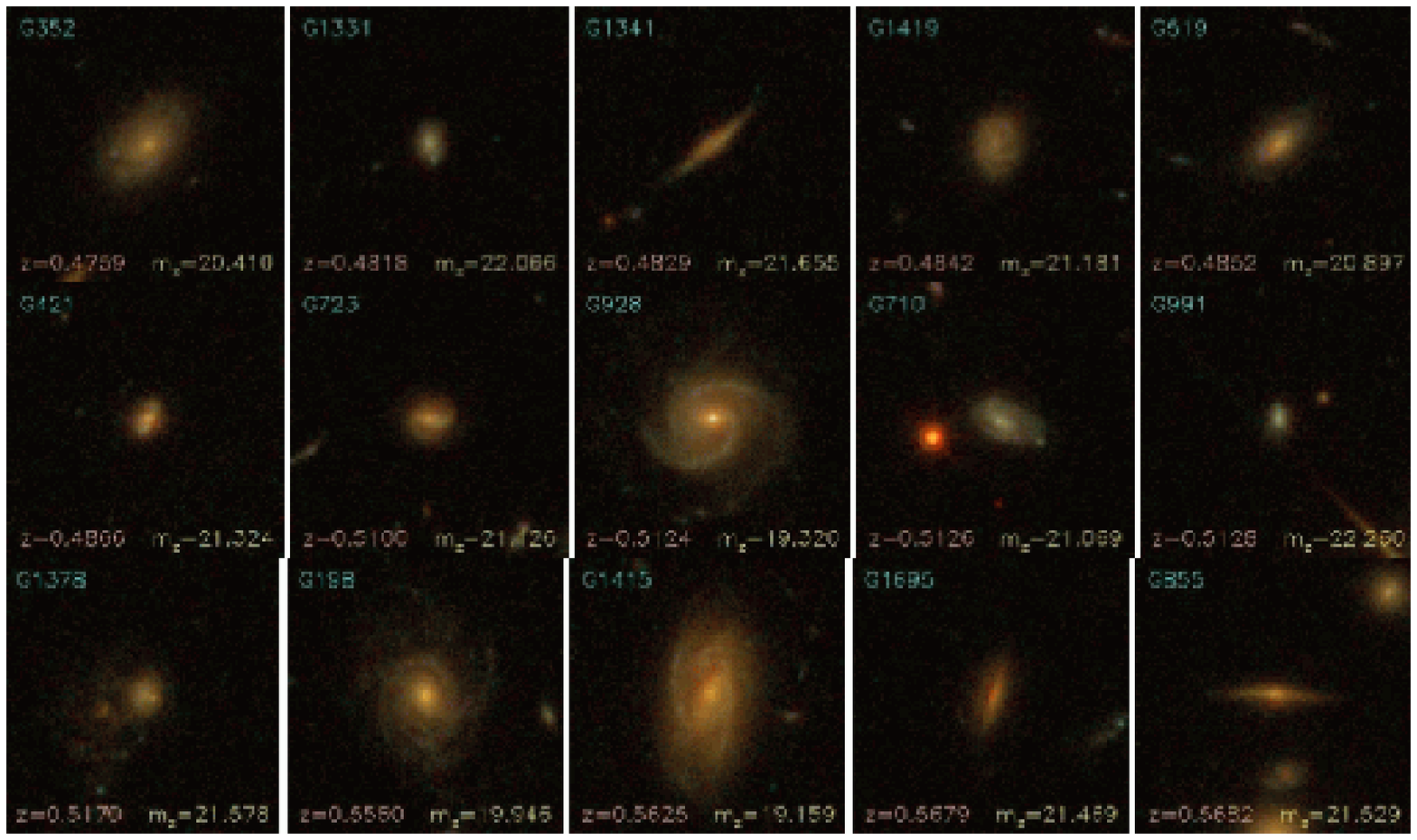}
    \figurenum{\ref{fig:mosaic}}
    \caption{Continued.}
\end{figure*}

The two spectroscopic subsets are naturally somewhat different. The first, 
from T05, is skewed to spheroid-dominated systems considered to be E/S0s in 
the originally-released v0.5 images.  Regardless of the original T-type 
classification (see T05 for details), all galaxies in the combined sample 
were independently visually examined as part of this analysis, in order to 
determine whether they were better described by 1 (pure spheroidal) or 
2-component (spheroid plus disk) systems.  Occasionally, particularly at high 
redshift, the visual inspection was not definitive.  In such cases, the 
shape of the light profile was used as an additional guide by comparing 
one-component \sersic\ fits to B/D decompositions (see \S\ref{sec:BDdecomps}). 
The different fits and their residuals were compared to single versus 
two-component fits of the bona-fide spirals.  
Since we only consider the radial component of the light profiles, 
the 25 high-inclination ($i$\,$\gtrsim$\,70$^\circ$) spirals in both 
samples are excluded from further analysis.  Furthermore, a kinematic 
decomposition into separate components for the observed central 
velocity dispersion is not feasible. We thus restrict ourselves to 
galaxies with bulge-to-total ratio ($B/T$) greater than 0.2 for 
the FP analysis to ensure that the effects of disk contamination is 
minimal (see \S~\ref{sec:sigmacor}).  
This excludes a further 21 galaxies from the recent campaign, and 25 from T05.

\begin{figure}[htbp]
  \centering
  \includegraphics[width=0.48\textwidth,bb=18 260 592 718]{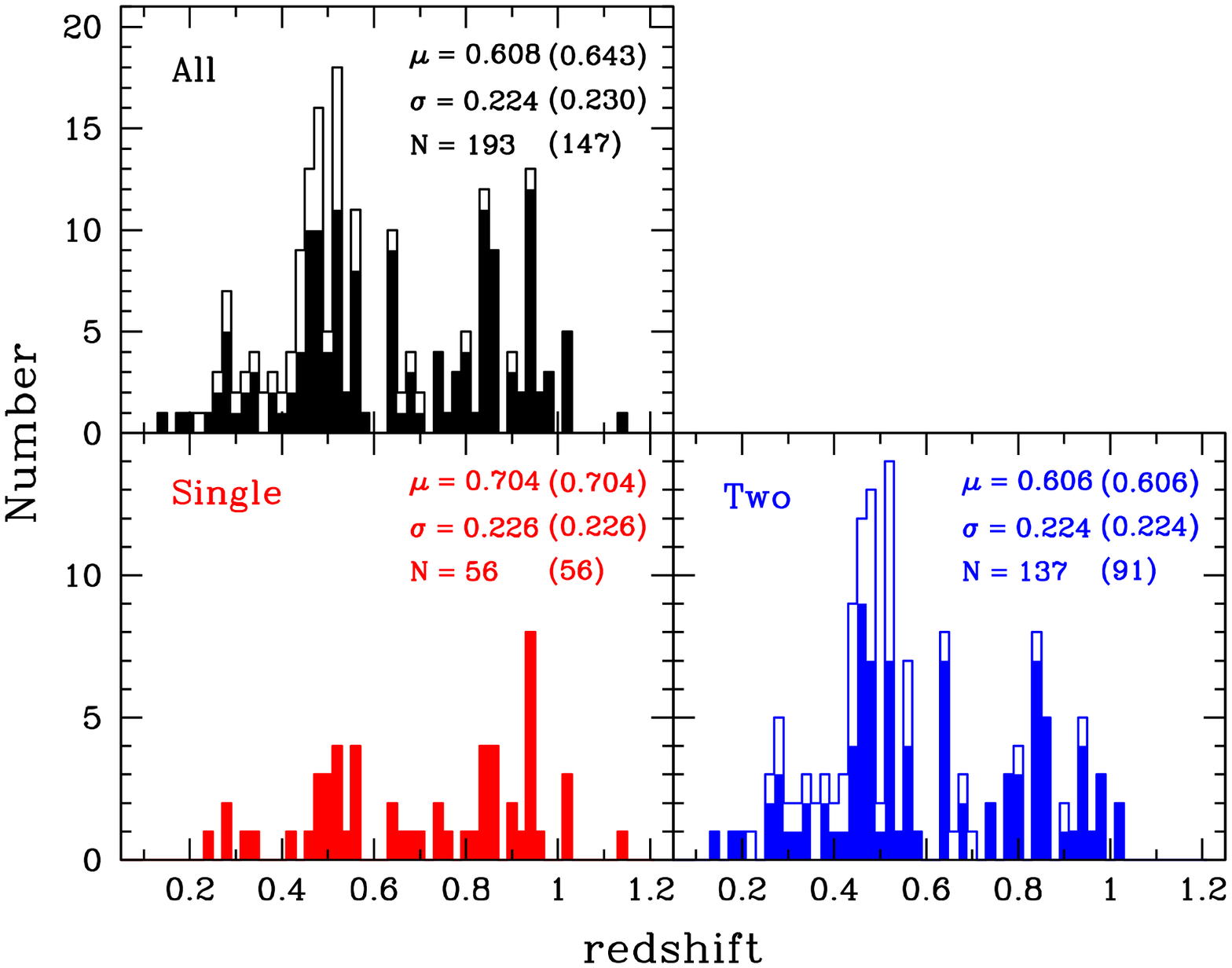}
  \caption{Histograms of sample redshift distribution. {\it Top}: full sample;
          {\it Bottom left}: single component galaxies;
          {\it Bottom right}: two-component galaxies.
          Open histograms refer to the photometric sample, filled histograms
          and numbers in parentheses refer to the spectroscopic sample. }
  \label{fig:redshift}
\end{figure}

In summary, therefore, the {\it photometric sample} for this study comprises 
193 galaxies in the redshift range 0.1\,$<$\,$z$\,$<$\,1.2 of which 56 were
modeled as single-component \sersic\ profiles and 137 were decomposed into 
two components: exponential disk plus \sersic\ bulge.
%The morphological breakdown is 23 Es, 33 S0's, 93 Sa-b and 44 Sc-Irr. 
The {\it spectroscopic sample}, from which the FP will be constructed,
comprises 147 galaxies, 56 pure (single-component) spheroidals
and the spheroidal component of 91 two-component galaxies. Eight bulges
are common to both spectroscopic sub-samples, thereby offering a check
on systematic errors (see \S~\ref{subsec:veldisp}).
Figure~\ref{fig:redshift} shows a histogram of the redshift
distributions for both the photometric and spectroscopic samples.
Finally, Table~\ref{tab:galpars} lists positions, redshifts, T-type,
number of fit components, $B/T$, and total observed magnitudes (see
\S~\ref{sec:totmagcol}) for all of our sample galaxies.

\section{Photometric Analysis}
\label{sec:photom}

We begin with an examination of the deep ACS images in the GOODS v1.0 data 
release to derive the structural and luminosity parameters of the spheroidal 
components (defined here as the entire galaxy for single-component galaxies, 
or the isolated bulge for the two-component galaxies) of the full photometric 
sample.  A key goal is the comparison of the photometric properties of our 
bulges with the spheroidals analyzed by T05.

Azimuthally-averaged surface brightness (SB) profiles are extracted from the
$z$-band images using the XVISTA package\footnote{Developed at Lick 
Observatory.  As of writing, XVISTA is maintained and distributed by Jon 
Holtzman at New Mexico State University and can be downloaded from 
{\rm http://ganymede.nmsu.edu/holtz/xvista/}.}.  Galaxy centers are determined 
interactively and fixed for the isophotal fit, but variable 
position angles (PA) and ellipticities ($\epsilon=1-b/a$, where $a$ and $b$ 
are respectively the semi-major and semi-minor axes) are permitted at each 
isophote. SB profiles are traced to $\sim$\,24--26 \magarc, corresponding to 
a systematic  SB error of $\lesssim$\,0.1\,\magarc.  Profiles were extracted 
in at least $z$ \& $i$ for each galaxy in our sample, and 
most $z$\,$\lesssim$\,0.5 also have adequate $V$ \& $B$ profiles.

\subsection{Light Profile Modeling}
\label{sec:BDdecomps}

Surface brightness profiles are modeled using techniques described in detail 
in MacArthur, Courteau, \& Holtzman (2003, hereafter Mac03).  A brief 
description is provided here.

Our profile modeling algorithm reduces one-dimensional (1D) projected
galaxy luminosity profiles into either a single component \sersic\
profile, or decomposes bulge and disk components simultaneously using
a nonlinear Levenberg-Marquardt least-squares fit to the logarithmic
intensities (\ie\ magnitude units).  Random SB errors are accounted
for in the (data$-$model) minimization, whereas systematic errors such
as uncertainties in the sky background and point spread function
measurements are accounted for separately by performing decompositions
with the nominal values as well as $\pm$ their respective errors (see
\S~\ref{subsec:skysing}).

A fundamental aspect of the profile decompositions is the choice of fitting
functions.  The exponential nature of disk profiles has been
established observationally both locally (Freeman 1970; Kormendy 1977; 
de~Jong 1996; Mac03), and at high-$z$ (Elmegreen \etal\ 2005).
Hence, here we model the disk light, in magnitudes, as:
\begin{equation}
\label{eq:expmag}
\mu_d(r)=\mu_0 + 2.5\log_{10}(e){\left\{{r\over{h}}\right\}},
\end{equation}
where $\mu_0 (\equiv -2.5\log_{10} I_0$) and $h$ are the disk central surface
brightness (CSB) and scale length respectively, and $r$ is the galactocentric
radius measured along the major axis.  

The appropriate form for spiral bulges and spheroidal galaxies is, however, 
less clear. Historically, all spheroids were modeled with
the de~Vaucouleurs $r^{1/4}$ profile which has proven to be a good
description for the shape of luminous elliptical galaxies.  However, higher 
resolution studies have revealed a range of profile shapes for both
E/S0s and spiral bulges, which generally correlates with the
luminosity (Caon \etal\ 1993; Andredakis, Peletier, \& Balcells 1995; 
Courteau \etal\ 1996; Graham 2001; Mac03).  Such a range is also
supported by numerical simulations of two component (halo plus stars) 
dissipationless collapse (Nipoti \etal\ 2006) when mergers and secular 
evolution of disk material are accounted for (Scannapieco \& Tissera 2003; 
Debattista \etal\ 2005).

To ensure the most general form for the bulge SB profile, we adopt the
formulation of \sersic, which, in magnitudes, becomes:
\begin{equation}
\label{eq:sersicbnmag}
\mu_{b}(r)=\mu_{e} +
   2.5\log_{10}(e)\,b_{n}{\left[\left({r\over{r_{e}}}\right)^{1/n} - 1\right]},
\end{equation}
where $\mu_e (\equiv-2.5\log_{10}(I_e))$ is the SB at the effective radius, 
$r_e$, enclosing half the total extrapolated
luminosity\footnote{For a pure exponential disk, $r_e = 1.678 h$.}, and
$b_n$ is chosen to ensure that
\begin{equation}
\int_{0}^{\infty}I_{b}(r)\,2\pi r\,dr = 2\int_{0}^{r_{e}}I_{b}(r)\,2\pi r\,dr.
\label{eq:bnderv}
\end{equation}

We use the approximation for $b_n$ given in Appendix A of Mac03 which is good
to one part in 10$^4$ for $n$\,$>$\,0.36 and two parts in 10$^3$ for 
$n$\,$\leq$\,0.36.

It has become customary to express the disk parameters in terms of scale
length and CSB ($h$ and $\mu_0$), while the spheroid is described
in terms of {\it effective} parameters ($r_e$ and $\mu_e$).  We adopt this
formalism, thus parameters with subscript ``$e$'' refer to the bulge or
spheroid.  We will use a capital $R_e$ to indicate radii expressed in
physical units (usually kpc).

For the Fundamental Plane analysis in the following sections, we also
need to compute the average SB within the effective radius.  For the
\sersic\ profile this is computed as:
\begin{equation}
\label{eq:avgSB}
{\rm SBe} = -2.5\log_{10}(\langle I_e \rangle),
\end{equation}
where
\begin{equation}
\label{eq:avgIe}
\langle I_{e} \rangle= {{I_{e}\,{\exp(b_{n})\,n\,\Gamma(2n)}}\over{b_{n}^{2n}}}.
\end{equation}

The best-fit parameters of the (data$-$model) comparison are then those which
minimize the reduced chi-square merit function, 
\begin{equation}
\label{eq:chi}
\chi^2_{\nu}={1\over{\rm{N-M}}}\,\,{\sum_{i=1}^{\rm{N}}}
\,\left[{\mu_{gal}(r_i)- \mu_{model}(r_i;h,\mu_{0},r_e,\mu_{e},n)\over
 \sigma_i }\right]^2,
\end{equation}
where N is the number of radial data points used, M is the number free 
parameters (\ie\ N~$-$~M = $\nu \equiv$ Degrees of Freedom), and $\sigma_i$ 
is the statistical error at each SB level.  From here on 
the $\nu$ subscript will be omitted and the $\chi^2$ variable refers to 
a $\chi^2$ per degree of freedom.

\subsection{Fitting Procedure}
\label{sec:BDfits}

Constraining the five structural parameters of spiral galaxies is
a challenge even at low-redshift.  At higher-$z$, the lower (relative) 
resolution and depth limitations of the images renders the challenge 
even greater.  In order to avoid discrepant fits between different
passbands, some workers have opted to perform fits simultaneously in all
bands.  The scale parameters are held fixed for the decompositions in all 
passbands, and only the flux levels are allowed to vary to obtain colors 
(\eg\ Koo05).  This procedure implies there are no significant 
color gradients and that the bulge shapes are identical at all wavelengths.  
However, just as any morphological description of galaxies (\eg\ Hubble 
types) depends on the wave band, intrinsic structural parameters have
also been shown to vary with wavelength as a result of stellar population 
and dust extinction effects (\eg\ MacArthur \etal\ 2004, hereafter Mac04).  
Thus, multi-wavelength information and independent fits at each wavelength
are required for any accurate description of galactic structural parameters.

An additional narrowing of the parameter space is often accomplished
by fixing the bulge shape to be that of a de~Vaucouleurs profile
(\sersic\ $n$\,=\,4).  While this may be justified for luminous spheroidal
galaxies and earliest-type spiral bulges, our sample contains a range
in spheroidal galaxy luminosity and extends to late-type spirals which
have been shown to have bulge profiles closer to that of an
exponential (\eg\ Mac03), thus such a constraint would be
inappropriate.  Physical differences in the shape and size of
spheroids among galaxies are also expected depending on how they were
formed.  Formation by dissipationless collapse (Nipoti \etal\ 2006) 
or accretion processes (\eg\, major/minor mergers)
can account for steeply rising (high-$n$) light profiles in the
central parts of galaxies (\eg\, Aguerri \etal\ 2001), while secular
evolution of disk material (possibly triggered by a satellite) would yield 
shallower distributions ($n$\,$\simeq$\,1--2) of  the central light 
(Scannapieco \& Tissera 2003; Eliche-Moral \etal\ 2006).  The formation 
of small bulges is indeed largely attributed to secular processes and 
redistribution of disk material (see Kormendy \& Kennicutt 2004 and 
references therein).

An important constraint in analyzing our present dataset is the need
to compare the evolutionary characteristics of our new bulge sample
with a similar sample of E/S0s (from T05). Ideally, the analysis
techniques should be identical.  In T05, structural homology was
assumed and all profiles were modeled with a fixed $n$\,=\,4 profile, to
conform with the standard practice for traditional studies of the
Fundamental Plane of early-type galaxies in the local universe (\eg\,
Dressler \etal\ 1987; Djorgovski \& Davis 1987; J{\o}rgensen \etal\ 1996). 
However, the goal of this paper is to extend the study to
bulges, which are typically best modeled as lower $n$ \sersic\
profiles. Thus, to ensure homogeneity between the analysis of bulges and 
spheroidals, we have re-modeled the light profiles of the  T05 galaxies 
using \sersic\ profile for the spheroid component and adding a disk
component when necessary.  We find a range of best fit \sersic\ parameters 
for these early-type galaxies spanning $\sim$\,0.9\,$<$\,$n$\,$<$\,3.4 
(\ie\ it does not extend as high as the de~Vaucouleurs $n$\,=\,4 shape, see 
panel (d) in Fig.~\ref{fig:Corr_Local}).

The following sections detail the fitting techniques which are
specific to our intermediate redshift sample and are based exclusively
on results from 1D B/D decompositions.  The fact that we do not
attempt to model non-axisymmetric shapes (bars, rings, oval
distortions) lessens the need for more computationally intensive 2D
B/D decompositions (extensive simulations in Mac03 showed no
improvements using the 2D over the 1D decomposition method when
modeling axisymmetric structures.)  The measured parameters and
correlations among them are discussed in \S\ref{sec:corr}.

\subsubsection{Initial Estimates}

In order to determine the range of best-fitted bulge and disk parameters,
we need to assist the minimization program in finding the lowest possible
(data$-$model) \chisqr\ (c.f.\@ Mac03). We base our initial estimates for the 
disk parameters $h$ and $\mu_0$ on a  ``marking the disk'' technique, where 
the linear portion of a luminosity profile is ``marked'' and the selected 
range is fit using standard least squares techniques to determine its slope.  
The baseline adopted for these fits is 0.25$\,r_{max}$ to $r_{max}$.  The 
inner boundary is chosen to exclude the major contribution of a putative bulge 
or a Freeman (1970) Type-II profile dip, and $r_{max}$ is the radius at 
which the surface brightness error is greater than 0.1\,\magarc.  

Flexibility in the {\it bulge} initial parameters is, however, limited
by point spread function and sampling effects; thus the \sersic\ $n$
exponent cannot be fit as a free parameter.  We therefore hold $n$ fixed in 
the decompositions and explore the full range of 0.1\,$<$\,$n$\,$<$\,6.0
in steps of $n$\,$=$\,0.1.  A grid search is performed to select the best
fit. As in Mac03, for each decomposition (at fixed $n$) we explore
four different sets of initial bulge parameter estimates to protect
against local minima in the parameter space.

\subsubsection{Point Spread Function and Sky Treatment}
\label{subsec:skysing}

Bulges of late-type spirals are small and their luminosity profiles
can be severely affected by the point spread function (PSF). The PSF
is accounted for by convolving the model light profiles with a 
radially-symmetric Gaussian PSF prior to comparison with the observed
profile.

The PSFs in the GOODS ACS images were estimated using a modified version 
of the Tiny Tim {\it HST} PSF modeling software\footnote{\rm
http://www.stsci.edu/software/tinytim/tinytim.html} (Rhodes \etal\
2007). For each of the 4 GOODS filters, 2500 artificial stars were
inserted across the ACS WFC field with a telescope focus value of
$-2\,\mu$m (\ie\ a primary/secondary spacing 2$\,\mu$m smaller than
nominal).  SExtractor (Bertin \& Arnouts 1996) was then used to measure 
the FWHM of those PSFs\footnote{Note that this procedure does not account 
for the long-wavelength ``halo'' observed in the PSF of the reddest ACS 
bands (Sirianni \etal\ 1998).
%caused by red photons passing through the CCD and scattered to large angles 
%within the glass mounting substrate and then back into the detector.  
As such, whenever we consider observed spheroid colors, we restrict ourselves
to the minimally affected F606W ($V$) and F775W ($i$) filters.}.
Table~\ref{tab:ACS} lists the mean and standard deviation of the
2500 PSF measurements for each of the 4 ACS filters.
To account for the spread in these PSF distributions, each profile is modeled 
with three different values of the PSF FWHM: the nominal mean value 
and $\pm$5\% of that value.

The GOODS ACS images are already sky subtracted, so no further attempt at
measuring a residual sky was made.  However, we do allow for 
sky subtraction errors in the decompositions, in a similar manner as the PSF 
errors, by using three different sky levels: the measured profile as is
and at $\pm$0.5\% of a nominal space-based value for each band. The 
adopted sky levels are listed in Table~\ref{tab:ACS}.

The final step is to determine the best fit from the resulting 2160 
decompositions.  We follow
Mac03 ($\S$4.3) and undertake a grid search for the minimum of two \chisqr\ 
merit function distributions; the global $\chi^2$ (eq.\@~[\ref{eq:chi}]) 
which is dominated by the contribution from the disk, and a separate 
{\it inner} $\chi^2$ statistic computed to twice the radius where the bulge 
and disk contribute equally to the total luminosity 
($r_{b=d} \equiv 2r(I_b=I_d)$).  
We label this statistic as \chisqrin.  For cases where the bulges are so 
small that they never truly dominate the light profile (\ie\ $r_{b=d}$ is 
undefined), we compute \chisqrin\ out to the radius at which $\nu$\,=\,1.
Only the global \chisqr\ is considered for single component fits.
\chisqrin\ was adopted to increase the sensitivity of the 
goodness-of-fit indicator to the bulge area\footnote{Note that our algorithm 
minimizes the \chisqrgl\ only. The \chisqrin\ is calculated and used as a 
discriminator only after the algorithm has converged.}.  

In cases where the bulge component is weak, fitting becomes more difficult, 
particularly where the rest-frame wavelength range extends into the 
UV for our highest-$z$ galaxies.  In order to avoid unrealistically large 
differences in the fits between the different bands, constraints were placed 
based on the $i$ and $z$-band fits (where the SB profiles are closest to 
rest-frame $R$-band).  These physical constraints are rather generous and do 
not contribute any subjective bias.

\begin{figure*}
\centering
\includegraphics[width=0.47\textwidth]{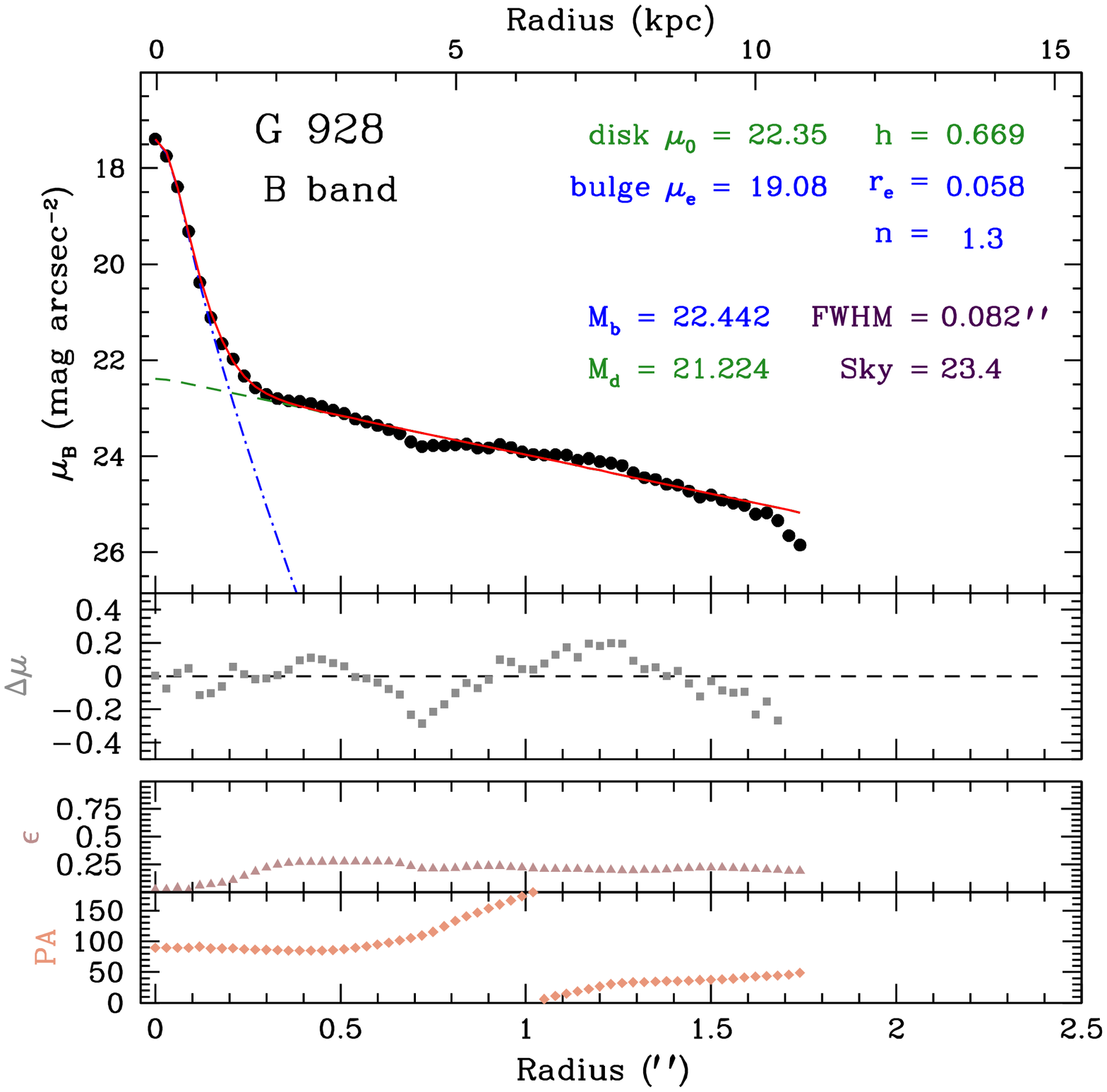}
\includegraphics[width=0.47\textwidth]{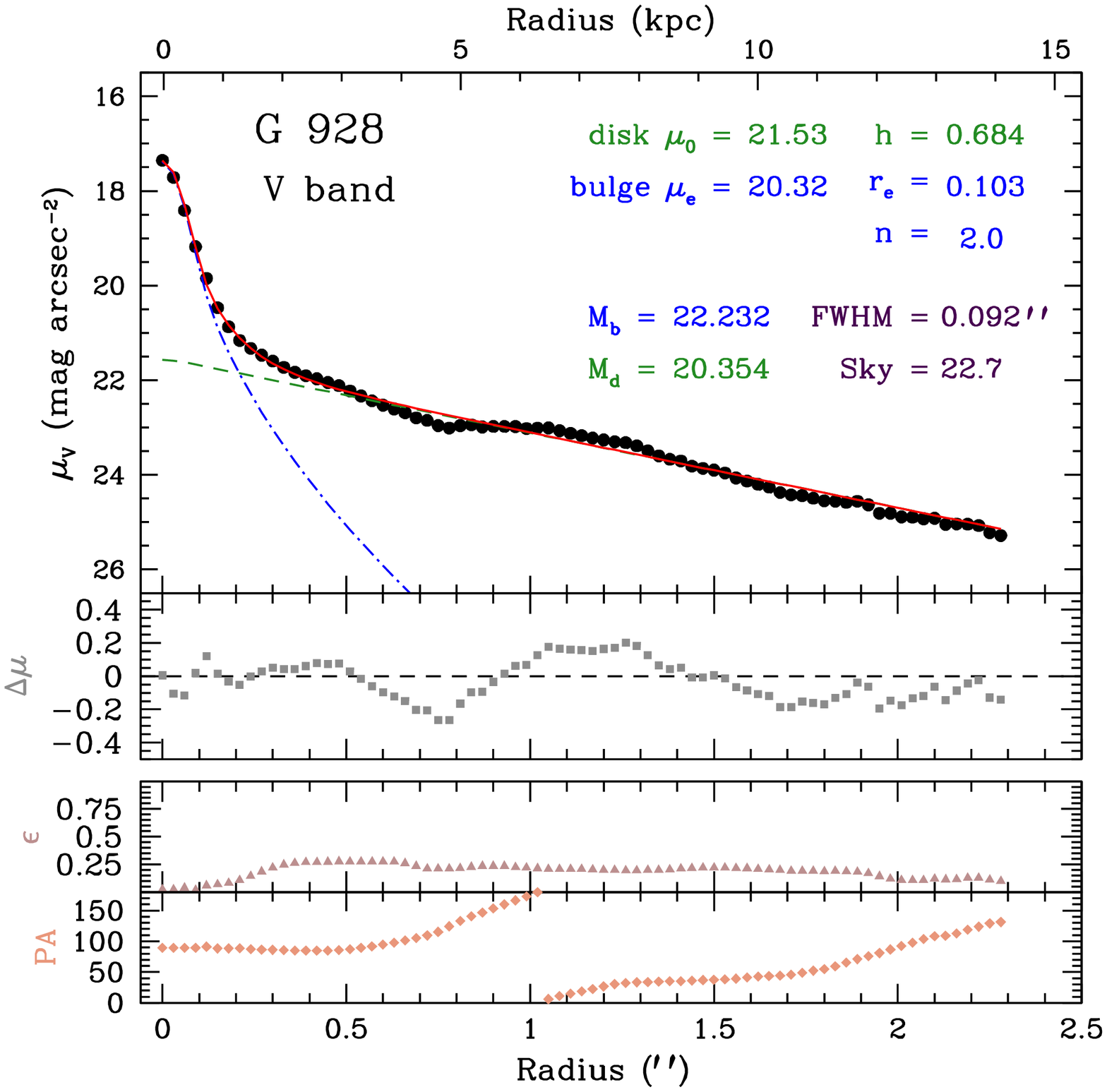}
\includegraphics[width=0.47\textwidth]{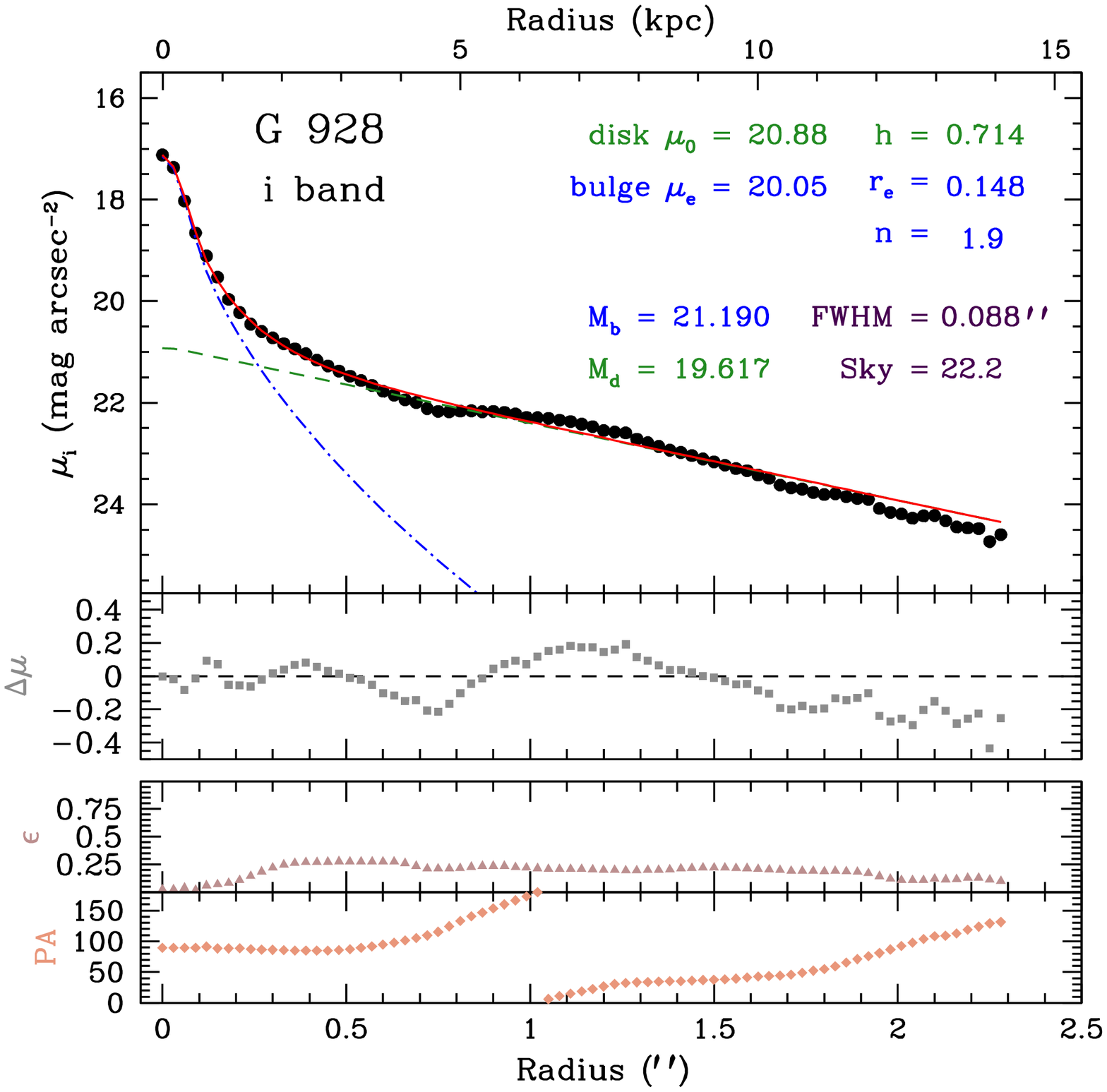}
\includegraphics[width=0.47\textwidth]{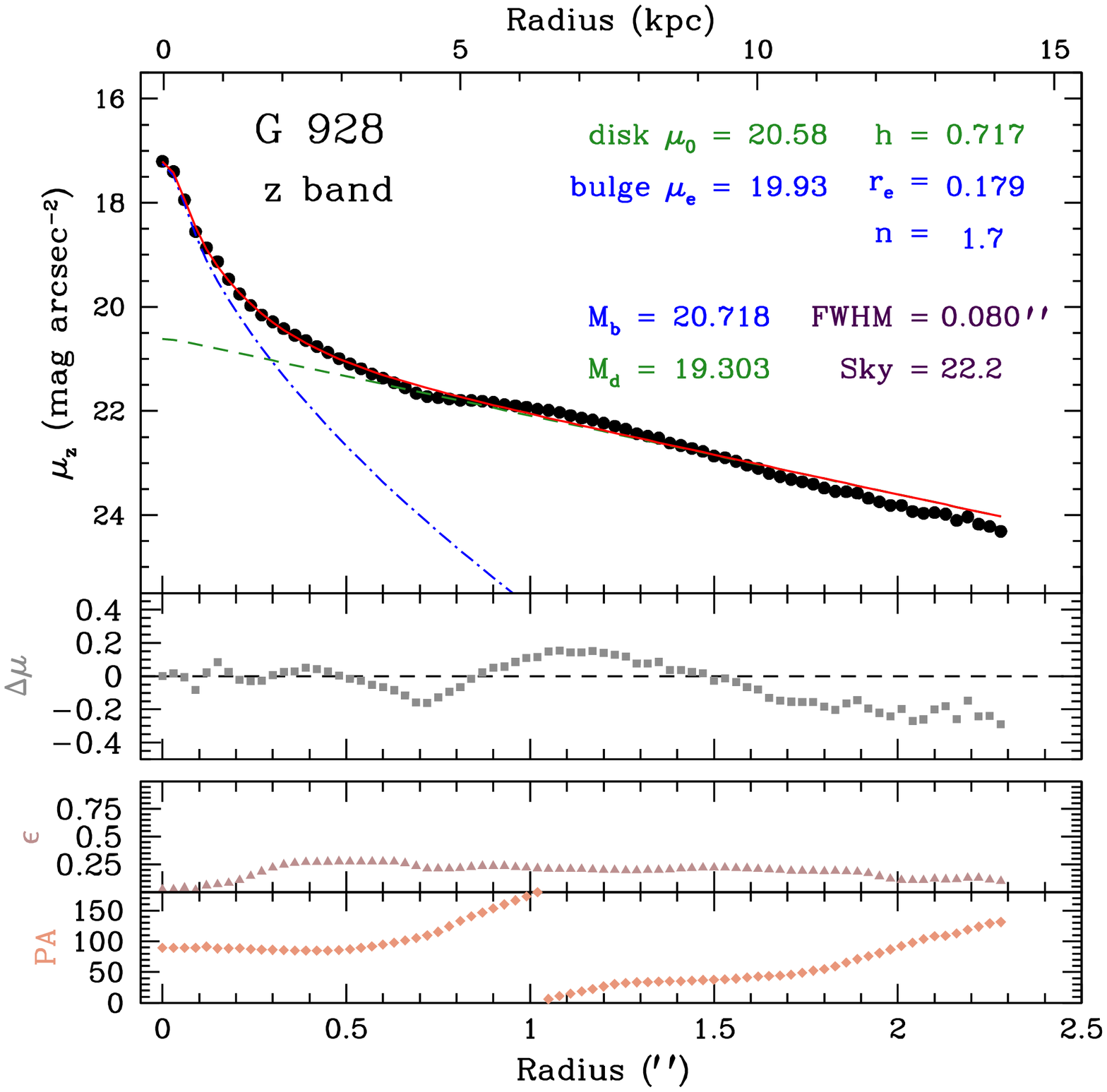}
 \caption{Example of B/D decompositions for the $z$\,=\,0.512 spiral galaxy 
          G928 in all 4 GOODS bands.  In each figure, the upper panel shows 
          the measured SB profile (solid black circles), the bulge fit 
          (blue dashed-dotted line), the disk fit (green dashed line), 
          and the total bulge+disk fit (solid red line).  All fits are 
          seeing-convolved using the best selected PSF values (see text).
          The five structural fit parameters are shown in the upper right
          corner, where the scales ($h$ \& $r_e$) are in arcseconds.
          The second panel shows the fit residuals where $\Delta \mu(r)
          \equiv fit(r)-data(r)$.  The bottom two panels show the
          run of ellipticity and position angle from the isophotal fits.  
          Note that these are the same for
          all four bands as the $z$-band fits are used for the profile
          extraction of the other 3 bands.}
\label{fig:BDdecomps}
\end{figure*}

Figure~\ref{fig:BDdecomps} shows an example of the best fit profiles
and the fit residuals derived for all four bands of the $z$\,=\,0.512 spiral 
galaxy G928. The five structural fit parameters are shown in the upper right
corner, the bottom two panels show the run of ellipticity and position angle 
from the isophotal fits.  Note that these are the same for all four bands as 
the $z$-band fits are used for the profile extraction of the other 3 bands.

\subsection{Total Magnitudes and Spheroid Colors}
\label{sec:totmagcol}

A key goal of our study is construction of the color-redshift
relation for bulges in our sample. We will thus explore in detail the
relationship between the colors of the spheroidal component and those
determined using apertures. We likewise wish to derive integrated
magnitudes so that we can more accurately determine the bulge/total 
ratios implied by our photometric analysis.

Magnitudes computed involving extrapolations will be highly sensitive to 
the actual profile shape. For example, outer disk (anti-)truncations which 
are often observed in spiral galaxies would not be accounted for with 
a single exponential disk fit.  We thus compute total galaxy magnitudes from
the photometry to the maximum observed radius with the addition of the 
extrapolation of a linear fit to the outer 20\% of the profile.  
The extrapolation typically increases the magnitude by $\simeq$\,0.1 mag, 
except when the profile is very shallow.  In cases where the extrapolation 
adds more than 0.5 mag, the galaxy is tagged and the tabulated error is 
increased accordingly.  
We correct total magnitudes for (small) Galactic foreground extinction using 
the reddening values, $A_{\lambda}$, of Schlegel, Finkbeiner, \& Davis (1998)
and interpolating to the effective wavelength of the GOODS-ACS filters.  
The adopted values for both the Northern and Southern fields are listed in
Table~\ref{tab:ACS}.

Similarly, the contribution to the total light at large radii (\ie\ 
extrapolated to infinity) in the \sersic\ profile changes significantly as 
a function of $n$ (see Fig.~2 of Mac03), thus a small error on the fitted 
$n$ could lead to an error in the total magnitude.  While this error will 
typically not be significant for the total spheroid magnitude, the relative 
differences between passbands can propagate to large errors on the colors
whose dynamical ranges are small (of order $\sim$\,1.5~mag for rest-frame 
$B-R$ and smaller for shorter wavelength baselines).  To minimize the 
potentially large errors due to extrapolation, while still allowing 
for color gradients, we compute the colors of our fitted spheroids using the 
best fit model parameters, but only integrating out to the radius corresponding
to the best fit $r_e$ in the $z$-band.  These colors should be equivalent to 
aperture colors within 1\,$r_e,z$ for the single component galaxies, but 
can differ significantly from aperture colors for spiral bulges as they 
account for the (differential) contamination from disk light.  

Bulge-to-total ($B/T$) luminosity ratios can be computed either from the 
fits alone, or as a combination of the bulge fit and the (nearly) 
non-parametrically measured galaxy magnitude discussed above.  We adopt 
the latter approach as we have found that extrapolating a disk fit often 
overestimates its contribution. 

\begin{deluxetable}{c|cc|cc|c}
\tabletypesize{\normalsize}
\tablewidth{0pt}
\tablecaption{Adopted quantities specific to the four GOODS-ACS filters
\label{tab:ACS}}
\tablehead{
\multicolumn{1}{c|}{Filter} &
\multicolumn{2}{c|}{$A_{\lambda}$ (mag)\tablenotemark{a}} &
\multicolumn{2}{c|}{PSF (arcsec)\tablenotemark{b}} &
\multicolumn{1}{c}{Sky\tablenotemark{c}} \\
\multicolumn{1}{c|}{} &
\multicolumn{1}{c}{North} &
\multicolumn{1}{c|}{South} &
\multicolumn{1}{c}{$\mu$} &
\multicolumn{1}{c|}{$\sigma$} &
\multicolumn{1}{c}{\magarc} 
}
\startdata
F435W  & 0.0531 & 0.0336 & 0.071 & 0.006 & 23.4 \\
F606W  & 0.0354 & 0.0233 & 0.080 & 0.006 & 22.7 \\
F775W  & 0.0254 & 0.0165 & 0.088 & 0.003 & 22.2 \\
F850LP & 0.0165 & 0.0114 & 0.093 & 0.003 & 22.2
\enddata
\tablenotetext{a}{{\small Galactic foreground extinction using the reddening 
                  values, $A_\lambda$, of Schlegel, Finkbeiner, \& Davis (1998)
                  and interpolating to the effective wavelength of the 
                  GOODS-ACS filters.}} 
\tablenotetext{b}{{\small ACS-PSF FWHM mean and standard deviations from 
                  SExtractor (Bertin \& Arnouts 1996) measurements of Tiny Tim 
                  simulations (Rhodes \etal\ 2007) of 2500 artificial stars in 
                  the GOODS field.}}
\tablenotetext{c}{{\small Nominal sky values for space-based observations.}}
\end{deluxetable}

\subsection{K-correction}
\label{sec:kcor}

The final step in deriving useful photometric quantities is the application 
of a $k$-correction to enable the compilation of rest-frame measures.

Often, a single function K($z$,$X-Y$), where $X-Y$ is the observed color
in (optical) bands $X$ and $Y$, is derived and applied to entire samples, 
regardless of the galaxy type.  For example, T05 provide
transformations from the GOODS ACS filters to rest-frame Landolt $B$ and
$V$ magnitudes (to within a few hundredths of a magnitude) for the redshift 
range 0\,$\le$\,$z$\,$<$1.25.  Two filters are used in each transformation 
depending on where the 4000\,\AA\ break falls at the given redshift.  
Gebhardt \etal\ (2003) present a similar transformation, derived from 43 
empirical SED templates, to convert from observed {\it HST} F606W and F814W 
magnitudes to rest-frame $U$ and $B$. 

Alternatively, Blanton \& Roweis (2007) provide an IDL code, {\tt kcorrect},
for computing $k$-corrections which fits linear combinations of a set
of model-based templates, including provision for emission lines and 
dust extinction, to the observed galaxy colors.  {\tt kcorrect} infers the 
underlying SEDs for galaxies over a range of redshifts by requiring that 
their SEDs be drawn from a similar population.  The resulting 
reconstructed SEDs can then be used to synthesize the galaxy's rest-frame 
magnitude in any bandpass. 

\begin{figure}
\begin{center}
\includegraphics[width=0.48\textwidth]{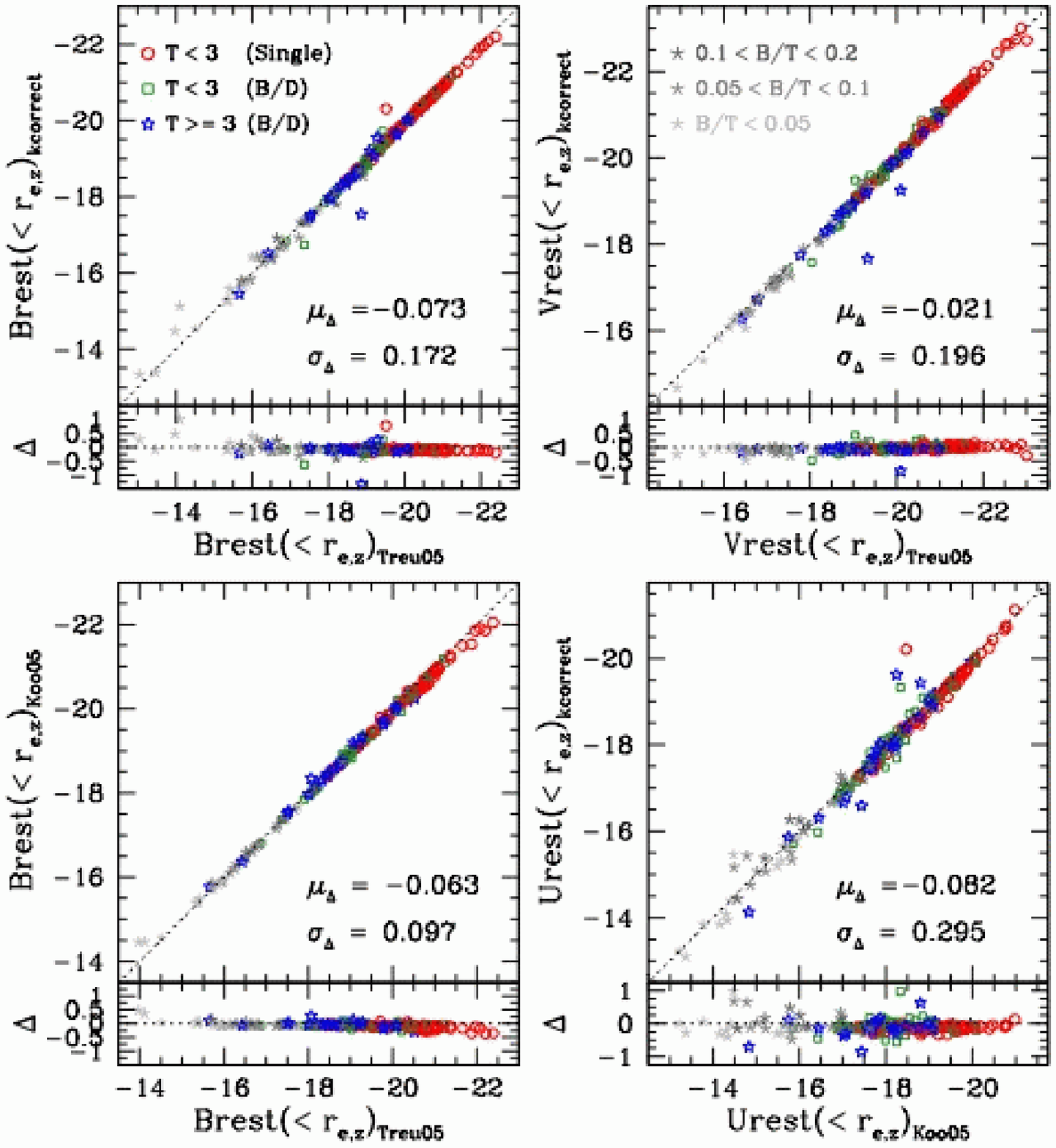}
\caption{Comparison of rest-frame spheroid magnitudes within the $z$-band
         effective radius using three
         different methods for deriving k-corrections: Treu05, 
         Gebhardt \etal\ (2003) (also provided by Koo05 and labeled
         as such in the figure), and {\tt kcorrect} (Blanton \& Roweis 2007, 
         see text for descriptions).  Point types and colors are 
         indicated in the upper left corners of the top panels.
         Gray shaded asterisks indicate galaxies with $B/T$\,$<$\,0.2 which 
         are excluded from the FP analysis in \S~\ref{sec:spec_results}.  
         The bottom panel in each figure shows the residuals and their
         respective mean and standard deviations are printed in
         the lower right corner of the upper panels.
         \label{fig:kcorrect}}
\end{center}
\end{figure}

The different methods for deriving k-corrections will have their own set of 
advantages and drawbacks depending on the application.
As a consistency check, in Figure~\ref{fig:kcorrect} we compare the results 
from all three methods mentioned above.  T05 provide transformations to 
rest-frame $B$ and $V$ while Gebhardt \etal\ (2003) give transformations to 
rest-frame $B$ and $U$.  The comparison between the three methods give 
reasonably consistent results for the $B$ and $V$-band k-corrections.  
The results are less consistent in the $U$-band.  This is not surprising
as it is more strongly dependent on the observed $B$-band magnitudes
which have the least-well-determined profiles.  An additional investigation
into the templates used in the {\tt kcorrect} code implies that the dust
extinction prescription employed also adds to this scatter.  Regardless,
we can be fairly confident in the absolute $B$ and $V$ rest-frame magnitudes
derived from any of the three methods for the current sample.

\subsection{Structural Parameters}
\label{sec:corr}

\begin{figure}
\begin{center}
\includegraphics[width=0.48\textwidth]{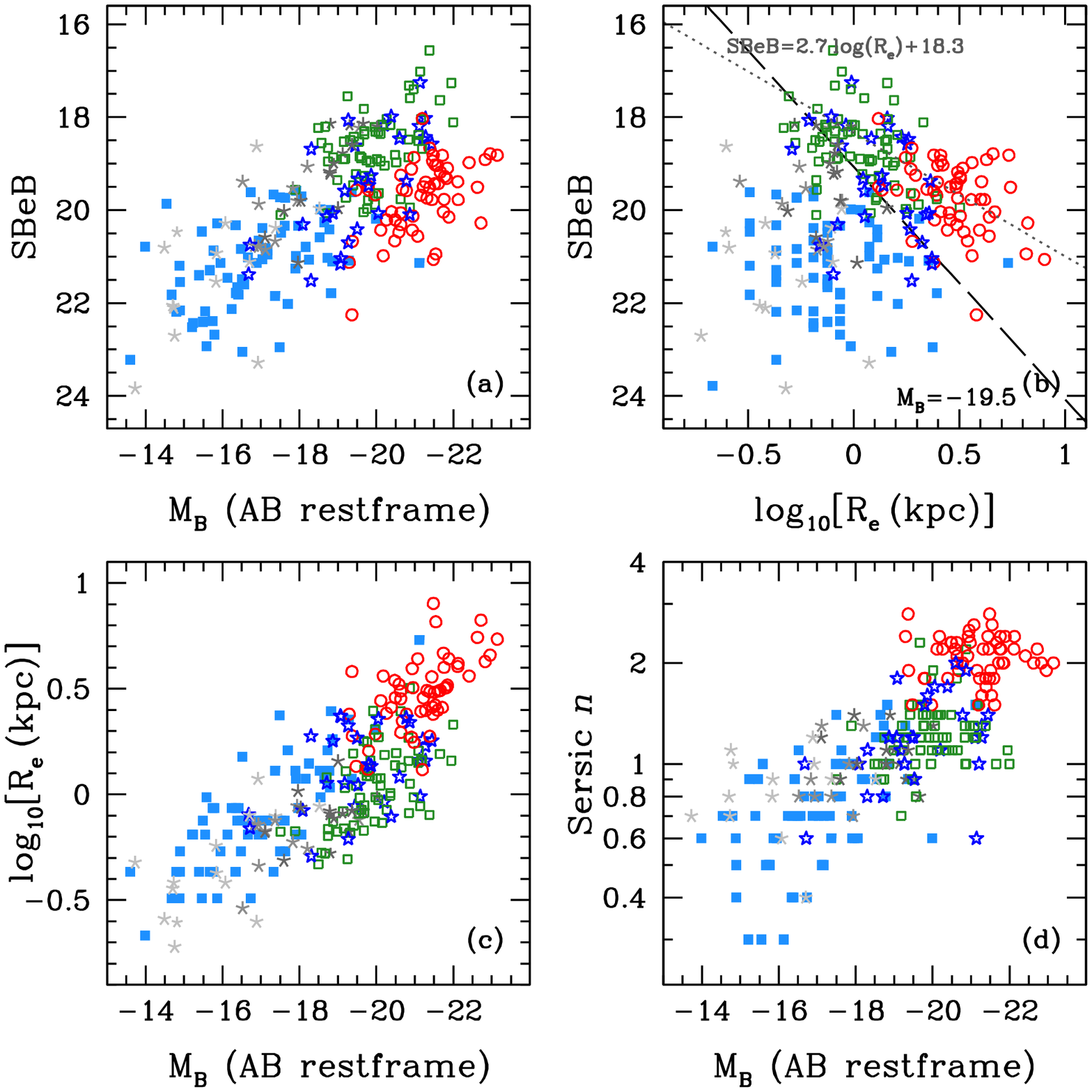}
\caption{Correlation of photometric parameters for the spheroidal 
         components of our intermediate-$z$ galaxies (point types and colors 
         are as in Fig.~\ref{fig:kcorrect}), and the local, mostly late-type,
         bulge sampleof Mac03 (light blue squares).  The dotted line in 
         panel (b) is a linear fit to the galaxies with M$_{B}<-19.5$ only 
         (\ie\ the Kormendy relation for this sample) and the
         dashed black line is the line of constant magnitude M$_{B}=-19.5$
         (see text for discussion).}
         \label{fig:Corr_Local}
\end{center}
\end{figure}

As a final check on our decompositions and sample characteristics, we
examine here the structural parameters of our sample of spheroids with
comparisons to local data.  Figure~\ref{fig:Corr_Local} compares
several correlations between our sample and the Mac03 sample of local
bulges. The reader is advised that evolutionary effects (\eg\
luminosity and/or size evolution) and selection effects (redshift dependent
limiting magnitude) have the be kept in mind when interpreting these
correlations and comparing them to local samples. Strong correlations
are seen between luminosity and both the effective radius [panel (c)]
and \sersic\ $n$ parameter [panel (d)].  At a given magnitude, bulges
in low-$B/T$ spirals have smaller effective radii [panel c], and
there is some indication that local bulges have slightly larger $R_e$
at a given luminosity than those at intermediate redshift.  Trujillo
\etal\ (2006) have likewise claimed significant size evolution in the
rest-frame optical for bulges in low-concentration galaxies between
$z$\,=\,2.5 and today.

The correlation between luminosity and the effective SB within $R_{e}$ 
[panel (a)] is weaker and it appears that the single component Es have lower 
SBe for a given 
M$_B$ than the spiral bulges. Graham \& Guzman (2003) point out that this is a 
manifestation of variations in profile shape as a function of total luminosity
[shown in panel (d) here and see Fig.~12 of Graham \& Guzman 2003]. 
If one plots the central surface brightness of the bulge, $\mu(0)_b$ rather 
than SBe, the relation becomes more linear with significantly less scatter 
over the full luminosity range.

%The local bulges do tend to
%have lower SB but similar $r_e$ which could indicate passive fading...***).
%This could also be a result of the biases between the two samples, the local
%sample being heavily weighted towards smaller $B/T$ galaxies.

Finally, panel (b) plots the so-called Kormendy (1977) relation between 
effective radius and SBe.  At first glance, there does not appear to be any 
correlation.  However, for the most luminous spheroids with 
M$_B$\,$<$\,$-$19.5, denoted by the dashed line, a reasonably tight relation 
emerges.  A linear fit (dotted line) with a slope of 2.7 is in rough agreement 
with previously determined values (\eg\ La~Barbera \etal\ 2003 find a 
constant slope of 2.9 for spheroidal galaxies out to $z$\,=\,0.64).  The 
difference in slope and the large scatter observed here might be expected, 
given the broad redshift range.  Fainter spheroids deviate from this relation 
having smaller $R_e$ and/or lower SBe.  The transition at M$_B$\,=\,$-$19.5 
roughly corresponds to 
the luminosity below which spheroid profiles are best characterized by a 
\sersic\ $n$\,$\lesssim$\,1 (exponential) shape.  Thus this represents the 
transition between cored ($n$\,$<$\,1) and cuspy ($n$\,$>$\,1) profile shapes.

Deviations in the Kormendy relation for smaller systems have been
noted before in the context of differences between Es and dEs (\eg\
Kormendy 1985).  More recently this observation has extended to the
bulges of local spiral galaxies in Ravikumar \etal\ (2006), who note
that the fainter bulges appear more like the dEs, while the brightest
bulges appear to follow the relation defined by the E/S0s.  This also
complies with the observation that both dEs and small bulges are more
supported by rotation than their larger counterparts.  The temptation
to interpret the Kormendy relation as an evolutionary diagram or an
indication of different formation mechanisms, however, has been
disregarded due to the mixture between different types, and the
changes in the profile shapes discussed above.

%One possibility is that the fundamental parameter tracking the
%evolutionary (merger/interaction) history of spheroids is the \sersic\
%$n$.  {\bf what does the previous sentence exactly mean, do we need to
%keep it in? } 
%Panel (d) indicates that, at a given luminosity, our
%bulges tend to have lower $n$ than pure spheroidal galaxies, but with
%a larger range.  This could be because Es are at a more advanced stage
%of evolution, while the bulges with large luminosities for their
%profile shape are still in transition due to recent merger/interaction
%activity.

Prior to our photometric analysis below, we summarize in 
Table~\ref{tab:photpars} our derived photometric parameters for the 
spheroidal components of our sample galaxies. 

\section{Spheroid Colors}
\label{sec:colors}

We now turn to the first component of the analysis of our dataset which is 
concerned with addressing the photometric evolution of our intermediate 
redshift bulges. Both EAD and Koo05 presented color-redshift diagrams but 
using different analysis methods and different sample selection criteria.  
Perhaps as a result of this, they arrived at rather different conclusions.  
Our goal is to understand the source of any discrepancy as well as to 
intercompare the various approaches.

The dispersion in color at a given redshift is an important indicator
of the diversity of the bulge population and hence the likelihood of continued
star formation and mass assembly. Moreover, a comparison with similar data for
low luminosity spheroidal galaxies will give an indication of the extent to
which growth is primarily driven by processes internal to spiral galaxies. 
%In the analysis below, following EAD, we will use the observed colors.

EAD based their analysis on aperture colors (within a fixed size of 5\% 
relative to the isophotal radius) and found that their bulges displayed a 
large scatter toward the blue in observed $V-I$, with few being as red as 
E/S0 counterparts at a given 
redshift.  They concluded that bulges underwent recent periods of growth and 
associated rejuvenation.  However, it is possible that the blueward scatter 
arises, in part, from disk contamination.  As we have undertaken careful
decomposition of the light profiles with the improved ACS data, here 
we examine the equivalent color trends with redshift for our sample.

\subsection{Comparison with Aperture Colors}
\label{sec:apmags}

\begin{figure*}
\begin{center}
\includegraphics[width=0.87\textwidth,bb= 18 414 592 718]{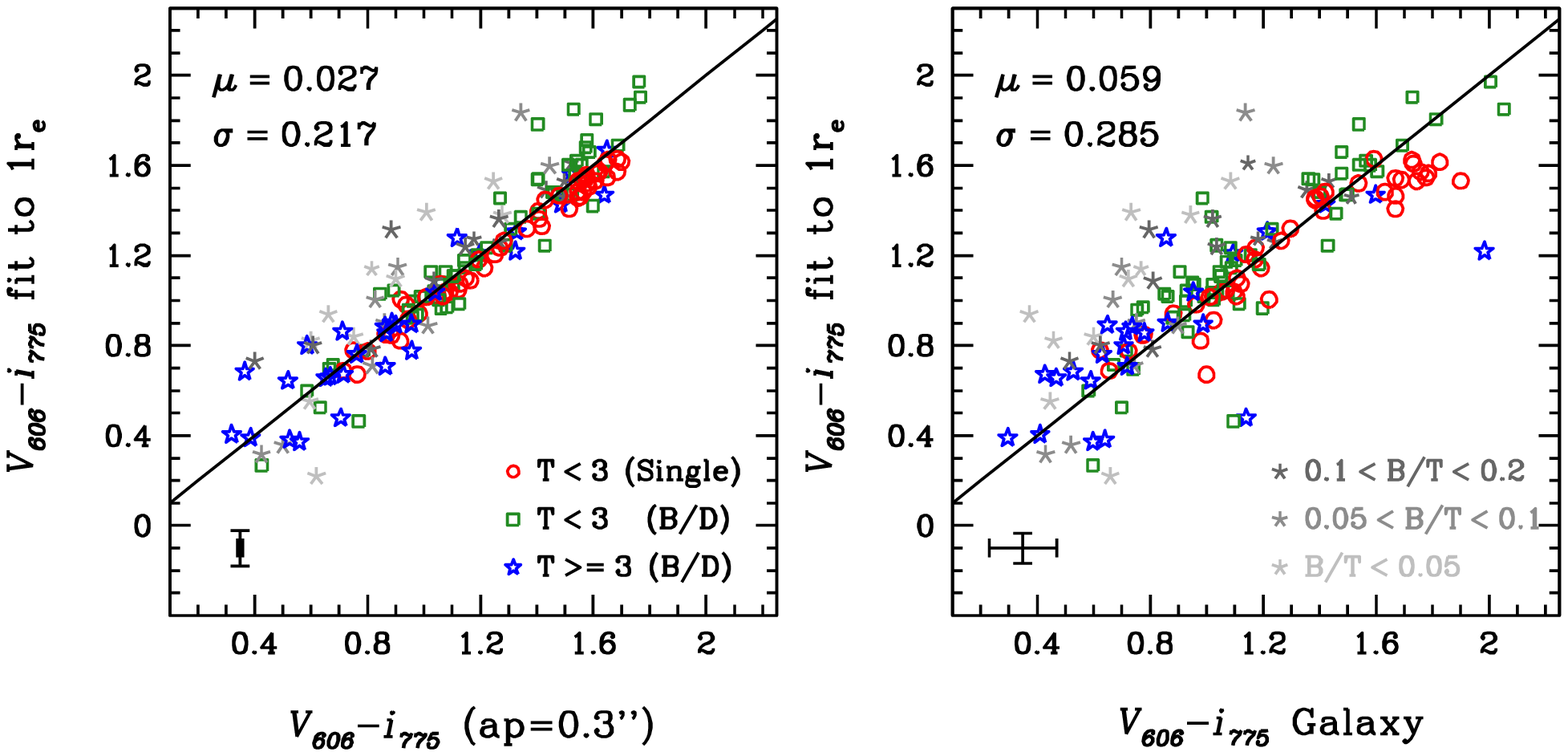}
\caption{Comparison of observed F606W$-$F775W ($V-i$) color 
         from the model fits with the 0\farcs3 aperture color [{\it left}] and 
         total galaxy color [{\it right}].  Point types and colors are 
         indicated 
         [{\it bottom right corners}] and the average errors are shown 
         [{\it bottom left corners}].  The mean difference and standard 
         deviation are given [{\it top left corners}]         
         and the solid line is the one-to-one relation.}
\label{fig:apmag}
\end{center}
\end{figure*}

Bulge colors computed from B/D decompositions are extremely sensitive to
the fits themselves. To understand some of these effects, and to calibrate
some of the contamination that might have affected the EAD analysis, in
Figure~\ref{fig:apmag} we compare our fitted bulge $V-i$ colors (measured
out to 1\,$r_e$) to the central 0\farcs3 aperture colors  [{\it left panel}] 
and those of the total galaxy [{\it right panel}].   The ACS PSF is 
$\lesssim$\,0\farcs09, so the 0\farcs3 aperture colors should not be 
seriously affected by PSF mismatch between bands (which are $\Delta$(PSF) 
$\lesssim$\,0\farcs008).  
%The mean value of $r_e$ is $\sim$0\farcs3). 
Bulge fits are not shown if the fit magnitudes in either band 
were fainter than 30\,mags (as these galaxies are fundamentally bulgeless).
The fit and aperture $V-i$ colors are very similar, but there is some 
scatter such that the model fit colors are often somewhat redder than the
aperture colors, particularly for galaxies that have a disk component
(squares, stars, and asterisks).  This probably arises from (bluer) disk 
contamination 
within the aperture.  The fit colors are also often redder than the total 
galaxy, consistent with negative (bluing outward) gradients.
Nonetheless, the above comparison confirms that our photometric 
decomposition is stable and yields reasonably-consistent bulge colors.

\subsection{Colors vs. Redshift}
\label{sec:colorvsz}

\begin{figure}
\begin{center}
\includegraphics[width=0.48\textwidth]{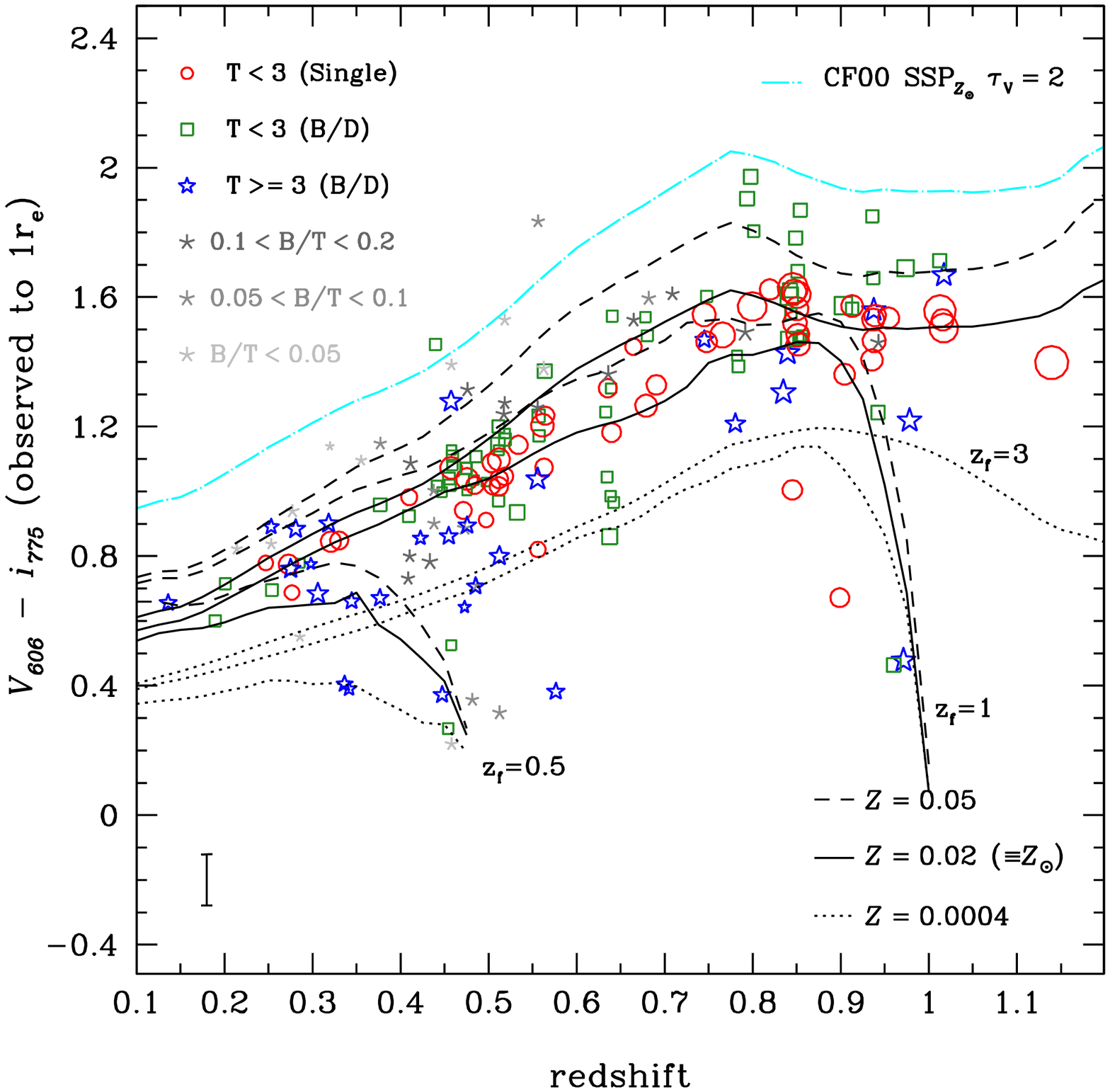}
\caption{Observed $V-i$ spheroid color (measured to $1\,r_e$ from the light
         profile modeling) versus redshift.  
         Point types and colors are indicated in upper left and the
         point size is proportional to the rest-frame $B$ magnitude of the
         spheroidal component.  The black
         lines represent BC03 models of single burst stellar populations (SSPs)
         formed at three redshifts ($z_f$\,=\,0.5,1,3) that evolve
         passively.  SSPs for three metallicities are shown
         ($Z$\,=\,0.0004, 0.02, \& 0.05, where $Z_{\odot}$\,=\,0.02).  The cyan
         dashed-dotted line is a $z_f$\,=\,3 model with solar metallicity that
         has extinction included using the dust models of
         Charlot \& Fall (2000) and 
         $\tau_V$\,=\,2.  The average error is shown as the error bar
         [{\it bottom left corner}].
         \label{fig:ImZz}}
\end{center}
\end{figure}

We now return to the redshift dependence of our observed spheroidal component 
colors. Figure~\ref{fig:ImZz} plots the observed $V-i$ colors as a function of 
redshift. The black lines represent model tracks of a passively-evolving 
single burst population for three formation redshifts ($z_f$\,=\,$0.5,1,3$) 
and three metallicities ($Z$\,=\,$0.004,0.02,0.05$).  All stellar population 
models considered are those of Bruzual \& Charlot (2003, hereafter BC03).  The 
dash-dotted cyan line is a $z_f$\,=\,3 model with solar metallicity that has 
extinction included using the dust models of Charlot \& Fall (2000) 
with $\tau_V$\,=\,2 (corresponding to an $A_B$\,$\sim$\,1.5).  

As observed by many authors (\eg\ T05), most of the elliptical (\ie\ single 
component) galaxies (circles) are consistent with a single burst model with 
$z_f$\,$\simeq$\,3.  Those Es which deviate from this trend, indicating a
more recent formation/activity, tend to be the lower-luminosity systems
(as indicated by the point size).  Many of the two-component galaxies 
(squares, stars, and asterisks) also follow this
trend, but a more significant number scatter strongly to the blue and a 
few lie redwards of the dust-free models (even if a very high metallicity is 
assumed).  The latter may arise from dust extinction, but the large scatter 
to the blue is consistent with the observations of EAD (see their Fig.~4).  
However, unlike EAD we do see a number of bulges that are just as red as 
the E/S0s.  These could represent the red bulges observed by Koo05 and,
indeed, if a similar cut is made on the observed $i$-band bulge magnitude
as that of Koo05, the remaining spheroids in the same redshift range 
($z$\,$\sim$\,0.8) are red and consistent with old single-burst populations.

Thus it seems that there is no real disagreement between the studies
of EAD and Koo05.  While there are differences in their analysis techniques,
our own internal tests suggest these are unlikely to be significant to a
level that could cause such a discrepancy (\S~\ref{sec:apmags}).  The  
fundamental difference is in the respective bulge samples: EAD sampled bulges 
drawn from a full range of Hubble types with the morphologically-classed 
spirals likely skewed towards later-type bulges.  Koo05 selected bulges only 
from the bright end of the bulge luminosity function which, according to 
Figure~\ref{fig:ImZz}, are predominantly red and passively-evolving.

In summary, therefore, our results confirm the diversity of bulge colors
at intermediate redshift originally identified by EAD, although the bulk of 
the more luminous systems more closely track the relationship defined by the 
larger spheroidal galaxies. The question thus arises whether we are looking 
at two different populations each following a distinct formation path, or 
rather a more continuous mode of assembly for the spheroidal components
where, for example, additional growth is occurring but it has proportionally
less effect for the more massive systems.

Phrased another way, we can ask whether the  spheroidal components of 
all galaxies (spirals, S0s, and Es) {\it at a given mass} are consistent in 
terms of their evolutionary paths.  It is hard to address such a question from 
colors alone as the bulge luminosity or $B/T$ ratio represent poor proxies for 
the mass of the system. To address this more fundamental question we 
now turn to considering the combined set of dynamical and photometric 
information for our sample. By constructing the Fundamental Plane for
a subset of our bulge components and comparing the relation to that
now well-mapped for spheroidal galaxies of various masses (T05), we can 
more accurately determine the mass growth rate of luminous bulges.

\section{Spectroscopic Data and Analysis}
\label{sec:spec}

Our DEIMOS/Keck spectroscopic data are used to derive central stellar
velocity dispersions which, together with the photometric parameters
derived in \S~\ref{sec:photom}, enable us to construct the FP for the
spheroidal components of E, S0, and spiral galaxies and to compare
their recent stellar mass growth rates.

As discussed earlier, the spectroscopic data were taken in two
distinct campaigns.  The first is that described in detail in T05,
relating to both E/S0s and bulge-dominated spirals. The second
campaign was designed to increase the sample of spiral bulges for the
current study, particularly in the redshift range
0.1\,$<$\,$z$\,$<$\,0.7.  Targets were selected with Sa+b (T=3)
morphologies from Bundy \etal\ (2005)'s catalog to a limit of
$z'_{AB}$\,=\,22.5 corresponding to spirals with luminosities
$-18.5$\,$<$\,$M_B$\,$<$\,$-20.5$.

In the following sections we describe how the new spectroscopic observations
were undertaken and discuss the extraction of central velocity dispersions 
in the context of the earlier work by T05.

\subsection{Observations \& Data Reduction}

The spectroscopic observations used here were taken with the Deep
Imaging and Multi-Object Spectrograph (DEIMOS) on the Keck II
Telescope.  The data set comprises a total exposure time of 5\,--\,7
hours for each of 4 masks (three in the Northern GOODS field, one in
the Southern GOODS field) spanning three observing campaigns in March
2004, December 2004, \& March 2005.  A further campaign on GOODS-S in
October 2005 was unsuccessful due to bad weather and yet another
campaign scheduled in October 2006 was canceled due to a major
earthquake which ceased all observatory operations. Most of the data
was obtained in seeing conditions in the range of 0.8--1.0\arcsec\ 
(FWHM).

We used the 1200 line gold coated grating blazed at 7500\,\AA\ and
centered at 8000\,\AA, which offers high throughput in the red, a
pixel scale of 0\farcs1185\,$\times$\,0.33\,\AA\ and a typical
wavelength coverage of 2600\,\AA.  A resolution of $\sim$\,30~\kms\
was determined from arc lines and sky emission lines for the
1\arcsec-wide slitlets. In deriving velocity dispersions, the
resolution was measured independently for each slit as discussed
below.

The raw spectra were reduced using the DEIMOS reduction pipeline as
developed by the DEEP2 Redshift Survey Team (Davis \etal\ 2003).  After 
extracting
one-dimensional sky-subtracted spectra, we estimated the redshift and
spectral type for each galaxy.  {\tt Zspec}, an IDL-based software
package also written by the DEEP2 Team, was used to evaluate the
accuracy of the assigned redshift and a quality criterion $q$ ($-$1
for stars, 2 for possible, 3 for likely, and 4 for certain) was
assigned.  All spectra in our sample have a redshift quality $\ge$\,3
and most have $q$\,=\,4.

Noting that the minimum S/N/\AA\ ratio for accurate central
dispersions (hereafter $\sigma_0$) is $\gtrsim$\,10 (see, \eg, Treu
\etal\ 2001), coupled with the desire to extract radial kinematic
profiles, the spectra were coadded in radial bins to ensure a minimum
S/N/\AA\ of 5.  Failure to obtain accurate $\sigma_0$ measurements
occurred due to low S/N spectra, bad sky subtraction, unfortunately
placed absorption features with respect to the atmospheric A\&B
absorption bands, or bulges whose velocity dispersions probably lie
below our resolution limit ($\sim$\,30~\kms). We detail below our
methodology for deriving these dispersions.

% Note to self:  Mgb enters/exits Bband at z=0.314/0.355
%                Mgb enters/exits Aband at z=0.454/0.505
%                NaD enters/exits Bband at z=0.153/0.190
%                NaD enters/exits Aband at z=0.276/0.322

\subsection{Measuring Velocity Dispersions}\label{subsec:veldisp}

The velocity dispersion and rotation profiles of our spectra were
measured using the well-tested Gauss-Hermite Pixel Fitting algorithm
provided by van~der~Marel (1994)\footnote{Obtained at {\tt
http://www-int.stsci.edu/$\sim$marel/software/}}.

Prior to analysis, the 1D galaxy and template spectra are matched in
rest-frame wavelength range and re-binned onto a logarithmic wavelength
scale.  After convolving a suitable template spectrum with Gaussian
velocity profiles of various dispersions, the best-fitting dispersion
for each template is found (see Fig.~\ref{fig:kinem} [{\it middle panels}]).
The goodness-of-fit is measured via the \chisqr\ statistic, weighted
by the observational errors. The parameters determined for a given
spectrum include the model-to-galaxy relative line strength
($\gamma$), the mean velocity ($v$), the stellar velocity dispersion
($\sigma$), and their respective errors. An advantage of the pixel
fitting technique is the ability to mask, via zero weighting,
undesirable regions of the spectrum, for example those containing
emission lines, gaps between CCDs, and atmospheric features.

Ideally, the template spectra should be an excellent match to the spheroid 
stellar population (Gonz{\'a}lez 1993).  From our previous analysis of
GOODS E/S0s (T05), we have an appropriate set of high-resolution stellar 
templates that range from spectral types G0III to K5III.  Recognizing that one
of the goals of this study is to address the conjecture that bulges contain
younger stellar populations, potentially spanning a range of metallicities, 
in addition to the observed red giant templates, we also include a selection 
of 36 templates from the high-resolution synthetic spectra of 
Coelho \etal\ (2005)\footnote{Obtained at 
{\rm http://www.mpa-garching.mpg.de/PUBLICAT-IONS/DATA/SYNTHSTELLIB/synthetic\_stellar\_spectra.html}} 
spanning the range T$_{eff}$\,=\,4000--7000\,K, log$_{10}$(g)\,=\,15--45, 
[Fe/H]\,=\,$-$2.0 to $+$0.5.  The best-fit template is selected primarily on 
the \chisqr\ figure of merit, but some weight is given to the line-strength 
parameter $\gamma$ (which should equal unity for a perfect spectral match 
between galaxy and template) and the formal error of the fit parameters.  
Indeed, in many cases, the bulge spectra are best matched by the ``younger'' 
model template spectra.

Out of 45 spiral bulges observed in this second observing campaign, 
successful central dispersions were obtained for 23 galaxies. 7 had 
marginal measurements, 13 failed to yield dispersions, and 2 had 
corrupt spectra.  Of the 23 galaxies with successful $\sigma_0$ measurements,
only 8 had $B/T$\,$>$\,0.2 and thus could be included in our FP sample.
The velocity dispersion fits for these 8 galaxies are shown 
in Figure~\ref{fig:kinem} [{\it middle panels}].

Finally, of the 8 galaxies observed in common in both campaigns, 5 had
successful $\sigma$ measurements in both and can be compared as a consistency 
check.  Four of the galaxies had $\delta\sigma/\sigma$\,$<$\,0.08 with no 
systematic trend.  The fifth galaxy (G928) had $\delta\sigma/\sigma$\,=\,0.54, 
but this is not surprising nor worrisome, given that the spectrum is dominated 
by young stellar populations and emission lines, while the measurement
procedure in T05 was only appropriate for older stellar populations,
starting with the choice of red giant stellar templates.  We thus conclude
that the measurements are consistent within the error bars, when they
are comparable.

\subsection{Aperture Corrections \& Disk Contamination}\label{sec:sigmacor}

Disk contamination could potentially present a source of uncertainty
when interpreting the central velocity dispersions for these spiral
galaxies as representative of the bulge component. In addition to
template mismatch biases, even a modest systematic rotation might lead
to an overestimated value for the smaller bulges.  The spatial
resolution of the observations are limited by the pixel scale
(0\farcs1185 per pixel), the physical size of the slit width
(1\arcsec) and the typical seeing FWHM during the time of the
observations ($\sim$0.8--1\arcsec).  As mentioned in
\S~\ref{sec:sample}, we have limited our bulge sample to those with
$B/T$\,$>$\,0.2 to ensure that the bulge light dominates in the center.
However, the presence of a disk could still influence the measurement
and interpretation of $\sigma_0$.

A proper kinematic decomposition of our data into contributions from
the bulge, disk, and dark halo components would require full dynamical
modeling (see Widrow \& Dubinsky 2005).  Conceivably this could be
possible for a few galaxies where we have adequately sampled kinematic
profiles, but is beyond the scope of the current analysis.  Our main
goal is therefore to estimate the likely degree to which disk light is
influencing our measurements.

For those spectra where resolved kinematic data can be extracted (see
Fig.~\ref{fig:kinem} [{\it right panels}]), we can attempt to estimate the
disk contribution to the true bulge dispersion, $\sigma_{0,b}$ from
the rotation measured across the bulge (assuming this arises entirely
from disk component). We model this as
$$\sigma_{0,b} = \sqrt{\sigma_0^2 - (V(r_{ap})/\sin\,i)^2},$$ where
$V(r_{ap})$ is the velocity at 1\,$r_e$ or at a radius of 2 pixels (at
0.1185\arcsec/pix), whichever was larger. The velocity was typically
very small, with the correction factor $\sigma_{0,b}/\sigma_{0}$
ranging from 0.7--1, with a mean value of 0.95.  We can also monitor
the possible effects of disk contamination in the bulge spectra by
considering the inner dispersion gradients. Inspection of
Figure~\ref{fig:kinem} [{\it right panels}] reveals that the $\sigma(r)$
profiles are largely flat within the central regions
($\lesssim$\,1\,$r_e$).  Although we do not have kinematic profiles
for all galaxies, given the bulk have near face-on inclinations, we
conclude contamination by a rotating disk component is not
significant.

Finally, we need to standardize the aperture within which the
dispersion measurement refers. Several studies have considered
$\sigma$ gradients in early-type galaxies and find
$\sigma_{r}/\sigma_{r_e} = (r/r_e)^{\alpha}$ with $\alpha = -0.04$ to
$-0.06$ (\eg\ J{\o}rgensen, Franx, \& Kjaergaard 1996, Cappellari
\etal\ 2006).  For our observations, we have a fixed slit-width of
1\arcsec, the DEIMOS scale is 0.1185\arcsec\ per pixel, and the
effective radii of our galaxies range from 0--1\arcsec, with an
average value of 0\farcs3.  As a result, although our extracted
central spectrum will be dominated by the highest SB central portion,
there will be some contribution from beyond 1--2\,$r_e$.  Adopting an
observed effective aperture of 0\farcs35, the average correction to an
$r_e/8$ aperture using $\alpha\,=\,-0.06$ is $\sigma_{r_e/8} =
1.16\,\sigma_{ap}$, where we have chosen the standard correction to an
aperture of $r_{e}/8$ for comparison of our results with other
studies.  Adopting $\alpha=-0.04$ would reduce the correction
factor to 1.10. Our correction takes into account the measured $r_e$, which
has the drawback of correlating $\sigma$ and $r_e$, but is important
given the wide range of spheroid sizes in our sample relative to the 
constant effective aperture of our observations.

\begin{figure*}
\centering
  \includegraphics{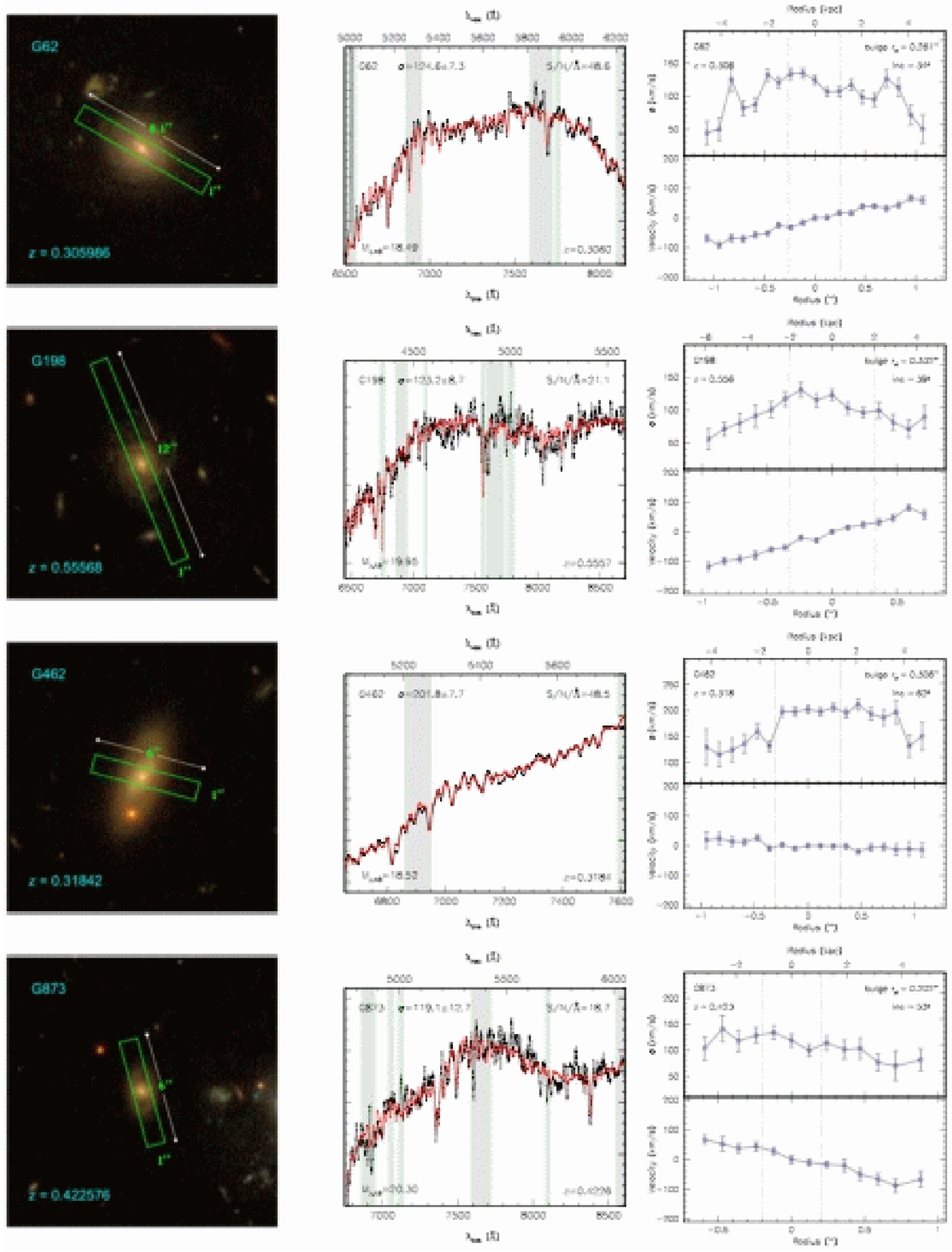}
  \caption{Spectroscopic measurements for all galaxies from the new bulge 
           sample with $B/T$\,$>$\,0.2.  {\it Left panels}: 3-color ACS image 
           and slit orientation.  {\it Middle panels}: observed central 
           galaxy spectrum (black lines) and the best-fitting stellar template 
           (red lines) convolved to the measured velocity dispersion. 
           Shaded regions (flanked by vertical green lines) indicate regions
           masked out during the fit due to an overlap with 
           either: A \& B atmospheric absorption bands, emission lines 
           (Balmer and \oiii $\lambda\lambda$5007,4959\,\AA\ in particular), 
           bad regions in the stellar templates, or the gap between the blue 
           and red spectra.  The spectra have been smoothed with a 
           10\,\AA\ boxcar filter.  Galaxy ID, $\sigma$ (in \kms), S/N/\AA\ of
	   the observation, observed $z$ magnitude, and redshift are indicated.
	   {\it Right panels}: kinematic profiles. {\it Top}:
           velocity dispersion as a function of the
           light-weighted radius (all radial bins were coadded to
           S/N/\AA\,$\ge$\,5).  {\it Bottom}: radial velocity 
           profiles (\ie\ rotation curves), shifted to zero velocity at the 
           center (determined from the peak of the spectroscopic light
           profile).  Solid black lines connect the points to guide the eye.  
           The vertical dashed lines are located at $\pm1\,r_e$ as 
           determined by the B/D decompositions.}
    \label{fig:kinem}
\end{figure*}

  \begin{figure*}[htbp]
    \centering
  \includegraphics{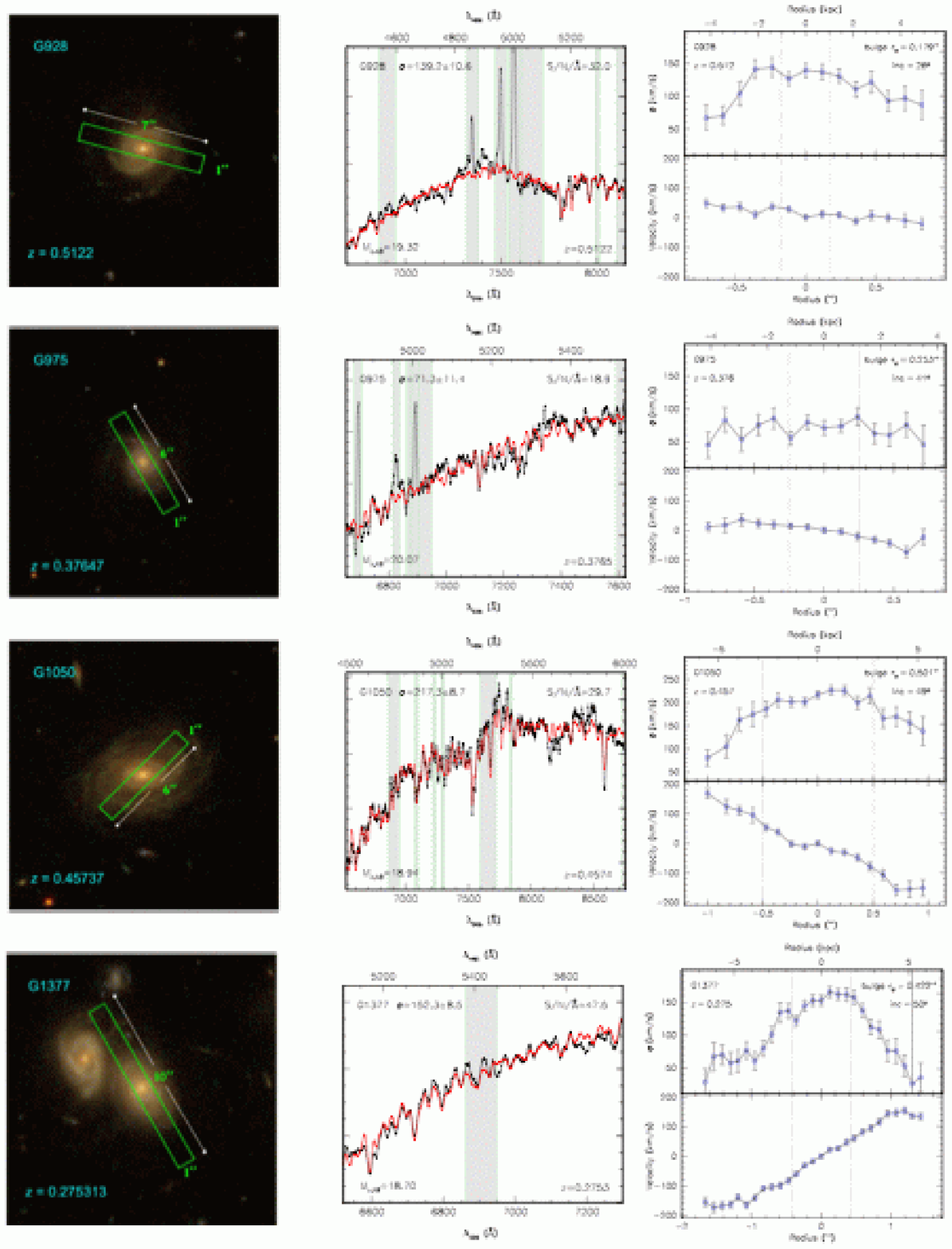}
    \figurenum{\ref{fig:kinem}}
    \caption{Continued.}
  \end{figure*}
  
\subsection{Fundamental Plane Parameters}

We now use the derived photometric and kinematic parameters to
construct the Fundamental Plane (FP) for our intermediate-$z$ sample
of the spheroidal components of galaxies with $B/T>0.2$.  The FP is
traditionally expressed in terms of galaxy effective radius, $R_e$,
the average surface brightness within the effective radius, SBe, and
the central velocity dispersion, $\sigma_0$ (\eg\ Dressler \etal\
1987; Djorgovski \& Davis 1987), related via:
\begin{equation}\label{eq:FP}
\log_{10}(R_e) = \alpha\,\log_{10}(\sigma_0) + \beta\,{\rm SBe} + \gamma,
\end{equation}
\noindent with $R_e$ in units of kpc, SBe in \magarc, and $\sigma_0$ in \kms.

\begin{figure}
\centering
\includegraphics[width=0.48\textwidth,bb=58 184 552 658]{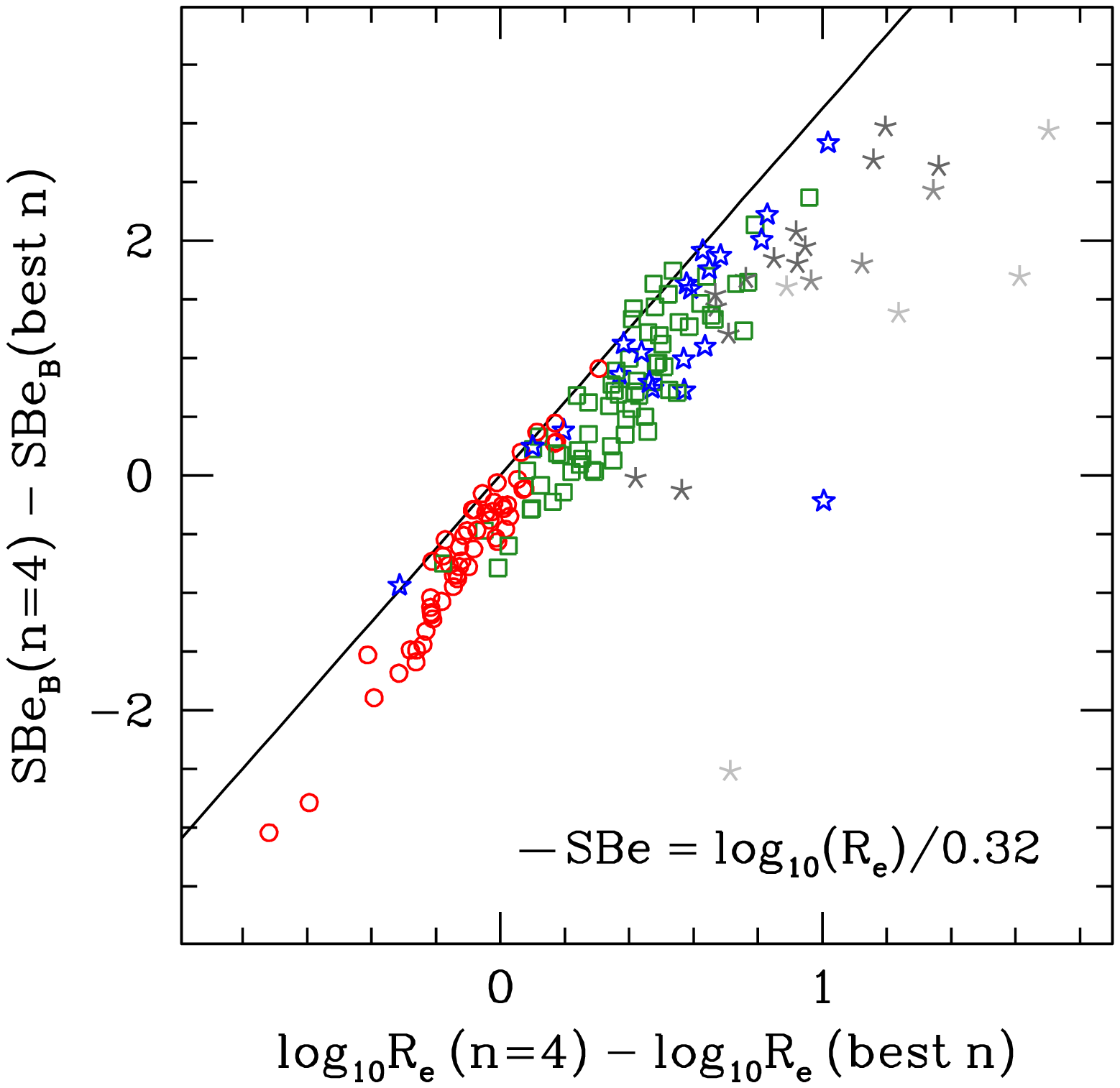}
\caption{Comparison of \sersic\ fit parameters derived here with the T05
         fixed de~Vaucouleurs fits shown as the difference in SBe
         versus the difference in log$_{10}(R_e)$.  Point types and colors are 
         as in Figure~\ref{fig:kcorrect}.  The solid black line
         delineates the combination of these parameters that enter the
         J96 Coma FP relation.}
\label{fig:FPconsp}
\end{figure}

In order to address the evolutionary history of mass build-up in
galaxies, it is necessary to convert these measured parameters into
masses and mass-to-light ($M/L$) ratios.  An effective dynamical
estimate of the stellar mass, $M$, can be defined using the scalar
virial theorem for a stationary stellar system as
\begin{equation}\label{eq:mass}
M \equiv \frac{c_{2}(n)\,\sigma_{0}^{2}\,R_e}{G},
\end{equation}
where $c_{2}(n)$ is the virial coefficient.  In the case of structural
homology, the virial coefficient is a constant for all galaxies, thus
the FP maps directly into a $M/L$ ratio (see, \eg\ T05) using the
usual definition for luminosity,
\begin{equation}
L \equiv 2 \pi\,\langle I_{e} \rangle\,R_e^2,
\end{equation}
\noindent
where $\langle I_{e} \rangle$, the average intensity within the
effective radius, $R_e$, is given in equation~(\ref{eq:avgIe}).  However, 
in the case of varying $n$, the profile shape does affect the measured
$M/L$ through a variation in the velocity dispersion profile (even
though these changes in profile shape do not significantly alter the
FP or its tilt, see Fig.~\ref{fig:FPconsp}).

A number of authors have constructed spherical dynamical models to
derive $c_2(n)$ for different profile shapes (\eg\ Ciotti \etal\ 1996;
Prugniel \& Simien 1997; Bertin \etal\ 2002; Trujillo \etal\ 2004,
hereafter Truj04), with generally consistent results.  We use the
derivation of Truj04 who constructed non-rotating isotropic spherical
models, and take into account the projected velocity dispersion over
an effective aperture\footnote{Note that models based on a single
\sersic\ profile, such as Truj04, neglect the effects of a dark matter
component, which could be important to understand other aspects of the
tilt of the Fundamental Plane, c.f.\@ Bolton \etal\ (2007).}.

For the purpose of assessing evolutionary trends, we need a local 
relation (at $z$\,=\,0) against which we can compare higher-$z$ galaxies. 
%The commonly used $B$-band local relation for early-type galaxies in the
%Coma cluster of J96 translates to: $\alpha$=1.25, $\beta$=0.32,
%and $\gamma$=$-$8.97, for the units used here 
%(AB magnitudes, \hub\,=\,65~\hunit).
Ideally, one would like to have a local comparison sample that is
large in numbers and homogeneous in measurement techniques to the
distant sample.  We are currently undergoing a major observational
project at the Palomar Observatory to construct such a large, homogeneous, 
local sample that extends to spiral galaxies and includes both field and 
cluster samples (MacArthur \etal, in prep).

While we await the results from that campaign, we derive a local FP
comparison from a sample of early-type galaxies in the Virgo Cluster.  
We start from the structural analysis of Ferrarese \etal\ (2006) of
{\it HST} images of early-type galaxies from the ACS Virgo Cluster
Survey (ACSVCS) of C{\^o}t{\'e} \etal\ (2004).  Precise distance
measurements come from the surface brightness fluctuation analysis of
Mei \etal\ (2007), and velocity dispersions are carefully collected
from the literature.  A detailed description of our derivation of the
local FP relation for Virgo cluster early-type galaxies, which relaxes
the assumption of structural homology, is provided in the Appendix.
The result is,
\begin{equation}\label{eq:MLMlocal}
\log_{10}(M/L_B)_{0} = 0.225\,\log_{10}(M) - 1.63,
\end{equation}
and we measure the offset for galaxy $i$ from this local relation as,
\begin{equation}\label{eq:DML}
\Delta\log_{10}(M/L)^i = \log_{10}(M/L)^i - \log_{10}(M/L)_0^i.
\end{equation}

Finally, since we have not applied the usual formalism found in the
literature of imposing a fixed profile shape to parametrize the light
profiles, some care in the analysis and, in particular, its comparison
with previous works must be taken.  The correlation between the
\sersic\ $n$ and total luminosity seen in panel (d) of
Figure~\ref{fig:Corr_Local} indicates weak homology among the spheroidal
components of galaxies.
This begs the question of whether this new parametrization of the
light profiles has an effect on their location in the FP.  To
illustrate the effect of using fixed versus varying profile shapes, in
Figure~\ref{fig:FPconsp} we plot the difference between the two
structural parameters entering the FP using the best fit \sersic\
profile from the current analysis and the fixed de~Vaucouleurs ($n$\,=\,4)
fits from T05.  The straight line is the best-fit
$\log_{10}(R_e)$\,--\,SBe combination from the J{\o}rgensen, Franx, \&
Kjaergaard (1996, hereafter J96) local relation for Coma.  The E/S0s
lie fairly close to the local line, indicating that the FP location of
these galaxies is not strongly affected by the shape of the profile.
This is due to the fact that, for a given total magnitude, a change in
the shape parameter $n$ will result in a different $r_e$ which is
compensated by a change SBe in the opposite sense.  The relation is
not precisely one-to-one though, but this would not be expected based
on the observed evolution of the FP tilt with redshift (\eg\ T05).  On
the other hand, there is a significant offset from the $n$\,=\,4 fits for
the spiral bulges.  This is entirely expected as T05 consider single
component $n$\,=\,4 fits whereas we have not only allowed $n$ to vary, but
we have also decomposed the profile into bulge and disk components.

Table~\ref{tab:FPpars} summarizes the FP parameters derived in the preceding 
sections for our spectroscopic sample of 147 galaxies.

\section{Spectroscopic Results}

\label{sec:spec_results}

This section presents our results based on the spectroscopic
sample. We first discuss the evolution of the Fundamental Plane in
\S~\ref{sec:FP} and then interpret it in terms of cosmic evolution of
the mass-to-light ratio of bulges and spheroids in \S~\ref{sec:MLM}.
\S~\ref{ssec:cavecanem} discusses a number of caveats that should be 
kept in mind through the discussion of the results and their
interpretation in \S~\ref{sec:discuss}.

\subsection{Fundamental Plane Evolution}
\label{sec:FP}

\begin{figure*}
\centering
\vskip -1.5in
\includegraphics[width=0.75\textwidth]{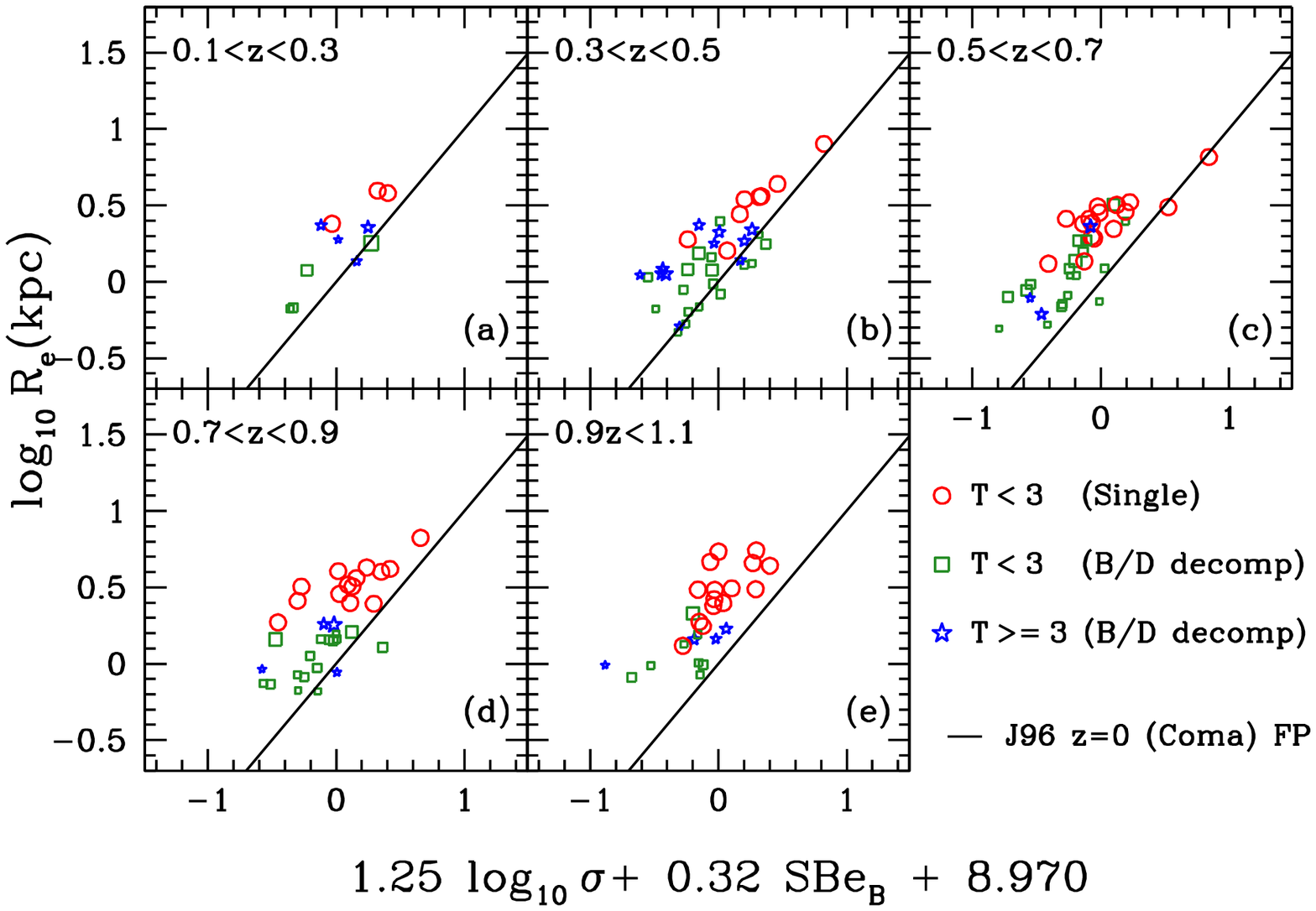}
\caption{Fundamental Plane of our spheroids, divided in to 5 redshift
bins, with respect to an edge-on view of the FP set by the local Coma
zeropoint of J96 (solid line in each panel).  Point types are
indicated and their size is proportional to the galaxy $B/T$ ratio.}
\label{fig:FPev5}
\end{figure*}

With central velocity dispersion measurements in addition to accurate
photometric parameters, we now discuss the Fundamental Plane for our
sample of intermediate-$z$ spheroids.  In Figure~\ref{fig:FPev5} we show
the location of our spheroids in the face-on projection of the FP
plane according to the J96 local relation for Coma.  We divide the
sample into five redshift bins (with constant $\Delta(z)$, thus some
additional scatter is expected due to the different age ranges in the
different bins).

%The black points are those from T05: solid pentagons are
%E/S0s with $\sigma\ge100$\,\kms, open pentagons are E/S0s with
%$\sigma<100$\,\kms, starred squares are Sa/b galaxies.  The colored points are
%our newly derived parameters for the T05 sample and our new bulge sample
%(including only those with $B/T>0.2$).

The most notable trend is that with redshift. As redshift increases, bulges 
and spheroids alike move further away from the local relation. This is 
consistent with the trend noted previously for spheroids and can be 
qualitatively understood in terms of the reduced age of the stellar 
populations.  As cosmic time goes by, stars are on average
older, and thus for a given size and velocity dispersion the surface
brightness is lower. It is important to keep in mind when comparing
with previous studies of spheroids (such as T05) that we have isolated
the spheroidal component from the surrounding disk, when present. As a
result, galaxies in common with T05 have generally moved downwards in
size (and in luminosity).  As expected, the scatter for the spiral
bulges is reduced with respect to T05, because we have taken care of
separating the pressure supported component from the rotation
supported disk.

Looking at the trends inside each bin, the overall effect seems to
imply a continuation in the FP space from the higher
mass (E/S0) to the lower mass (bulge) spheroids. Also, the FP
at intermediate redshift appears to be tilted with respect to
the local relation. The evolution of the tilt seems to be less
pronounced than that observed for spheroids considered as a single
component. A confirmation and quantification of this trend awaits a
larger sample.

Finally, we note that many of the smallest bulges appear to lie too
far to the right of the relation defined by the bigger spheroids.
While this effect could be real, we speculate that this is due to the fact
that, in the case of two-component galaxies, our central velocity
dispersions are sensitive to the potential from the total galaxy
(bulge/disk/halo) and thus are overestimates of the ``bulge-only''
dispersions.  This could be a partial explanation for the above mentioned
apparent weaker FP tilt evolution in comparison with the pure
spheroids.

\subsection{Mass-to-Light Ratio Evolution}
\label{sec:MLM}

\begin{figure*}
\centering
\vskip -1.5in
\includegraphics[width=0.75\textwidth]{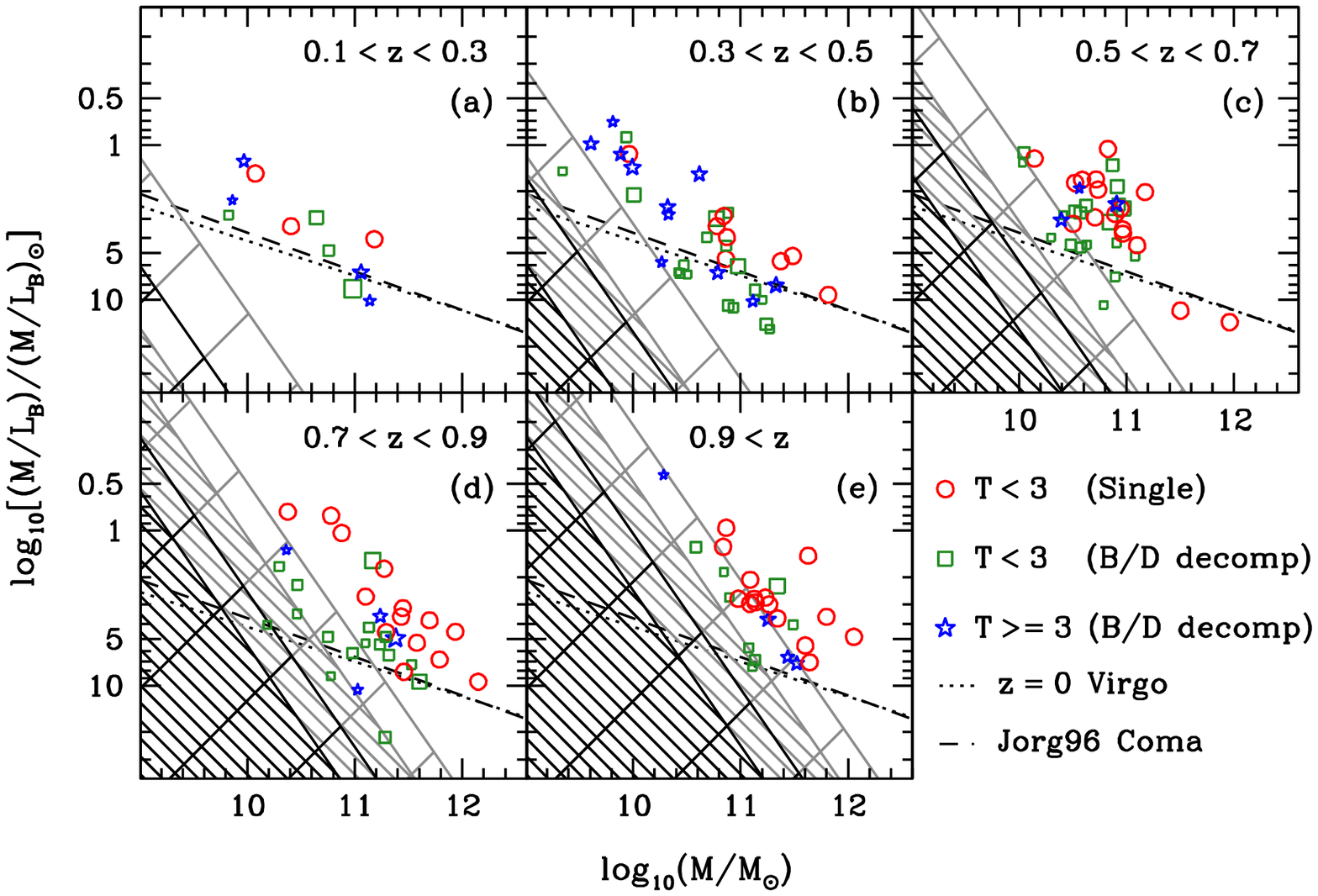}
\caption{$M/L$ vs.\@ $M$ projection of the FP divided into five redshift 
   bins.  Panels and symbols are as in Figure~\ref{fig:FPev5}.
   The dotted line is our Virgo local relation (eq.\@~[\ref{eq:MLMlocal}])
   and the dashed line is that of J96 for Coma.  Point size is
   proportional to $B/T$ ratio.  The gray hatched regions correspond
   roughly to our total galaxy magnitude limit of $z_{AB}$\,$<$\,22.5.  
   The more
   densely hatched regions apply to the lower ends of the redshift
   limits while the sparsely hatched regions correspond to the upper
   redshift end.  The black hatched regions are the corresponding
   $z_{AB}$\,$<$\,24.25 limits on our bulges (\ie\ for $B/T$\,=\,0.2).}
\label{fig:MLM}
\end{figure*}

We now use equation~(\ref{eq:mass}) to derive masses for our spheroids and
test some of the conjectures above of a correlation between
evolutionary state with spheroid mass.  In Figure~\ref{fig:MLM} we plot
the rest-frame $B$-band mass-to-light ratio ($M/L_B$; note the inverted
scale on the y-axis) as a function of mass (both expressed in solar
units), separated into the same redshift bins as in
Figure~\ref{fig:FPev5}.  The dotted line indicates our Virgo local
relation (eq.\@~[\ref{eq:MLMlocal}]) and, for comparison, the dashed line
shows the J96 relation for Coma which was constructed under the
assumption of structural homology\footnote{These two relations are remarkably
consistent with each other, and would argue against structural
non-homology as a significant contributer to the FP tilt.  Again,
however, a confirmation of this awaits a larger sample.}.
The shaded areas indicate the selection limits due to our total galaxy
magnitude limit of $z_{AB}$\,$<$\,22.5 mag.  The more densely hatched regions
correspond to the lower limit of the redshift bin, while the sparsely
hatched regions are for the upper redshift limit.  The gray hatches
apply to the pure spheroidal galaxies and the black hatches regions
are the corresponding $z_{AB}$\,$<$\,24.25 limit for the two-component galaxy
sample (\ie\ according to our $B/T$\,$>$\,0.2 limit).  This selection limit
imposes some restrictions on our interpretation of evolution,
particularly at the highest-redshift and low-mass ends.

Several trends are noticeable upon examination of Figure~\ref{fig:MLM}.
First, the slope of our relation is somewhat steeper than the local
relation slope, and there is some indication that the steepening
increases with redshift (although the latter is not conclusive due to
our selection limits).  Broadly speaking, it appears that the bulges
follow a similar relation to that defined by the pure spheroidal
galaxies.  In particular, there is no evidence of an offset to lower
$M/L$ ratios (due to increased $B$-band luminosity) which may be
expected if the bulges are undergoing a more continuous mode of SF, as
is expected in the secular formation scenario (\eg\ Kormendy \&
Kennicutt 2004).  If anything, some bulges are offset to higher $M/L$.
However, as mentioned above we believe this could be due to the fact
that our central velocity dispersions are sensitive to the total 
galaxy potential, but our derived masses are based on a formulation for 
the total mass of a single \sersic\ profile.  We may, therefore, be 
overestimating the mass of the spheroid for multiple-component systems.  
The effect does appear to be stronger for the lower $B/T$ bulges, thus
strengthening the support for this hypothesis.

\begin{figure*}
\centering
\includegraphics[width=0.95\textwidth]{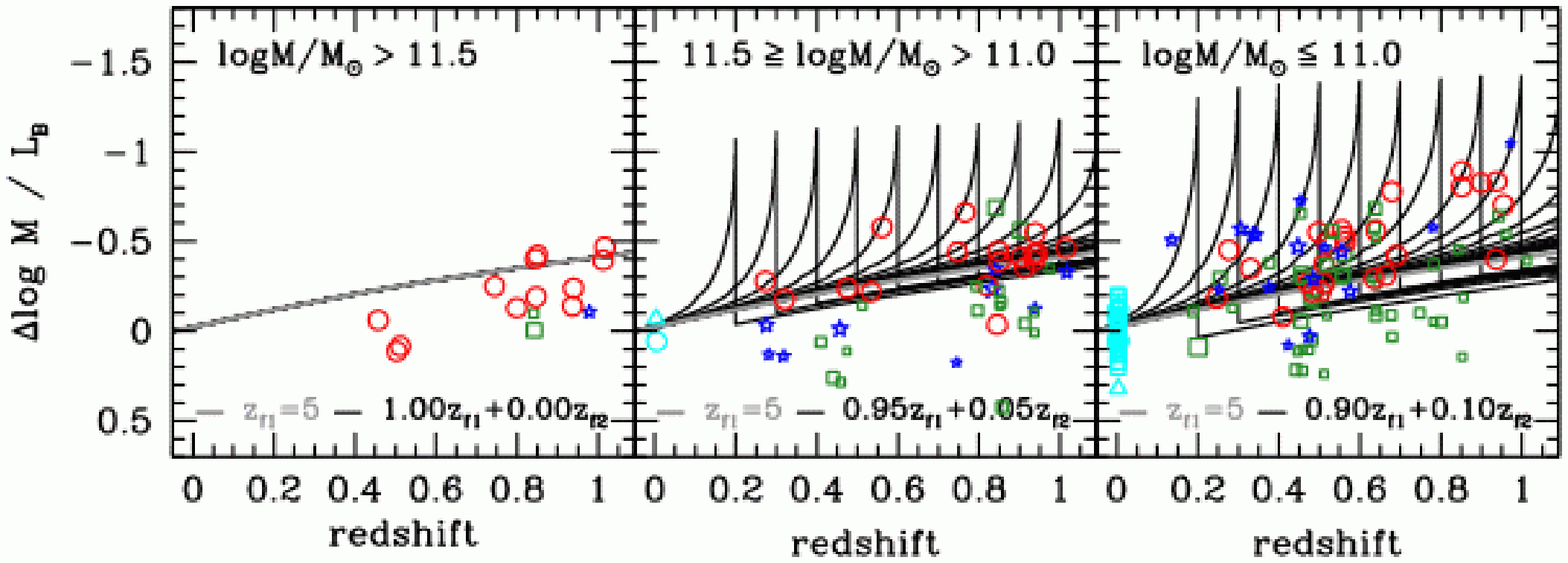}
\caption{Offset of $B$-band $M/L$ ratio, $\Delta\log_{10}M/L_B$ 
         (eq.\@~[\ref{eq:DML}]) versus redshift and divided into
         3 dynamical mass bins.  Point size is proportional to $B/T$ ratio.  
         The gray line indicates the evolutionary
         trend for a system that formed in a single burst at $z_f$\,=\,5 and
         evolves passively (from the stellar population models of BC03).  
         Black lines illustrate the effects of secondary burst of 
         star formation (5\% or 10\% by mass) at $z_{f2}$\,=\,$0.1,0.2...$
         added to the initial burst. The light blue points at 
         $z$\,$\sim$\,0 are the local Virgo galaxies (see Appendix).}
\label{fig:DML}
\end{figure*}

Also evident in Figure~\ref{fig:MLM} is that the deviation from the
local relation increases as the spheroid mass decreases.  We can
represent this offset directly using equation~(\ref{eq:DML}).  This is shown
in Figure~\ref{fig:DML} which plots $\Delta\log_{10}\,M/L_B$ as a
function of redshift and divided into three spheroid mass bins.  This
representation can be easily interpreted in terms of star formation
histories in an analogous procedure to that presented in T05 (their
Fig.~21).  Overplotted on Figure~\ref{fig:DML} are $M/L$ evolution
models from BC03.  The gray lines represent a single burst of star
formation at $z_f$\,=\,5 that evolves passively.  The most massive
spheroids (including E/S0s and bulges) in the
$\log_{10}M/M_{\odot}>$\,11.5 bin are well described by this old
population formed in a single burst.  In the lower mass bins, however,
many galaxies deviate significantly from this relation.  These
deviations can be explained in the context of more recent spheroid
building in the form of subsequent bursts of star formation on top of
an underlying old population, where the old population dominates the
total stellar mass.  The black lines in the two lower mass bins
represent the $M/L$ evolution for such models.  For the intermediate
mass range (11.5\,$\le$\,$\log_{10}M/M_{\odot}$\,$<$\,11.0), the data are
consistent with a more recent burst of SF that represents $\sim$\,5\% of 
the total stellar mass, and in the lowest mass bin, recent bursts
involving $\sim$\,10\% of the total mass are required.  Due to the
short timescales of the initial $M/L$ decline after a burst of SF,
these mass fractions are likely underestimates.  These
results echo those found in T05 for E/S0s.

Of particular note here is that the distribution of bulges in the
representation shown in Figure~\ref{fig:DML} is largely
indistinguishable from that of the pure spheroidal galaxies.  Thus, we
conclude that, for host systems with $B/T$\,$>$\,0.2, all spheroids 
{\it at a given mass} appear to follow the same evolutionary path.  The 
smaller mass spheroids -- whether isolated or embedded in a disk -- must 
have more recent stellar mass growth, so the recent observations of 
galaxy ``downsizing'' in spheroidal galaxies, both empirically 
(\eg\ Bundy \etal\ 2005; T05; di Serego Alighieri \etal\  2005; 
van~der~Wel \etal\ 2005), and in cosmological N-body + semi-analytical 
simulations (De~Lucia \etal\ 2006), 
extends to the bulges of spiral galaxies.

\subsection{Caveats}
\label{ssec:cavecanem}

Before proceeding to a comparison of our results with other studies in
the literature and to a physical interpretation of our spectroscopic
and photometric results in the next section, it is important to list a
number of caveats that should be kept in mind throughout the
discussion.

\begin{itemize}
\item  Most local FP comparisons are not measured in
       quite the same way as our distant sample. For example, $r_e$ is
       typically not computed from \sersic\ fits, nor are two
       component fits generally applied to lenticular galaxies.  This
       may introduce some systematic bias in our determination of the
       evolution of the mass-to-light ratio. For example, if the slope
       is steeper locally than we estimate, the deviations in
       Figure~\ref{fig:DML} will be tempered, and vice versa. We
       partially addressed this issue by compiling the FP
       of Virgo galaxies based on \sersic\ fits (Ferrarese \etal\
       2006). However, the Virgo sample is small (smaller in
       number than the intermediate redshift sample!) and lenticular
       galaxies are not decomposed into their spheroidal and disk
       constituents. Our ongoing Palomar program will hopefully
       provide us with a more suitable local comparison sample in the
       near future.

\item Our $\sigma_{0}$ (or $\sigma_{r_e/8}$) measurement is going to
      be sensitive to the total galaxy potential (bulge + disk +
      halo), so the bulge mass may be systematically over-estimated
      the less prominent the bulge becomes.  This could help explain
      the high $M/L$ ratios measured for our bulges, which appears to
      be more pronounced for the low $B/T$ systems, as
      expected. Although this is beyond the scope of the present
      observational work, multicomponent dynamical models are needed
      to determine a more accurate mapping between measured $\sigma_0$
      effective radius, and {\it bulge} stellar mass for low-$B/T$
      systems.

\item Significant amounts of dust could also bias high our $M/L$
      ratios.  If present, one would expect it to be more of an issue
      the later the Hubble type, and therefore dust could also explain
      in part the high $M/L$ ratios found for low-$B/T$
      systems. A simple way to check for the presence of dust is to
      study evolutionary trends as a function of
      inclination. Unfortunately the present sample is too small to be
      divided up in bins of the relevant controlling parameters,
      Hubble type, redshift, inclination and mass.

\item The main limiting factor of our study is sample size.  While 
      collecting data for hundreds of spheroids has been a major effort and
      represents significant progress with respect to earlier work, in
      order to detect conclusive trends and address the caveats listed
      in this section, larger samples are needed at each redshift bin,
      covering wider -- and overlapping across bins -- ranges in mass
      (and luminosity) and Hubble type. It would also be interesting
      to extend the samples to smaller bulges to verify how far down
      the mass function the downsizing trend extends.
\end{itemize}

\section{Discussion}\label{sec:discuss}

Having discussed our photometric and spectroscopic results separately
in Sections 4 \& 5, we are now in a position to combine the
inferences from the two diagnostics, compare them with previous
studies in the literature, and discuss the physical interpretation.

Colors and mass-to-light ratios provide a consistent picture. The
stellar populations of the most massive bulges are homogeneously old
(formed at $z$\,$>$\,2--3), while an increasingly larger fraction of younger
stars (formed below $z$\,$\sim$\,1) is required as one moves down the mass
function.  This ``downsizing'' trend is consistent within the errors
with that observed for the pure spheroidals and for the spheroidal
component of lenticular galaxies.  As discussed earlier, this finding
allows us to reconcile the results of previous studies of intermediate
redshift bulges based on WFPC2 data (EAD and Koo05).

The results of our study are also consistent with recent studies of
the ``fossil record'' from local samples. Recently -- based on the
analysis of absorption features -- Thomas \& Davies (2006) found that,
at a given velocity dispersion, local bulges are indistinguishable
from elliptical galaxies in terms of their correlations with age and
alpha-element enhancement ratios (as a proxy for star formation
timescales).  Their sample is similarly limited to large $B/T$
spirals.

What does this all imply about bulge formation? It is important to
emphasize that different scenarios could lead to the same star
formation history. For example, for the most massive spheroids
analyzed in T05, one cannot distinguish based on stellar populations
alone whether the mass assembly happened at high redshift, perhaps
concurrent with the major episode of star formation, or whether they
were assembled at a later time via dry mergers (\eg\ van~Dokkum 2005;
Bell \etal\ 2006). Independent data, such as those on the evolution
of the stellar and dynamical mass functions (\eg\ Bundy, Treu \& Ellis
2007), are needed to break the degeneracy and conclude that the {\it
assembly} happened at high redshift as well. For the less massive
spheroids, a larger amount of recent star formation is required, but
understanding whether this happened in situ or via the accretion of
gas reach satellites of mass of order a few $10^9$\,$M_{\odot}$
(\eg\ T05), again requires external information. The study by Bundy 
Treu \& Ellis (2007), points toward minor mergers or internal/secular 
process as being important alongside major mergers.
As far as bulges are concerned then, the similarity in star formation
need not necessarily reflect a similarity in mechanisms/timing of
assembly. At the high mass end, the homogeneously old stellar
populations of bulges seem hard to reconcile with recent formation
from existing disk stars or gas.  Therefore it appears that we can
safely conclude that massive bulges have been {\it in place} well
before the redshift range probed by this study, all but ruling out
significant secular evolution (see also Thomas \& Davies 2006). Hence
the similarity with spheroids for the most massive bulges appears to 
be both in the star formation history and in the mass assembly history.

At lower masses, secular process may become more important and it is
difficult to disentangle recent in situ star formation, mass transfer
from the disk to the bulge, and merging with satellites. External
evidence suggests that at this redshift and mass scale, bulges are
growing significantly via mergers (Woo \etal\ 2006; Treu \etal\ 2007),
although this may not be the only evolutionary mechanism. If bulges at
masses below $10^{11}$\,$M_{\odot}$ grow by mergers, than the rightmost
plot in Figure~\ref{fig:DML} must be interpreted with caution: galaxies
of a given mass at one redshift are not the progenitors of objects
with the same mass at a lower redshift. Perhaps a way to identify the
spheroid progenitors is to study bulges at fixed central black hole
mass, assuming that growth by accretion is mostly negligible. This is
difficult to do in practice and is limited at intermediate redshifts
to samples of spheroids hosting active nuclei. However -- albeit with
large uncertainties -- such an exercise (Treu \etal\ 2007) suggests that 
spheroids of a few $10^{10}$\,$M_{\odot}$ may grow in stellar mass 
by $0.20\pm0.14$\,dex in the past four billion years, while keeping the 
stellar mass-to-light 
ratio approximately constant as a result of adding young stars
to an aging resident population.  The uncertainties are currently too
large and the samples too small to perform a quantitative test of this
scenario, but it is clear that the evolution of spheroids in this mass
range is not purely passive (see also Hopkins \etal\ 2006) and that 
complementary information from stellar populations studies, spheroid 
demographics, and scaling relations between host galaxy and central black 
hole are needed to disentangle star formation, assembly history and black 
hole growth and feedback. The line of studies presented in this work is 
crucial to accurately pinpoint the star formation history and could in 
future provide additional independent information if extended to include
tracers of chemical enrichment.

\section{Summary}
\label{sec:summary}

We have presented a comprehensive study of field galactic bulges, and
of a comparison sample of lenticular and elliptical galaxies, in the
redshift range $z$\,=\,0.1--1.2. The galaxies are selected from the GOODS
survey based on visual classification and luminosity
($z_{AB}$\,$<$\,22.5). We measure accurate colors and structural parameters
for 193 galaxies, fitting two components (bulge and disk) for 137
galaxies and a single \sersic\ profile for the remaining 56, classified
as pure spheroids. Stellar velocity dispersions are measured from deep
spectroscopic observations using DEIMOS on the Keck-II Telescope. By
combining new observations with the measurements derived in T05, we
compile a spectroscopic subsample of 147 galaxies suitable for
Fundamental Plane analysis, \ie\ reliable stellar velocity dispersion
and bulge to total luminosity ratio greater than 20\%, to minimize
disk contamination. The spectroscopic subsample includes 56 one
component galaxies (pure spheroids) and 91 two component (bulge + disk)
galaxies.

We use the two complementary diagnostics -- colors and mass-to-light
ratios as determined from the evolution of the Fundamental Plane -- to
derive the star formation history of bulges and compare it to that of
the spheroidal component of lenticular galaxies and of elliptical
galaxies. Although the uncertainties are larger for bulges than for
pure spheroids (\S~\ref{ssec:cavecanem}) -- and more data both at
intermediate redshift and in the local universe are needed to confirm
the trends, the two diagnostics give consistent results that can be
summarized as follows:

\begin{itemize}

\item The stellar populations of the more massive bulges 
($M$\,$>$\,$10^{11}$\,$M_{\odot}$) are homogeneously old, consistent with 
a single major burst of star formation at redshift $\sim$\,2 or higher, 
and only minor episodes of star formation below $z$\,$\sim$\,1 
($\lesssim$\,5\% in mass).

\item The colors and mass-to-light ratios of smaller bulges  
($M$\,$<$\,$10^{11}$\,$M_{\odot}$) span a wider range, consistent with an
increasing fraction of younger stars going to smaller masses
($\sim$\,10\% in mass).

\item The assembly history of bulges is consistent within the error with
that of the spheroidal component of lenticular and elliptical galaxies
of comparable mass.

\end{itemize}

Our detection of a mass dependent star formation history allows us to
reconcile the findings of two earlier studies of intermediate redshift
bulges based on WFPC2 data (EAD and Koo05). The more massive bulges
are as old and red as massive spheroids, but smaller bulges have quite
diverse star formation histories, with significant star formation
below $z$\,$\sim$\,1.

The similarity between the mass assembly history of massive bulges and
that of spheroids appears quite naturally explained in a scenario
where the former are assembled at high redshift, rather than recently
via secular process within spiral galaxies. At the lower mass end of our 
sample ($M$\,$\sim$\,$10^{10}$\,$M_{\odot}$), the evolution of bulges below
$z$\,$\sim$\,1 becomes much more diverse, requiring perhaps a combination 
of merging and secular processes as observed for spheroids of similar
mass (Bundy, Treu, \& Ellis 2007).

This ``downsizing'' (\eg\ Cowie \etal\ 1996; Bundy \etal\ 2005; 
De~Lucia \etal\ 2006; Renzini 2007) picture for bulges extends previous
findings based on the integrated properties of galaxies. In addition, this 
picture is qualitatively consistent with recent results on the co-evolution of
black holes and spheroids (Treu, Malkan, \& Blandford 2004; 
Walter \etal\ 2004; Woo \etal\ 2007; Peng 2007; 
Treu \etal\ 2007), which suggests that bulges in the mass range considered 
here may have completed their growth after that of the central black hole.

\acknowledgements

The authors would like to extend their appreciation to Jason Rhodes
for running the Tiny Tim PSF simulations, and to St{\' e}phane
Courteau and Michael McDonald for sharing their XVISTA routines.
Thanks also to Ignacio Trujillo, David Koo, Mark Dickinson, and 
Bob Abraham for useful discussions.  
LAM acknowledges financial support from the National Science and
Engineering Council of Canada.  TT acknowledges
support from the NSF through CAREER award NSF-0642621, by the Sloan
Foundation through a Sloan Research Fellowship, and by the Packard
Foundation through a Packard Fellowship. 
Some of the data presented herein were
obtained at the W.M.\@ Keck Observatory, which is operated as a
scientific partnership among the California Institute of Technology,
the University of California and the National Aeronautics and Space
Administration. The Observatory was made possible by the generous
financial support of the W.M.\@ Keck Foundation.
This work is partly based on archival data from the 
{\it Hubble Space Telescope}, obtained from the data archive at the 
Space Telescope Institute, which is operated by the association of 
Universities for Research in Astronomy, Inc.\@ for NASA under 
contract NAS5-26555.

\appendix
\section{Derivation of Local Virgo Fundamental Plane}

A crucial aspect for the above evolutionary study is a local zeropoint
against which we can compare our higher-$z$ galaxies.  In order to
obtain a meaningful comparison, strict homogeneity in the data and
analysis techniques are required.  There do exist a few local FP
standards in the literature.  One of the most commonly used is the FP
of J96 based on Coma cluster early type galaxies.  This FP was derived
under the assumption of structural homology, fitting all galaxies with
a single de~Vaucouleurs profile.  In this case, the
virial coefficient is a constant for all galaxies and the FP maps
directly into a $M/L$ ratio (\eg\ T05).  However, in the case of
varying $n$, the profile shape does affect the $M/L$ through a
variation in the $\sigma(r)$ profile.  In the current study we have
relaxed the assumption of structural homology and fit all spheroids
with the more general \sersic\ profile with varying $n$.  The J96
relation is thus not appropriate as a local comparison relation here.

We seek a local sample for which galaxy profiles are modeled with the
\sersic\ profile and comprise a homogeneous database of velocity dispersion
measurements.  
As far as we are aware, the only analysis to date that provides a
local $M/L$ versus $M$ relation, taking into account varying
profile shapes is that of Truj04.  Their FP sample includes 45 cluster
ellipticals (from Virgo, Fornax, and Coma).  The photometric
parameters were obtained from the literature and the velocity
dispersions from the {\tt Hypercat}\footnote{http://leda.univ-lyon1.fr} 
database.  The sample in that analysis is quite inhomogeneous and could
not make use of more recent accurate distance measurements to Virgo
cluster galaxies, thus rendering it an unsatisfying zeropoint from
which to compare our distant sample.
%We note, however, that this sample is still not ideal for our purposes for
%a number of reasons.  First, their photometric parameters $R_e$ and 
%$\langle I_e \rangle$ are computed non-parametrically, rather than from
%\sersic\ fitting, as we do here.  Additionally, the velocity dispersions 
%were collected from a non-homogeneous set of observations (via the Hypercat 
%catalogue) making aperture corrections difficult.  Finally, their sample
%only includes cluster elliptical galaxies, whereas our sample includes 
%spiral and elliptical field galaxies.  

Recently, however, the ACS Virgo Cluster Survey (ACSVCS) of 
C{\^o}t{\'e} \etal\ (2004) have obtained ACS images in the F475W and F850LP 
bandpasses of 100 early-type galaxies in the Virgo Cluster.  
Ferrarese \etal\ (2006, hereafter Fer06) present a detailed structural 
analysis of these galaxies modeling their SB profiles with \sersic\ models.
Furthermore, Mei \etal\ (2007) present precise distance measurements
to 84 of these galaxies based on surface brightness fluctuations
(SBF).  Accurate distance measurements are crucial for conversion to
physical parameters for FP analyses.  Given the significant
line-of-sight depth and sub-clustering of the Virgo cluster, we
restrict ourselves to these 84 galaxies with accurate SBF distance
measurements.

While the F475W ($g$) band is close to $B$, they are not identical and
a calibration between the two is required.  Fer06 tabulate the total
$B$ magnitudes from the VCC survey of Binggeli \etal\ (1985) and in
their Fig.~115, plot these $B$ magnitudes against their total $g$,
computed by integrating the \sersic\ fit parameters to infinity.  The
fainter galaxies closely follow the one-to-one relation.
Moving to brighter systems, there is a systematic trend to brighter
$g$ magnitudes.  Fer06 associate this with observationally established
color-magnitude trends (becoming redder with higher luminosity).
However, at the brightest end, the ``core'' galaxies, whose light
profiles within the central few 100 pc fall below an inward
extrapolation of the ``outer'' \sersic\ fit (for which the central
points were omitted), deviate strongly in the sense that the fit
derived magnitudes are significantly brighter than those derived
``from the data''.  This deviation seems too strong to be attributed
to SP effects alone.

As a consistency check, we compared the $g-B$ versus $g-z$ colors with
the stellar population models of BC03.  The data span a reasonable range 
in $g-z$ in the context of SPs, but the $g-B$ colors 
scatter significantly off the model grids, indicating a calibration issue 
between the ACSVCS and Binggeli \etal\ (1985) photometry.  Most galaxies
scatter to the blue in $B-g$ but all of the ``core'' galaxies scatter to
the red (again indicating an overestimated $g$ magnitude).  This color plane 
is fairly degenerate in metallicity thus a simple linear relation can be used 
to convert from the observed $g$-band to $B$-band.  We adopted the following 
conversion: $B = g + ( -0.01*(g-z) - 0.105 )$ and refer to this as the
``$gB$'' magnitude.

An independent check on the ACSVCS photometric analysis was made by
comparing their results to 23 galaxies in common with the study of
Gavazzi \etal\ (2005; Gav05) who present $B$-band \sersic\ modeling of
226 giant and dwarf elliptical Virgo cluster galaxies located in the
North-East quadrant.  The agreement between structural parameters is
quite good for all non-core galaxies.  
Additionally, at observed magnitudes fainter than $B$\,$\sim$\,11, our
derived ``$gB$'' magnitudes agree very well with the Gav05 total
$B$-band magnitudes, which were derived ``from the data'' rather than
computed from the fit parameters.  The core galaxies show significant
discrepancies in total magnitude and the photometric parameters.

A number of studies have pointed out unique properties of core elliptical
galaxies when compared with the dE/E sequences.  They define a different 
and quite distinct photometric plane than dE/Es (\eg\ Graham \& Guzman 2003, 
Gav05, Fer06).  Many argue that the observed cores arise from 
the partial evacuation of the nuclear region by coalescing black holes
(\eg\ Merritt 2006).  Core ellipticals are also relatively rare (\ie\ they are 
the exception rather than the rule) even in a cluster environment, and
are presumably even less common in the field. As such, we 
likely do not have any core ellipticals in our field sample (although
we cannot tell for sure due to insufficient resolution for these high-$z$ 
galaxies.)  All of the above argues against including these outliers in
our local sample, thus we also exclude all galaxies identified as 
``core \sersic" in Fer06.  This leaves us with 74 Virgo early-type galaxies 
with accurate photometric decompositions and distance measurements.

\begin{figure*}
  \centering
  \includegraphics[width=0.48\textwidth]{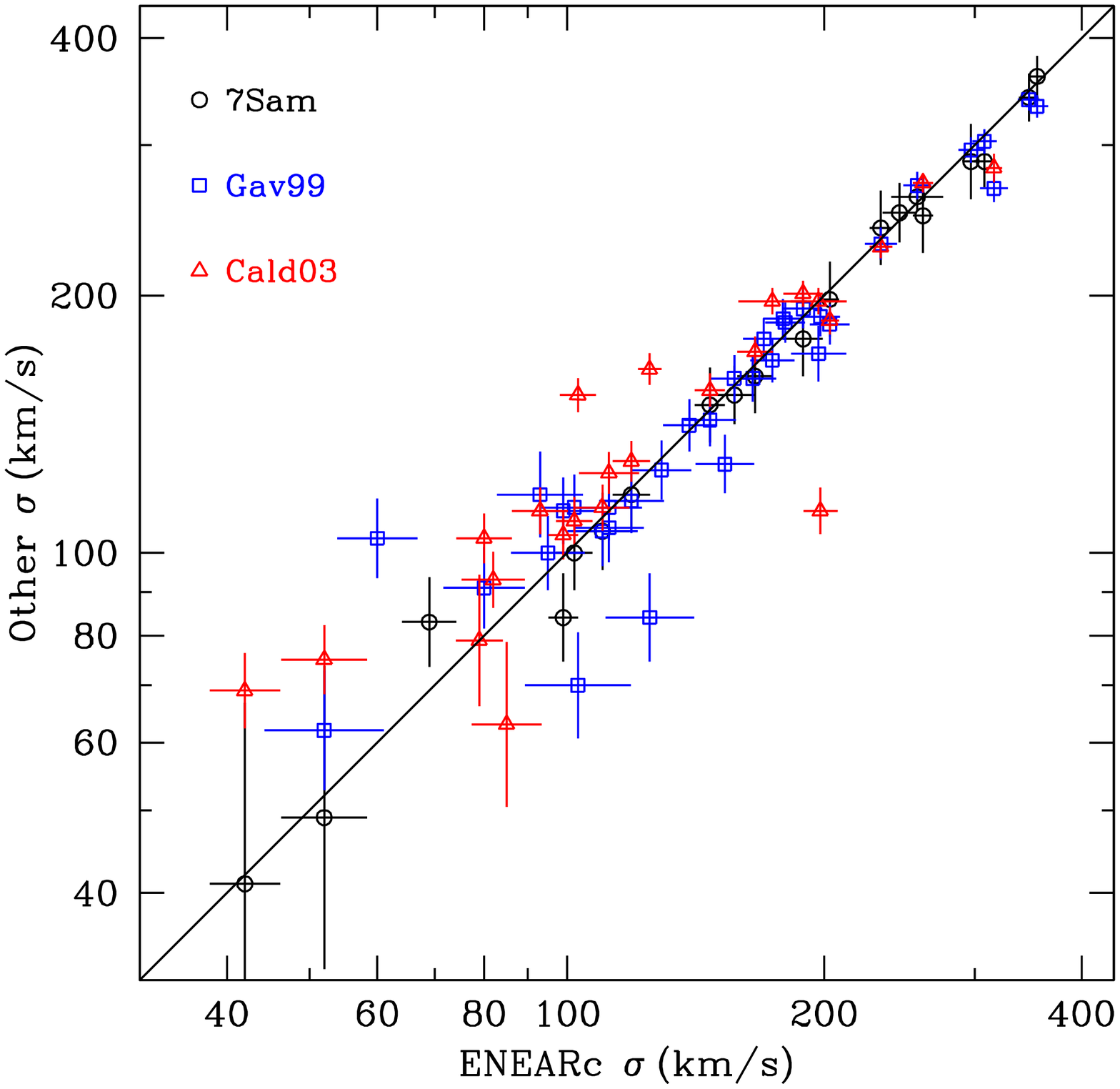}
  \includegraphics[width=0.48\textwidth]{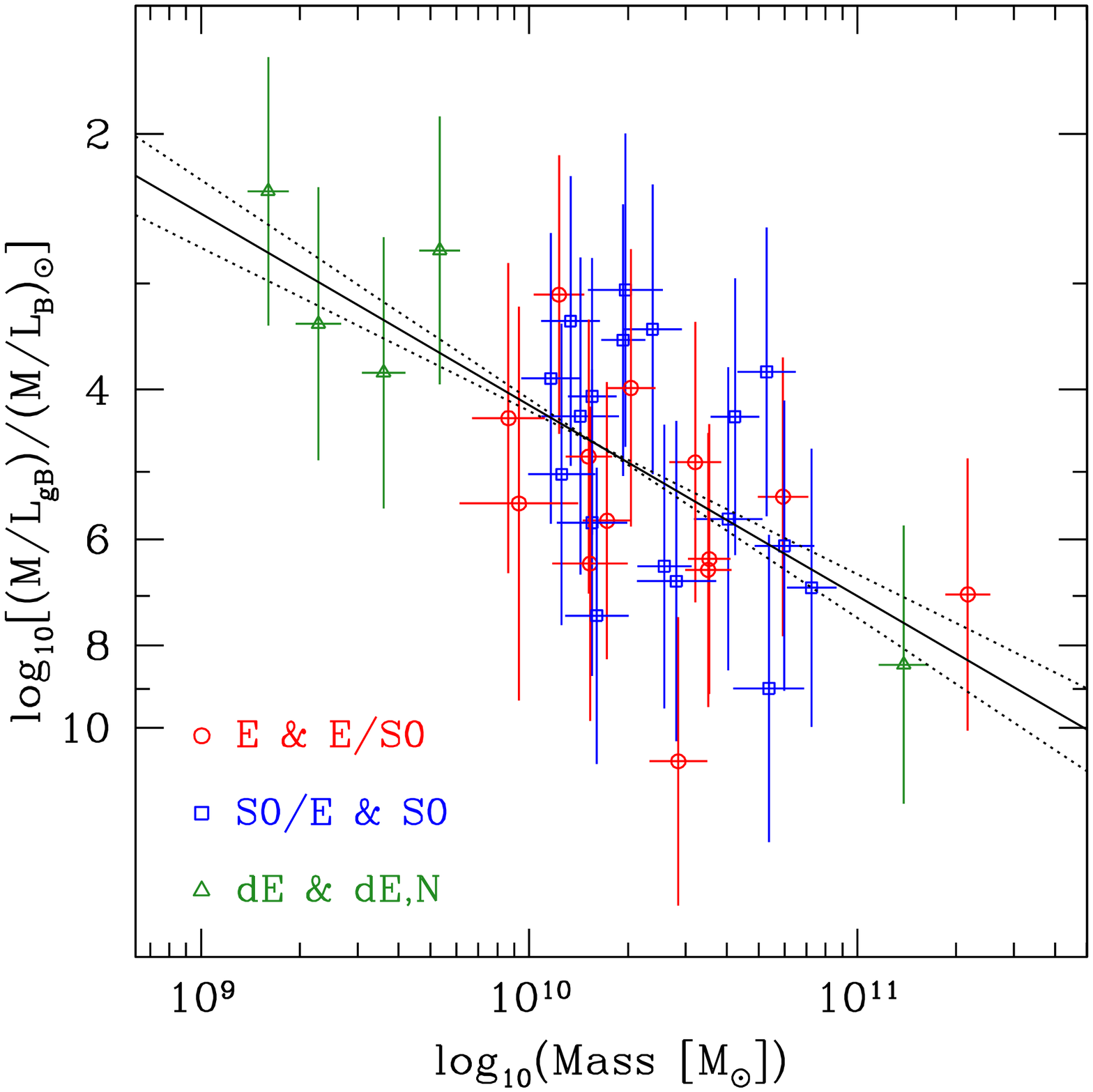}
  \caption{{\it Left}: Comparison of literature velocity dispersion 
   measurements used in our compilation of Virgo early-type galaxies.  The
   solid line it the one-to-one relation. {\it Right}: The Virgo FP projected 
   onto the $M/L$ 
   vs.\@ $M$ plane using equation~(\ref{eq:mass}).  The solid line is the 
   orthogonal fit to the data and the dotted lines represent the fit
   errors.}
  \label{fig:Virgo_sigmas}
\end{figure*}

We then searched the literature and {\tt HyperLeda} database for velocity 
dispersion measurements for this sample to construct a local FP.  Given the 
strict requirement of uniformity, we were scrupulous in culling the velocity 
dispersion measurement among different studies.  In the end, we included 
$\sigma$s from 5 different samples: Bernardi \etal\ 2002 [ENEARc], 
Faber \etal\ 1989 [7Sam], Gavazzi \etal\ 1999 [Gav99], 
Caldwell \etal\ 2003 [Cald03].
Velocity dispersions were selected using the following hierarchy
ENEARc if $\sigma$\,$>$\,75\,\kms\ else 7Sam if $\sigma$\,$>$\,75\,\kms\
else Gav99 if $\sigma$\,$>$\,75\,\kms\ else Cald03 if $\sigma$\,$<$\,100\,\kms.
The left panel of Figure~\ref{fig:Virgo_sigmas} compares the literature values
from the above studies for common galaxies.  The agreement is quite good at
large $\sigma$ but starts to break down at $\sigma$\,$\lesssim$\,100\,\kms.
Finally, we included all velocity dispersion measurements  for dEs from in
common with the study of Geha \etal\ (2003)  (without restrictions).  We 
are thus left with a final Virgo FP relation including 36 E/S0/dE galaxies.

Corrections to an aperture of $r_e/8$ were made as above (see 
\S~\ref{sec:sigmacor}) assuming a 2\arcsec\ aperture for all samples except 
that of Geha \etal\ (2003), for which we adopt a 1\arcsec\ aperture.  

With this final sample of 36 non-cored early-type Virgo galaxies we construct 
the $M/L_B$ vs.\@ $M$ projection of the FP using equation(~\ref{eq:mass}).  
An orthogonal fit to the data using the procedure of
 Akritas \& Bershady (1996) provides the relation
\begin{equation}\label{eq:MLM_Virgo}
\log_{10}(M/L_B)_{0} = 0.225 (\pm 0.033)  \,\log_{10}(M) - 1.63 (\pm 0.34).
\end{equation}
We use this relation as the local comparison for our higher-$z$ sample 
in \S~\ref{sec:MLM}.
%The slope and intercept of this relation are remarkably close to that of J96 
%projected into the same $M/L$ versus $M$ plane (see Fig.~\ref{fig:MLM}).  
%While we cannot draw any strong conclusions given our small number 
%statistics, it would seem from this result that structural non-homology of 
%early-type galaxies does not contribute to the tilt of the FP.  Alternative 
%explanations (\eg\ variations in stellar population parameters and/or 
%dark matter fractions) must be invoked to explain the observed tilt.

\clearpage
\LongTables
\begin{deluxetable*}{rrrrccccccc}
\tablecaption{Galaxy Sample Parameters. \label{tab:galpars}}

\tablehead{\colhead{ID} & \colhead{R.A.} & \colhead{Decl.} & \colhead{$z$} & 
\colhead{T}  & \colhead{N.cp} & \colhead{B/T} & \colhead{$B_{F435W}$} & 
\colhead{$V_{F606W}$} & \colhead{$i_{F775W}$} & \colhead{$z_{F850LP}$} \\ 
\colhead{} & \colhead{(J2000)} & \colhead{(J2000)} & \colhead{} & \colhead{} & 
\colhead{}  & \colhead{} & \colhead{(AB mag)} & \colhead{(AB mag)} & 
\colhead{(AB mag)} & \colhead{(AB mag)} \\
\colhead{(1)} & \colhead{(2)} & \colhead{(3)} & \colhead{(4)} & 
\colhead{(5)} & \colhead{(6)} & \colhead{(7)} & \colhead{(8)} & 
\colhead{(9)} & \colhead{(10)} & \colhead{(11)} }

\startdata
% This file has all the formatted data for the table
%Gal       RA           DEC         z      T   nc    B/T    BobsT   VobsT   IobsT   ZobsT
  62 & 189.469696 &  62.274593 & 0.3060 &  3 & 2 &  0.511 & 19.98 & 19.32 & 18.79 & 18.49 \\ 
 198 & 189.329727 &  62.192097 & 0.5557 &  3 & 2 &  0.564 & 22.30 & 21.06 & 20.11 & 19.95 \\ 
 280 & 189.317703 &  62.229740 & 0.4473 &  3 & 2 &  0.598 & 23.17 & 22.09 & 21.49 & 21.21 \\ 
 319 & 189.347794 &  62.246853 & 0.2531 &  3 & 2 &  0.035 & 21.23 & 20.07 & 19.47 & 19.21 \\ 
 352 & 189.326721 &  62.265003 & 0.4759 &  3 & 2 &  0.159 & 22.24 & 21.53 & 20.73 & 20.41 \\ 
 384 & 189.333603 &  62.275661 & 0.4378 &  3 & 2 &  0.054 & 23.13 & 21.91 & 21.25 & 20.92 \\ 
 421 & 189.442383 &  62.291962 & 0.4865 &  3 & 2 &  0.055 & 23.80 & 22.58 & 21.83 & 21.32 \\ 
 438 & 189.360596 &  62.307400 & 0.4379 &  3 & 2 &  0.085 & 22.37 & 21.46 & 20.71 & 20.33 \\ 
 462 & 189.373825 &  62.326393 & 0.3184 &  3 & 2 &  0.358 & 21.73 & 19.73 & 18.86 & 18.52 \\ 
 465 & 189.356293 &  62.328259 & 0.2773 &  3 & 2 &  0.032 & 19.58 & 18.65 & 18.27 & 18.10 \\ 
 519 & 189.360291 &  62.345501 & 0.4853 &  3 & 2 &  0.500 & 22.55 & 21.98 & 21.26 & 20.90 \\ 
 667 & 189.209091 &  62.145840 & 0.4335 &  3 & 2 &  0.198 & 23.07 & 21.51 & 20.70 & 20.32 \\ 
 710 & 189.159317 &  62.164963 & 0.5124 &  3 & 2 &  0.061 & 22.60 & 21.65 & 21.22 & 21.07 \\ 
 723 & 189.165512 &  62.169491 & 0.5100 &  3 & 2 &  0.000 & 24.01 & 22.52 & 21.60 & 21.13 \\ 
 790 & 189.174805 &  62.201591 & 0.4334 &  3 & 2 &  0.002 & 22.54 & 21.69 & 21.06 & 20.74 \\ 
 793 &  53.131104 & -27.829250 & 0.2140 &  3 & 2 &  0.025 & 19.86 & 19.15 & 18.69 & 18.48 \\ 
 873 & 189.252197 &  62.235050 & 0.4226 &  3 & 2 &  0.286 & 23.28 & 21.48 & 20.70 & 20.30 \\ 
 920 & 189.165421 &  62.257412 & 0.3769 &  3 & 2 &  0.071 & 22.53 & 21.10 & 20.40 & 20.07 \\ 
 928 & 189.261292 &  62.262199 & 0.5122 &  3 & 2 &  0.285 & 21.05 & 20.32 & 19.61 & 19.32 \\ 
 956 & 189.264297 &  62.275383 & 0.3575 &  3 & 2 &  0.012 & 22.68 & 21.63 & 21.15 & 21.04 \\ 
 974 & 189.243591 &  62.277031 & 0.2982 &  3 & 2 &  0.267 & 24.62 & 22.78 & 22.07 & 21.85 \\ 
 975 & 189.269287 &  62.281139 & 0.3765 &  3 & 2 &  0.345 & 22.03 & 20.81 & 20.39 & 20.07 \\ 
 989 & 189.249786 &  62.287785 & 0.3552 &  3 & 2 &  0.018 & 21.76 & 20.37 & 19.65 & 19.32 \\ 
 991 & 189.209198 &  62.288433 & 0.5127 &  3 & 2 &  0.000 & 23.83 & 22.84 & 22.43 & 22.26 \\ 
1050 &  53.163155 & -27.812397 & 0.4574 &  3 & 2 &  0.552 & 22.60 & 20.17 & 19.31 & 18.94 \\ 
1310 & 189.062897 &  62.164471 & 0.4537 &  3 & 2 &  0.016 & 23.51 & 22.60 & 22.01 & 21.78 \\ 
1331 & 189.126114 &  62.170731 & 0.4816 &  3 & 2 &  0.096 & 23.73 & 22.79 & 22.27 & 22.07 \\ 
1377 & 189.050095 &  62.194218 & 0.2753 &  3 & 2 &  0.529 & 20.94 & 19.62 & 18.99 & 18.70 \\ 
1378 & 189.115891 &  62.190289 & 0.5172 &  3 & 2 &  0.073 & 24.05 & 22.75 & 21.83 & 21.58 \\ 
1413 & 189.066269 &  62.210491 & 0.2856 &  3 & 2 &  0.005 & 20.90 & 19.88 & 19.44 & 19.21 \\ 
1415 & 189.145691 &  62.206814 & 0.5624 &  3 & 2 &  0.039 & 22.24 & 20.50 & 19.56 & 19.16 \\ 
1419 & 189.151428 &  62.207962 & 0.4842 &  3 & 2 &  0.003 & 22.96 & 22.34 & 21.55 & 21.18 \\ 
1422 & 189.081116 &  62.214703 & 0.4727 &  3 & 2 &  0.094 & 22.58 & 21.31 & 20.41 & 20.01 \\ 
1429 & 189.092896 &  62.211666 & 0.4577 &  3 & 2 &  0.047 & 22.80 & 21.56 & 20.83 & 20.48 \\ 
1471 & 189.041290 &  62.235031 & 0.4727 &  3 & 2 &  0.202 & 22.48 & 22.99 & 22.40 & 22.33 \\ 
1489 & 189.024902 &  62.243362 & 0.4084 &  3 & 2 &  0.142 & 23.22 & 22.17 & 21.66 & 21.48 \\ 
1690 & 188.990997 &  62.180950 & 0.3415 &  3 & 2 &  0.559 & 23.38 & 22.08 & 21.79 & 21.61 \\ 
  92 & 189.452744 &  62.289963 & 0.4385 &  3 & 2 &  0.032 & 22.86 & 21.62 & 20.89 & 20.53 \\ 
  94 & 189.483551 &  62.290199 & 0.6817 &  3 & 2 &  0.084 &  \nod & 22.88 & 21.65 & 21.26 \\ 
 170 & 189.320251 &  62.168732 & 0.4111 &  3 & 2 &  0.152 & 22.52 & 21.16 & 20.35 & 19.98 \\ 
 211 & 189.359985 &  62.203800 & 0.3443 &  3 & 2 &  0.431 & 21.56 & 20.48 & 20.21 & 19.95 \\ 
 254 & 189.327591 &  62.222668 & 0.4757 &  3 & 2 &  0.460 &  \nod & 21.98 & 20.99 & 20.55 \\ 
 289 & 189.322052 &  62.232246 & 0.3363 &  3 & 2 &  0.480 & 22.52 & 21.50 & 21.09 & 20.87 \\ 
 318 & 189.324020 &  62.244164 & 0.4104 &  3 & 2 &  0.114 & 23.16 & 21.91 & 21.29 & 20.93 \\ 
 548 & 189.351547 &  62.365982 & 0.2809 &  3 & 2 &  0.317 & 20.54 & 18.89 & 18.15 & 17.87 \\ 
 698 & 189.243088 &  62.166161 & 0.1360 &  3 & 2 &  0.415 & 18.99 & 18.10 & 17.63 & 17.40 \\ 
 704 & 189.173203 &  62.163338 & 0.5183 &  3 & 2 &  0.018 &  \nod & 21.46 & 20.38 & 19.87 \\ 
 750 & 189.264557 &  62.183910 & 0.5176 &  3 & 2 &  0.000 & 24.18 & 22.44 & 21.52 & 21.16 \\ 
 891 & 189.226608 &  62.242840 & 0.5763 &  3 & 2 &  0.490 & 23.59 & 22.78 & 22.14 & 21.93 \\ 
1008 & 189.220154 &  62.302116 & 0.2529 &  3 & 2 &  0.224 & 21.50 & 19.77 & 19.12 & 18.87 \\ 
1037 & 189.189194 &  62.316879 & 0.4547 &  3 & 2 &  0.302 & 22.53 & 21.32 & 20.61 & 20.31 \\ 
1250 & 189.042969 &  62.141731 & 0.6358 &  3 & 2 &  0.174 & 22.81 & 21.74 & 20.72 & 20.39 \\ 
1452 & 189.152222 &  62.228001 & 0.5562 &  3 & 2 &  0.075 &  \nod & 22.86 & 21.72 & 21.18 \\ 
1473 & 189.088837 &  62.238129 & 0.5174 &  3 & 2 &  0.187 & 23.29 & 21.65 & 20.62 & 20.26 \\ 
1735 & 189.003113 &  62.204365 & 0.4579 &  3 & 2 &  0.022 & 23.38 & 22.05 & 21.39 & 21.00 \\ 
  27 & 189.484161 &  62.253712 & 0.1895 &  2 & 2 &  0.465 & 20.22 & 18.67 & 18.09 & 17.83 \\ 
 181 & 189.341080 &  62.176598 & 0.9782 &  3 & 2 &  0.465 & 25.56 & 23.80 & 21.81 & 21.12 \\ 
 272 & 189.415787 &  62.227119 & 0.8395 &  3 & 2 &  0.457 & 26.15 & 22.72 & 21.30 & 20.66 \\ 
 971 & 189.263702 &  62.275753 & 0.7450 &  3 & 2 &  0.268 & 29.22 & 23.44 & 21.84 & 21.30 \\ 
 973 & 189.241989 &  62.277451 & 1.0172 &  3 & 2 &  0.499 &  \nod & 24.03 & 22.31 & 21.32 \\ 
 446 & 189.356293 &  62.313828 & 0.8408 &  3 & 2 &  0.072 & 24.19 & 22.55 & 20.85 & 20.34 \\ 
 463 & 189.397217 &  62.321598 & 0.8348 &  3 & 2 &  0.735 & 24.38 & 22.51 & 21.29 & 20.81 \\ 
 509 & 189.393143 &  62.341839 & 0.9375 &  3 & 2 &  0.384 & 27.88 & 23.26 & 21.68 & 20.87 \\ 
 938 & 189.161774 &  62.263111 & 0.9037 &  3 & 2 &  0.138 & 23.24 & 22.97 & 22.16 & 21.79 \\ 
 661 & 189.194473 &  62.142559 & 0.9713 &  3 & 2 &  0.209 & 23.33 & 22.14 & 20.99 & 20.44 \\ 
 671 & 189.228439 &  62.146492 & 0.7913 &  3 & 2 &  0.187 & 25.66 & 23.01 & 21.65 & 21.12 \\ 
1478 & 189.040619 &  62.239578 & 0.7803 &  3 & 2 &  0.238 & 23.89 & 22.26 & 21.17 & 20.76 \\ 
 635 & 189.202972 &  62.128838 & 0.8195 &  2 & 1 &  1.147 &  \nod & 23.82 & 22.10 & 21.38 \\ 
 683 & 189.287384 &  62.152798 & 0.8544 &  2 & 2 &  0.296 &  \nod & 23.96 & 22.47 & 21.94 \\ 
 199 & 189.323608 &  62.187260 & 0.9028 &  2 & 2 &  0.765 &  \nod & 23.04 & 21.43 & 20.60 \\ 
 784 & 189.239410 &  62.202862 & 0.6647 &  2 & 2 &  0.189 & 26.12 & 22.63 & 21.19 & 20.72 \\ 
 273 & 189.359314 &  62.229668 & 0.4741 &  2 & 2 &  0.226 & 23.09 & 20.98 & 20.03 & 19.56 \\ 
 237 & 189.393051 &  62.211399 & 0.8506 &  2 & 1 &  1.242 & 27.29 & 23.01 & 21.28 & 20.45 \\ 
 292 & 189.434189 &  62.232899 & 0.5115 &  2 & 2 &  0.247 & 22.28 & 22.00 & 20.95 & 20.60 \\ 
 377 & 189.446106 &  62.275520 & 0.4397 &  2 & 2 &  0.511 & 23.46 & 20.70 & 19.72 & 19.28 \\ 
  18 & 189.472183 &  62.248280 & 0.5107 &  2 & 2 &  0.599 & 23.53 & 21.03 & 19.94 & 19.53 \\ 
  17 & 189.469177 &  62.246952 & 0.5111 &  2 & 2 &  0.314 & 23.21 & 21.21 & 20.05 & 19.64 \\ 
  47 & 189.481873 &  62.261803 & 0.4588 &  2 & 2 &  0.307 & 23.61 & 21.13 & 20.08 & 19.66 \\ 
 402 & 189.415833 &  62.283752 & 0.4579 &  2 & 2 &  0.274 & 23.80 & 21.71 & 20.80 & 20.38 \\ 
  38 & 189.477448 &  62.257439 & 0.4563 &  2 & 2 &  0.730 & 24.01 & 21.74 & 20.75 & 20.38 \\ 
  83 & 189.482544 &  62.282333 & 0.4415 &  2 & 2 &  0.450 & 24.88 & 21.84 & 20.98 & 20.54 \\ 
  63 & 189.506577 &  62.274181 & 0.4834 &  2 & 2 &  0.373 & 24.93 & 22.39 & 21.37 & 20.97 \\ 
 369 & 189.430466 &  62.270500 & 0.6329 &  2 & 2 &  0.529 & 25.16 & 22.59 & 21.56 & 21.12 \\ 
 115 & 189.454468 &  62.305649 & 0.9131 &  2 & 2 &  0.423 & 27.14 & 23.85 & 22.38 & 21.63 \\ 
 400 & 189.438248 &  62.282032 & 0.8517 &  2 & 2 &  0.245 &  \nod & 24.11 & 22.57 & 21.89 \\ 
1267 & 189.033600 &  62.147469 & 0.4096 &  2 & 2 &  0.431 & 23.13 & 20.76 & 19.88 & 19.50 \\ 
1373 & 189.056183 &  62.188499 & 1.1396 &  2 & 1 &  1.094 & 26.48 & 23.40 & 21.99 & 20.85 \\ 
1308 & 189.073105 &  62.162781 & 0.8434 &  2 & 2 &  0.342 &  \nod & 23.06 & 22.28 & 21.60 \\ 
 871 & 189.283920 &  62.239635 & 0.5638 &  2 & 2 &  0.712 & 24.36 & 20.88 & 19.87 & 19.43 \\ 
 879 & 189.262207 &  62.239880 & 0.5113 &  2 & 2 &  0.302 & 22.65 & 21.24 & 20.47 & 20.11 \\ 
 870 & 189.172882 &  62.234032 & 0.5568 &  2 & 2 &  0.750 & 24.73 & 21.90 & 20.81 & 20.46 \\ 
 895 & 189.195358 &  62.246452 & 0.5574 &  2 & 2 &  0.518 & 25.13 & 22.14 & 21.07 & 20.64 \\ 
 405 & 189.360703 &  62.287121 & 0.5643 &  2 & 1 &  1.091 & 25.48 & 22.34 & 21.17 & 20.72 \\ 
 399 & 189.309631 &  62.282314 & 0.4471 &  2 & 2 &  0.310 & 24.16 & 22.09 & 21.17 & 20.75 \\ 
 345 & 189.334961 &  62.255814 & 0.9369 &  2 & 2 &  0.266 &  \nod & 23.75 & 22.28 & 21.32 \\ 
 346 & 189.334961 &  62.256569 & 0.9368 &  2 & 1 &  0.964 &  \nod & 24.32 & 22.42 & 21.35 \\ 
 417 & 189.400208 &  62.290531 & 0.7089 &  2 & 2 &  0.194 & 26.38 & 22.60 & 21.45 & 21.00 \\ 
1743 & 188.956741 &  62.210548 & 0.4590 &  2 & 2 &  0.359 & 23.86 & 21.75 & 20.80 & 20.35 \\ 
1403 & 189.040039 &  62.199657 & 0.7836 &  2 & 2 &  0.415 & 25.52 & 23.54 & 22.08 & 21.53 \\ 
1385 & 189.025681 &  62.192234 & 0.7476 &  2 & 2 &  0.406 &  \nod & 23.64 & 22.06 & 21.38 \\ 
1463 & 189.082047 &  62.231449 & 0.2861 &  2 & 2 &  0.310 & 22.59 & 21.14 & 20.42 & 20.09 \\ 
1417 & 189.063751 &  62.206032 & 0.3197 &  2 & 2 &  0.036 & 23.32 & 21.60 & 20.83 & 20.50 \\ 
1506 & 189.080963 &  62.251549 & 0.7822 &  2 & 2 &  0.243 & 27.74 & 23.86 & 22.35 & 21.87 \\ 
1577 & 189.158295 &  62.291393 & 0.5557 &  2 & 2 &  0.150 & 23.51 & 22.12 & 20.88 & 20.45 \\ 
1032 & 189.282242 &  62.314278 & 0.2009 &  2 & 2 &  0.931 & 21.12 & 19.51 & 18.84 & 18.54 \\ 
1021 & 189.277252 &  62.305099 & 0.8563 &  2 & 2 &  0.510 &  \nod & 24.30 & 22.79 & 22.11 \\ 
1088 & 189.289215 &  62.352760 & 0.9439 &  2 & 1 &  1.054 &  \nod & 24.37 & 22.65 & 21.72 \\ 
 811 & 189.179886 &  62.211720 & 0.8480 &  2 & 1 &  1.183 &  \nod & 23.32 & 21.78 & 20.96 \\ 
1354 & 189.135315 &  62.184814 & 0.5183 &  2 & 2 &  0.187 & 23.07 & 20.99 & 19.81 & 19.33 \\ 
1364 & 189.146957 &  62.186081 & 0.4100 &  2 & 1 &  0.971 & 23.78 & 21.75 & 20.79 & 20.44 \\ 
1633 & 189.011459 &  62.158810 & 0.8405 &  2 & 2 &  0.332 & 25.54 & 22.37 & 20.87 & 20.31 \\ 
 655 & 189.195679 &  62.139729 & 0.6421 &  1 & 2 &  0.230 & 24.43 & 22.46 & 21.26 & 20.88 \\ 
 681 & 189.162018 &  62.151016 & 0.8422 &  1 & 2 &  0.728 & 28.95 & 23.54 & 21.98 & 21.23 \\ 
 664 & 189.241379 &  62.143196 & 0.6780 &  1 & 2 &  0.413 &  \nod & 23.25 & 21.87 & 21.42 \\ 
 752 & 189.281876 &  62.186661 & 0.4712 &  1 & 1 &  1.129 & 24.74 & 22.06 & 21.18 & 20.82 \\ 
 253 & 189.319580 &  62.219498 & 0.8989 &  1 & 1 &  1.031 & 24.03 & 23.43 & 22.43 & 22.13 \\ 
 256 & 189.438126 &  62.222931 & 0.9737 &  1 & 2 &  0.325 & 28.40 & 23.07 & 21.38 & 20.30 \\ 
1261 & 189.084595 &  62.145679 & 1.0168 &  1 & 1 &  1.105 & 32.31 & 23.02 & 21.90 & 20.83 \\ 
1285 & 189.086288 &  62.151867 & 1.0124 &  1 & 2 &  0.294 &  \nod & 24.27 & 22.15 & 21.35 \\ 
1336 & 189.083191 &  62.172337 & 0.4541 &  1 & 2 &  0.436 & 23.49 & 22.45 & 21.86 & 21.47 \\ 
1328 & 189.071228 &  62.169838 & 0.8450 &  1 & 1 &  0.984 & 25.39 & 23.59 & 22.37 & 21.50 \\ 
1297 & 189.128021 &  62.158115 & 1.0157 &  1 & 1 &  1.043 &  \nod & 24.61 & 22.87 & 21.86 \\ 
 876 & 189.263840 &  62.238331 & 0.5118 &  1 & 1 &  1.077 & 24.91 & 21.98 & 20.92 & 20.55 \\ 
 815 & 189.209549 &  62.212666 & 0.6803 &  1 & 2 &  0.360 & 27.10 & 22.89 & 21.49 & 20.96 \\ 
 930 & 189.217911 &  62.260303 & 0.9395 &  1 & 1 &  0.968 &  \nod & 24.59 & 22.26 & 21.08 \\ 
1706 & 189.007675 &  62.190731 & 0.9130 &  1 & 1 &  1.025 &  \nod & 23.85 & 21.88 & 21.32 \\ 
1741 & 189.002594 &  62.208782 & 0.4576 &  1 & 2 &  0.654 & 24.41 & 22.81 & 22.11 & 21.73 \\ 
1503 & 189.112671 &  62.252647 & 0.7937 &  1 & 2 &  0.388 & 26.56 & 22.64 & 20.91 & 20.14 \\ 
1464 & 189.069427 &  62.230881 & 0.6390 &  1 & 2 &  0.447 & 28.64 & 23.30 & 22.07 & 21.64 \\ 
 959 & 189.179977 &  62.273636 & 0.8511 &  1 & 2 &  0.248 &  \nod & 22.62 & 21.28 & 20.59 \\ 
1559 & 189.155685 &  62.279930 & 0.9428 &  1 & 2 &  0.190 &  \nod & 23.68 & 22.17 & 21.34 \\ 
1046 & 189.189758 &  62.327511 & 0.2766 &  1 & 1 &  1.117 & 22.79 & 21.17 & 20.51 & 20.20 \\ 
1020 & 189.248611 &  62.304062 & 0.4969 &  1 & 2 &  0.262 & 25.05 & 23.22 & 22.36 & 22.00 \\ 
1084 & 189.254608 &  62.352547 & 0.7995 &  1 & 1 &  1.157 & 27.32 & 22.51 & 20.75 & 19.87 \\ 
1076 & 189.251724 &  62.344421 & 0.8011 &  1 & 2 &  0.431 &  \nod & 24.10 & 22.28 & 21.60 \\ 
1066 & 189.305054 &  62.336098 & 0.6348 &  1 & 2 &  0.471 & 24.39 & 23.02 & 22.09 & 21.72 \\ 
 818 & 189.192215 &  62.212917 & 0.9043 &  1 & 1 &  1.110 & 26.66 & 23.74 & 22.33 & 21.74 \\ 
1399 & 189.153137 &  62.198898 & 0.5560 &  1 & 1 &  0.874 & 24.35 & 22.86 & 21.88 & 21.45 \\ 
1430 & 189.115784 &  62.211319 & 0.5185 &  1 & 2 &  0.242 & 25.30 & 22.37 & 21.27 & 20.86 \\ 
1287 & 189.132477 &  62.151817 & 0.8457 &  0 & 1 &  1.196 & 25.88 & 22.42 & 20.82 & 19.96 \\ 
 648 & 189.165482 &  62.138489 & 0.2469 &  0 & 1 &  1.050 & 22.53 & 20.83 & 20.21 & 19.95 \\ 
 609 & 189.191315 &  62.107990 & 0.7976 &  0 & 2 &  0.464 &  \nod & 22.98 & 20.97 & 20.48 \\ 
1236 & 189.134720 &  62.137367 & 0.8503 &  0 & 1 &  1.039 & 31.12 & 23.42 & 21.63 & 20.87 \\ 
 748 & 189.191391 &  62.183578 & 0.9358 &  0 & 2 &  0.336 &  \nod & 24.48 & 22.42 & 21.48 \\ 
 206 & 189.370728 &  62.191010 & 0.9355 &  0 & 1 &  1.113 &  \nod & 24.09 & 22.42 & 21.66 \\ 
 793 & 189.235962 &  62.205502 & 0.9543 &  0 & 1 &  1.038 &  \nod & 24.20 & 22.51 & 21.75 \\ 
 329 & 189.420547 &  62.254772 & 0.3771 &  0 & 2 &  0.408 & 21.13 & 19.52 & 18.77 & 18.42 \\ 
 264 & 189.339630 &  62.226200 & 0.4756 &  0 & 1 &  1.028 & 23.00 & 20.73 & 19.64 & 19.28 \\ 
 241 & 189.374634 &  62.216820 & 0.5117 &  0 & 1 &  1.137 & 23.11 & 21.06 & 19.95 & 19.49 \\ 
 244 & 189.411011 &  62.214752 & 0.4549 &  0 & 2 &  0.682 & 24.31 & 21.50 & 20.42 & 20.05 \\ 
 291 & 189.424408 &  62.232937 & 0.5118 &  0 & 1 &  1.028 & 24.05 & 21.80 & 20.77 & 20.42 \\ 
 302 & 189.403534 &  62.237892 & 0.6906 &  0 & 1 &  0.876 & 27.72 & 22.53 & 21.27 & 20.73 \\ 
  43 & 189.488739 &  62.263248 & 0.4574 &  0 & 1 &  0.847 & 23.44 & 20.15 & 19.25 & 18.90 \\ 
1346 & 189.033875 &  62.176636 & 0.6790 &  0 & 1 &  0.945 & 23.92 & 22.09 & 20.83 & 20.36 \\ 
1292 & 189.120224 &  62.156509 & 0.5184 &  0 & 1 &  0.984 & 25.14 & 22.21 & 21.14 & 20.72 \\ 
1303 & 189.094666 &  62.162849 & 0.7476 &  0 & 1 &  1.094 & 27.06 & 22.89 & 21.49 & 20.97 \\ 
1342 & 189.030716 &  62.175411 & 0.6397 &  0 & 1 &  0.944 & 25.59 & 22.62 & 21.45 & 20.77 \\ 
1353 & 189.017761 &  62.180531 & 0.8512 &  0 & 1 &  1.104 & 25.98 & 23.16 & 21.74 & 21.12 \\ 
1329 & 189.035889 &  62.170059 & 0.6354 &  0 & 1 &  1.102 & 26.41 & 22.97 & 21.67 & 21.14 \\ 
 884 & 189.220352 &  62.245541 & 0.3209 &  0 & 1 &  1.169 & 21.78 & 19.73 & 18.96 & 18.65 \\ 
 841 & 189.200409 &  62.219128 & 0.4745 &  0 & 2 &  0.382 & 24.14 & 21.46 & 20.43 & 20.04 \\ 
 885 & 189.289917 &  62.239880 & 0.4754 &  0 & 2 &  0.228 & 24.06 & 22.11 & 21.09 & 20.72 \\ 
 872 & 189.193176 &  62.234570 & 0.9602 &  0 & 2 &  0.277 & 24.19 & 22.69 & 21.59 & 20.81 \\ 
 912 & 189.220871 &  62.252369 & 0.9390 &  0 & 1 &  0.875 &  \nod & 23.96 & 22.29 & 21.36 \\ 
 804 & 189.252426 &  62.209660 & 0.5630 &  0 & 1 &  0.992 & 25.14 & 22.49 & 21.37 & 20.95 \\ 
 854 & 189.254532 &  62.222111 & 0.6645 &  0 & 1 &  1.206 &  \nod & 23.83 & 22.45 & 21.99 \\ 
 353 & 189.343216 &  62.262062 & 0.9379 &  0 & 1 &  1.048 &  \nod & 24.01 & 22.34 & 21.33 \\ 
 468 & 189.391724 &  62.327000 & 0.2728 &  0 & 1 &  1.099 & 21.38 & 19.42 & 18.69 & 18.42 \\ 
1718 & 189.002289 &  62.198837 & 0.5612 &  0 & 1 &  0.889 & 23.44 & 20.93 & 19.79 & 19.41 \\ 
1362 & 189.013672 &  62.186432 & 0.6376 &  0 & 2 &  0.550 & 22.48 & 21.14 & 20.21 & 19.88 \\ 
1709 & 188.952637 &  62.192177 & 0.8417 &  0 & 2 &  0.782 & 24.55 & 22.58 & 21.04 & 20.54 \\ 
1740 & 188.984238 &  62.208260 & 0.6385 &  0 & 2 &  0.216 & 24.92 & 22.46 & 21.35 & 21.03 \\ 
1454 & 189.073685 &  62.228970 & 0.5338 &  0 & 1 &  1.055 & 24.22 & 22.06 & 20.87 & 20.46 \\ 
1491 & 189.103775 &  62.244114 & 0.6404 &  0 & 2 &  0.312 & 26.27 & 22.65 & 21.29 & 20.84 \\ 
1440 & 189.104462 &  62.216911 & 0.5185 &  0 & 2 &  0.552 & 24.46 & 21.86 & 20.68 & 20.27 \\ 
 957 & 189.193176 &  62.274834 & 0.5033 &  0 & 1 &  0.932 & 24.85 & 21.43 & 20.52 & 20.16 \\ 
 933 & 189.165771 &  62.263260 & 0.8469 &  0 & 1 &  0.945 &  \nod & 23.30 & 21.47 & 20.39 \\ 
1558 & 189.155106 &  62.279419 & 0.8518 &  0 & 1 &  1.019 &  \nod & 22.97 & 21.58 & 20.90 \\ 
1543 & 189.122101 &  62.270439 & 0.8487 &  0 & 2 &  0.386 & 24.36 & 22.99 & 21.45 & 20.96 \\ 
 951 & 189.159271 &  62.268051 & 0.8542 &  0 & 2 &  0.433 &  \nod & 23.84 & 21.84 & 20.90 \\ 
 982 & 189.167389 &  62.282120 & 0.9422 &  0 & 2 &  0.452 & 25.87 & 23.53 & 22.10 & 21.41 \\ 
1061 & 189.212814 &  62.340851 & 0.7449 &  0 & 1 &  0.893 & 26.76 & 22.62 & 20.85 & 20.17 \\ 
 454 & 189.310379 &  62.316208 & 0.4971 &  0 & 1 &  0.950 & 24.62 & 22.95 & 21.92 & 21.53 \\ 
 505 & 189.326355 &  62.343208 & 0.2541 &  0 & 2 &  0.633 & 21.45 & 19.68 & 18.94 & 18.71 \\ 
1107 & 189.274551 &  62.364063 & 0.3303 &  0 & 1 &  1.056 & 22.57 & 20.39 & 19.62 & 19.25 \\ 
 506 & 189.308426 &  62.343449 & 0.5325 &  0 & 2 &  0.625 & 23.14 & 20.93 & 20.01 & 19.70 \\ 
 783 & 189.193878 &  62.197567 & 0.5041 &  0 & 1 &  1.022 & 25.65 & 22.95 & 21.94 & 21.56 \\ 
 761 & 189.184982 &  62.192547 & 1.0127 &  0 & 1 &  0.879 & 28.49 & 23.16 & 21.41 & 20.23 \\ 
 774 & 189.182571 &  62.195236 & 0.7657 &  0 & 1 &  1.181 & 25.86 & 22.42 & 20.79 & 20.37 \\ 
 827 & 189.167221 &  62.218197 & 0.4857 &  0 & 2 &  0.438 & 24.03 & 21.61 & 20.56 & 20.18 \\ 
1436 & 189.155685 &  62.214550 & 0.4843 &  0 & 1 &  0.987 & 25.13 & 22.14 & 21.03 & 20.56 \\ 
 738 & 189.240448 &  62.178017 & 0.9366 & -1 & 1 &  0.928 & 25.60 & 24.26 & 22.57 & 22.03 \\ 
1323 & 189.024017 &  62.167500 & 0.9362 & -1 & 1 &  1.094 & 24.99 & 23.67 & 22.59 & 22.04 

\enddata

%% Include any \tablenotetext{key}{text}, \tablerefs{ref list},
%% or \tablecomments{text} between the \enddata and 
%% \end{deluxetable} commands
\tablecomments{
%Table \ref{tab:galpars} will be published in its entirety 
%in the electronic edition of the {\it Astrophysical Journal}. A portion is 
%shown here for guidance regarding its form and content.  
Col.\@ (1): Internal galaxy ID.  
Cols.\@ (2) \& (3): right ascention \& declination in J2000 epoch coordinates. 
Col.\@ (4): spectroscopic redshift.  
Col.\@ (5): galaxy T type according to T05.  
Col.\@ (6): number of components in decomposition (1 $\equiv$ 
 single \sersic\ profile in fit; 2 $\equiv$ simultaneous \sersic\ bulge 
 plus exponential disk decomposition).  
Col.\@ (7): Galaxy bulge-to-total ratio.  
Cols.\@ (8--11): Total observed galaxy magnitudes corrected for Galactic
extinction (see Table~\ref{tab:ACS}) in all four GOODS-ACS filters.}
%% No \tablecomments indicated

%% No \tablerefs indicated
\end{deluxetable*}
\clearpage

\begin{deluxetable*}{rccccccccccccc}
%\tabletypesize{\scriptsize}
\tablecaption{Spheroid Photometric Parameters. \label{tab:photpars}}

\tablehead{\colhead{ID} & \colhead{$n_{B}$}  & \colhead{$r_{e,B}$}  & 
\colhead{$\mu_{e,B}$} & \colhead{$n_{V}$}  & \colhead{$r_{e,V}$} & 
\colhead{$\mu_{e,V}$} & \colhead{$n_{i}$}  & \colhead{$r_{e,i}$}  & 
\colhead{$\mu_{e,i}$} & \colhead{$n_{z}$}  & \colhead{$r_{e,z}$}  & 
\colhead{$\mu_{e,z}$} & \colhead{$V-i$} \\ 
\colhead{} & \colhead{} & \colhead{(\arcsec)} & \colhead{(\magsqarc)} & 
\colhead{} & \colhead{(\arcsec)} & \colhead{(\magsqarc)} & \colhead{} & 
\colhead{(\arcsec)} & \colhead{(\magsqarc)} & \colhead{} & 
\colhead{(\arcsec)} & \colhead{(\magsqarc)} & \colhead{(to $r_{e,z}$)} \\
\colhead{(1)} & \colhead{(2)} & \colhead{(3)} & \colhead{(4)} & 
\colhead{(5)} & \colhead{(6)} & \colhead{(7)} & \colhead{(8)} & 
\colhead{(9)} & \colhead{(10)} & \colhead{(11)} & \colhead{(12)} & 
\colhead{(13)} & \colhead{(14)} } 

\startdata
%Gal    Bn     Bre    Bme     Vn     Vre    Vme     In     Ire    Ime     Zn     Zre    Zme    V-I(re)
  62 & 1.9  & 0.10 & 19.10 & 2.6  & 0.28 & 20.67 & 2.2  & 0.26 & 19.80 & 2.0  & 0.26 & 19.33  & 0.68 \\
 198 & 0.9  & 0.31 & 23.95 & 1.3  & 0.34 & 22.42 & 1.4  & 0.32 & 21.34 & 1.4  & 0.33 & 21.08  & 1.04 \\
 280 & 0.9  & 0.15 & 22.37 & 0.9  & 0.18 & 21.65 & 1.0  & 0.19 & 21.38 & 1.1  & 0.23 & 21.36  & 0.37 \\
 319 & 0.9  & 0.08 & 23.57 & 0.8  & 0.11 & 22.05 & 0.9  & 0.13 & 21.43 & 0.9  & 0.14 & 21.32  & 0.84 \\
 352 & 1.1  & 0.13 & 25.05 & 1.1  & 0.14 & 23.18 & 1.3  & 0.21 & 22.50 & 1.4  & 0.28 & 22.53  & 1.31 \\
 384 & 3.0  & 0.00 & 28.56 & 0.5  & 0.05 & 21.28 & 0.7  & 0.05 & 20.48 & 0.8  & 0.06 & 20.56  & 1.00 \\
 421 & 0.5  & 0.07 & 22.87 & 0.4  & 0.07 & 21.86 & 0.5  & 0.07 & 21.27 & 0.2  & 0.06 & 20.64  & 0.71 \\
 438 & 0.1  & 0.01 & 22.13 & 0.9  & 0.11 & 22.71 & 1.0  & 0.13 & 22.19 & 1.3  & 0.20 & 22.37  & 0.90 \\
 462 & 1.0  & 0.14 & 22.45 & 1.6  & 0.30 & 21.17 & 1.6  & 0.29 & 20.21 & 1.5  & 0.30 & 19.98  & 0.90 \\
 465 & 1.2  & 0.20 & 24.03 & 1.0  & 0.16 & 21.74 & 0.9  & 0.16 & 20.71 & 0.9  & 0.16 & 20.57  & 0.94 \\
 519 & \nod & \nod & \nod  & 1.0  & 0.33 & 23.15 & 1.2  & 0.32 & 22.48 & 1.0  & 0.30 & 21.78  & 0.71 \\
 667 & 0.3  & 0.16 & 22.77 & 0.6  & 0.13 & 21.16 & 1.0  & 0.14 & 20.63 & 1.1  & 0.14 & 20.52  & 0.78 \\
 710 & 0.9  & 0.11 & 22.89 & 0.9  & 0.14 & 22.34 & 0.7  & 0.11 & 21.67 & 0.7  & 0.12 & 22.03  & 0.32 \\
 723 & 0.1  & 0.02 & 23.06 & 2.9  & 0.00 & 27.87 & 4.9  & 0.01 & 30.37 & 0.1  & 0.00 & 19.84  & 0.75 \\
 790 & 1.6  & 0.02 & 26.10 & 1.1  & 0.06 & 23.68 & 1.7  & 0.08 & 24.41 & 0.8  & 0.02 & 22.18 & -0.06 \\
 793 & 0.7  & 0.14 & 24.06 & 0.8  & 0.17 & 22.40 & 0.8  & 0.19 & 21.74 & 0.9  & 0.22 & 21.71  & 0.82 \\
 873 & 0.2  & 0.04 & 21.41 & 0.8  & 0.09 & 20.73 & 0.9  & 0.09 & 19.93 & 1.1  & 0.14 & 20.14  & 0.86 \\
 920 & 0.5  & 0.08 & 24.70 & 0.8  & 0.13 & 22.63 & 0.7  & 0.12 & 21.37 & 0.9  & 0.15 & 21.46  & 1.15 \\
 928 & 1.3  & 0.06 & 19.08 & 2.0  & 0.10 & 20.32 & 1.9  & 0.15 & 20.05 & 1.7  & 0.18 & 19.93  & 0.80 \\
 956 & 0.1  & 0.00 & 24.34 & 0.8  & 0.09 & 25.10 & 1.0  & 0.13 & 24.12 & 0.7  & 0.13 & 24.08  & 1.47 \\
 974 & 0.9  & 0.12 & 24.61 & 1.1  & 0.18 & 22.85 & 0.9  & 0.14 & 21.74 & 1.0  & 0.14 & 21.78  & 0.78 \\
 975 & 1.3  & 0.41 & 23.93 & 1.3  & 0.32 & 22.50 & 1.1  & 0.23 & 21.35 & 1.2  & 0.28 & 21.28  & 0.67 \\
 989 & 0.6  & 0.00 & 24.25 & 0.6  & 0.07 & 21.82 & 0.6  & 0.07 & 20.73 & 0.6  & 0.08 & 20.59  & 1.10 \\
 991 & \nod & \nod & \nod  & \nod & \nod & \nod  & \nod & \nod & \nod  & \nod & \nod & \nod   & \nod \\
1050 & 1.0  & 0.16 & 23.54 & 1.6  & 0.31 & 22.01 & 2.0  & 0.52 & 21.58 & 1.9  & 0.51 & 21.14  & 1.28 \\
1310 & \nod & \nod & \nod  & 0.2  & 0.01 & 24.70 & 0.4  & 0.02 & 23.53 & 1.0  & 0.12 & 24.42  & 2.43 \\
1331 & 0.7  & 0.09 & 22.88 & 1.0  & 0.11 & 22.79 & 1.1  & 0.13 & 22.63 & 0.8  & 0.12 & 22.57  & 0.36 \\
1377 & 2.1  & 0.55 & 23.52 & 1.8  & 0.49 & 21.79 & 1.8  & 0.46 & 20.95 & 1.7  & 0.44 & 20.58  & 0.76 \\
1378 & \nod & \nod & \nod  & \nod & \nod & \nod  & 0.5  & 0.24 & 24.41 & 0.4  & 0.24 & 23.63  & \nod \\
1413 & 0.1  & 0.01 & 21.60 & 0.3  & 0.10 & 22.64 & 0.6  & 0.14 & 22.51 & 0.7  & 0.10 & 22.51  & 0.55 \\
1415 & \nod & \nod & \nod  & 0.5  & 0.11 & 22.46 & 1.0  & 0.15 & 21.65 & 0.9  & 0.15 & 21.18  & 1.38 \\
1419 & 0.7  & 0.05 & 23.38 & 0.7  & 0.04 & 22.87 & 0.9  & 0.04 & 22.88 & 0.1  & 0.03 & 22.00  & 0.06 \\
1422 & 1.0  & 0.13 & 23.32 & 0.9  & 0.08 & 21.35 & 1.2  & 0.09 & 20.67 & 1.3  & 0.12 & 20.75  & 0.89 \\
1429 & \nod & \nod & \nod  & 0.2  & 0.14 & 24.27 & 1.5  & 0.37 & 24.55 & 1.0  & 0.25 & 23.50  & 1.39 \\
1471 & \nod & \nod & \nod  & 0.2  & 0.11 & 22.67 & 0.4  & 0.10 & 22.05 & 0.6  & 0.14 & 22.33  & 0.64 \\
1489 & 0.6  & 0.08 & 21.89 & 0.6  & 0.08 & 21.46 & 0.9  & 0.12 & 21.30 & 0.9  & 0.12 & 21.59  & 0.73 \\
1690 & 1.0  & 0.14 & 22.88 & 0.9  & 0.16 & 21.84 & 1.0  & 0.19 & 21.68 & 1.0  & 0.22 & 21.70  & 0.39 \\
  92 & 0.6  & 0.01 & 25.43 & 0.4  & 0.02 & 21.57 & 1.0  & 0.11 & 22.23 & 1.1  & 0.12 & 22.36  & 3.17 \\
  94 & \nod & \nod & \nod  & 0.9  & 0.07 & 23.16 & 0.8  & 0.08 & 21.65 & 1.0  & 0.09 & 21.49  & 1.60 \\
 170 & \nod & \nod & \nod  & 0.8  & 0.15 & 21.95 & 0.8  & 0.13 & 20.73 & 1.0  & 0.14 & 20.52  & 1.09 \\
 211 & 1.1  & 0.59 & 24.55 & 1.1  & 0.45 & 22.98 & 1.2  & 0.44 & 22.33 & 1.2  & 0.50 & 22.18  & 0.66 \\
 254 & \nod & \nod & \nod  & 1.1  & 0.28 & 22.77 & 1.3  & 0.31 & 22.06 & 1.2  & 0.31 & 21.64  & 0.90 \\
 289 & \nod & \nod & \nod  & 0.9  & 0.24 & 21.83 & 0.8  & 0.24 & 21.41 & 0.8  & 0.28 & 21.56  & 0.40 \\
 318 & 1.4  & 0.15 & 23.55 & 1.2  & 0.11 & 22.62 & 1.0  & 0.12 & 21.87 & 1.2  & 0.16 & 22.18  & 0.80 \\
 548 & 1.3  & 0.28 & 22.74 & 1.4  & 0.30 & 20.77 & 1.6  & 0.35 & 20.13 & 1.6  & 0.36 & 19.88  & 0.88 \\
 698 & 2.0  & 0.93 & 23.80 & 1.9  & 0.85 & 21.75 & 1.8  & 0.68 & 20.79 & 1.6  & 0.64 & 20.32  & 0.66 \\
 704 & \nod & \nod & \nod  & 0.3  & 0.12 & 23.57 & 0.4  & 0.11 & 22.09 & 0.4  & 0.12 & 21.89  & 1.53 \\
 750 & 0.3  & 0.05 & 23.89 & 1.1  & 0.00 & 21.97 & 0.2  & 0.00 & 18.91 & 0.4  & 0.00 & 19.12  & 5.71 \\
 891 & 0.8  & 0.10 & 21.22 & 0.8  & 0.09 & 20.53 & 0.8  & 0.09 & 20.12 & 1.0  & 0.11 & 20.67  & 0.38 \\
1008 & 0.5  & 0.43 & 24.44 & 0.7  & 0.45 & 22.56 & 0.8  & 0.47 & 21.77 & 0.6  & 0.44 & 21.21  & 0.89 \\
1037 & 0.8  & 0.13 & 21.81 & 1.0  & 0.17 & 21.51 & 1.2  & 0.20 & 20.86 & 0.9  & 0.20 & 20.75  & 0.86 \\
1250 & \nod & \nod & \nod  & 1.0  & 0.12 & 22.09 & 0.9  & 0.11 & 20.60 & 1.1  & 0.13 & 20.67  & 1.36 \\
1452 & \nod & \nod & \nod  & 0.7  & 0.10 & 23.89 & 0.5  & 0.11 & 21.94 & 0.8  & 0.11 & 21.82  & 1.83 \\
1473 & 0.8  & 0.10 & 23.87 & 1.0  & 0.13 & 21.98 & 1.0  & 0.12 & 20.69 & 1.0  & 0.13 & 20.45  & 1.24 \\
1735 & 0.6  & 0.04 & 21.05 & 1.0  & 0.04 & 20.71 & 1.2  & 0.04 & 20.65 & 1.2  & 0.03 & 20.57  & 0.22 \\
  27 & 1.3  & 0.20 & 20.51 & 1.4  & 0.19 & 18.91 & 1.5  & 0.20 & 18.40 & 1.4  & 0.20 & 18.24  & 0.60 \\
 181 & \nod & \nod & \nod  & 1.8  & 0.49 & 24.33 & 1.0  & 0.16 & 21.57 & 1.2  & 0.21 & 21.40  & 1.22 \\
 272 & \nod & \nod & \nod  & 1.1  & 0.21 & 22.82 & 1.0  & 0.22 & 21.41 & 1.0  & 0.22 & 20.94  & 1.43 \\
 971 & \nod & \nod & \nod  & 1.0  & 0.13 & 22.82 & 1.0  & 0.11 & 21.14 & 1.2  & 0.13 & 21.05  & 1.47 \\
 973 & \nod & \nod & \nod  & 1.0  & 0.12 & 23.15 & 0.9  & 0.11 & 21.32 & 1.3  & 0.17 & 21.13  & 1.67 \\
 446 & \nod & \nod & \nod  & \nod & \nod & \nod  & 1.1  & 0.09 & 21.20 & 0.9  & 0.09 & 20.66  & \nod \\
 463 & 1.6  & 0.52 & 24.98 & 1.4  & 0.24 & 22.97 & 1.2  & 0.22 & 21.44 & 1.4  & 0.26 & 21.07  & 1.31 \\
 509 & \nod & \nod & \nod  & 1.0  & 0.14 & 22.92 & 1.2  & 0.16 & 21.65 & 1.2  & 0.18 & 20.98  & 1.56 \\
 938 & \nod & \nod & \nod  & 1.2  & 0.40 & 28.61 & 1.2  & 0.18 & 23.45 & 1.1  & 0.15 & 22.64  & 1.97 \\
 661 & \nod & \nod & \nod  & 0.6  & 0.11 & 20.60 & 0.7  & 0.11 & 20.21 & 0.6  & 0.12 & 19.96  & 0.48 \\
 671 & \nod & \nod & \nod  & 0.9  & 0.10 & 22.35 & 0.9  & 0.10 & 20.87 & 0.8  & 0.09 & 20.43  & 1.49 \\
1478 & 0.4  & 0.13 & 23.38 & 0.8  & 0.13 & 22.03 & 1.1  & 0.12 & 20.82 & 1.1  & 0.12 & 20.54  & 1.21 \\
 635 & \nod & \nod & \nod  & 1.4  & 0.31 & 23.96 & 1.5  & 0.31 & 22.33 & 1.6  & 0.35 & 21.90  & 1.62 \\
 683 & \nod & \nod & \nod  & 1.1  & 0.08 & 22.69 & 1.3  & 0.10 & 21.58 & 1.5  & 0.13 & 21.67  & 1.47 \\
 199 & \nod & \nod & \nod  & 1.2  & 0.37 & 23.42 & 1.0  & 0.25 & 21.37 & 1.0  & 0.26 & 20.70  & 1.57 \\
 784 & \nod & \nod & \nod  & 0.8  & 0.09 & 21.83 & 0.8  & 0.09 & 20.23 & 0.9  & 0.09 & 20.09  & 1.53 \\
 273 & 0.8  & 0.10 & 23.27 & 2.0  & 0.32 & 22.77 & 2.0  & 0.30 & 21.61 & 1.8  & 0.23 & 21.01  & 1.07 \\
 237 & \nod & \nod & \nod  & 2.0  & 0.43 & 23.89 & 2.1  & 0.43 & 22.29 & 2.1  & 0.48 & 21.70  & 1.61 \\
 292 & 0.7  & 0.10 & 23.61 & 0.9  & 0.11 & 21.52 & 1.0  & 0.11 & 20.41 & 1.0  & 0.12 & 20.21  & 1.13 \\
 377 & \nod & \nod & \nod  & 1.1  & 0.29 & 22.33 & 1.2  & 0.30 & 20.97 & 1.3  & 0.36 & 20.61  & 1.45 \\
  18 & 1.4  & 0.27 & 24.17 & 1.4  & 0.28 & 21.76 & 1.4  & 0.27 & 20.56 & 1.3  & 0.29 & 20.23  & 1.15 \\
  17 & 1.1  & 0.14 & 23.40 & 1.3  & 0.17 & 21.31 & 1.2  & 0.15 & 19.95 & 1.1  & 0.16 & 19.62  & 1.20 \\
  47 & 1.0  & 0.15 & 24.03 & 1.2  & 0.21 & 21.79 & 1.2  & 0.21 & 20.64 & 1.2  & 0.20 & 20.28  & 1.11 \\
 402 & \nod & \nod & \nod  & 0.8  & 0.11 & 21.19 & 0.9  & 0.12 & 20.17 & 0.9  & 0.13 & 19.97  & 1.13 \\
  38 & 1.3  & 0.17 & 23.66 & 1.5  & 0.25 & 21.67 & 1.4  & 0.24 & 20.55 & 1.4  & 0.22 & 20.34  & 1.02 \\
  83 & \nod & \nod & \nod  & 1.2  & 0.14 & 21.26 & 1.2  & 0.14 & 20.25 & 1.2  & 0.15 & 20.06  & 1.02 \\
  63 & \nod & \nod & \nod  & 1.0  & 0.10 & 21.19 & 1.0  & 0.10 & 20.13 & 1.2  & 0.12 & 20.28  & 1.03 \\
 369 & \nod & \nod & \nod  & 1.8  & 0.28 & 23.73 & 1.5  & 0.17 & 21.74 & 1.5  & 0.23 & 21.59  & 1.25 \\
 115 & \nod & \nod & \nod  & 0.8  & 0.08 & 22.47 & 1.2  & 0.11 & 21.44 & 1.2  & 0.13 & 20.87  & 1.57 \\
 400 & \nod & \nod & \nod  & 0.9  & 0.07 & 22.65 & 1.1  & 0.08 & 21.34 & 1.2  & 0.09 & 21.06  & 1.48 \\
1267 & 0.8  & 0.13 & 22.25 & 1.2  & 0.17 & 20.53 & 1.2  & 0.16 & 19.56 & 1.1  & 0.17 & 19.34  & 0.92 \\
1373 & \nod & \nod & \nod  & 1.8  & 0.45 & 24.64 & 1.9  & 0.50 & 23.41 & 2.0  & 0.58 & 22.63  & 1.40 \\
1308 & \nod & \nod & \nod  & 0.8  & 0.10 & 22.49 & 0.8  & 0.09 & 20.83 & 1.0  & 0.11 & 20.68  & 1.55 \\
 871 & 0.8  & 0.07 & 23.72 & 2.2  & 0.47 & 23.53 & 2.1  & 0.43 & 22.00 & 2.2  & 0.65 & 21.99  & 1.37 \\
 879 & 0.2  & 0.03 & 24.50 & 1.5  & 0.43 & 22.99 & 1.4  & 0.26 & 21.50 & 1.2  & 0.22 & 20.88  & 0.97 \\
 870 & 1.1  & 0.16 & 23.92 & 1.4  & 0.18 & 21.73 & 1.6  & 0.22 & 20.76 & 1.4  & 0.22 & 20.34  & 1.23 \\
 895 & \nod & \nod & \nod  & 1.4  & 0.19 & 22.19 & 1.3  & 0.16 & 20.81 & 1.4  & 0.20 & 20.79  & 1.17 \\
 405 & \nod & \nod & \nod  & 2.2  & 0.40 & 23.42 & 2.3  & 0.41 & 22.23 & 2.1  & 0.38 & 21.64  & 1.23 \\
 399 & 0.9  & 0.10 & 22.91 & 0.9  & 0.08 & 20.76 & 0.9  & 0.09 & 19.82 & 1.0  & 0.10 & 19.86  & 1.00 \\
 345 & \nod & \nod & \nod  & 1.0  & 0.09 & 22.96 & 1.0  & 0.09 & 21.19 & 1.2  & 0.11 & 20.83  & 1.66 \\
 346 & \nod & \nod & \nod  & 1.0  & 0.29 & 24.13 & 1.3  & 0.33 & 22.86 & 1.5  & 0.39 & 22.29  & 1.53 \\
 417 & \nod & \nod & \nod  & 0.9  & 0.11 & 22.82 & 0.8  & 0.10 & 21.03 & 1.0  & 0.13 & 20.99  & 1.61 \\
1743 & 1.3  & 0.28 & 24.66 & 1.3  & 0.21 & 22.27 & 1.4  & 0.20 & 21.21 & 1.3  & 0.20 & 20.86  & 1.08 \\
1403 & \nod & \nod & \nod  & 1.3  & 0.11 & 22.27 & 1.2  & 0.09 & 20.63 & 1.4  & 0.12 & 20.75  & 1.39 \\
1385 & \nod & \nod & \nod  & 0.9  & 0.10 & 22.33 & 1.0  & 0.10 & 20.83 & 1.1  & 0.13 & 20.64  & 1.60 \\
1463 & 1.2  & 0.16 & 22.96 & 1.1  & 0.14 & 21.06 & 1.2  & 0.15 & 20.38 & 1.3  & 0.18 & 20.43  & 0.78 \\
1417 & 0.9  & 0.06 & 24.28 & 0.5  & 0.05 & 21.71 & 1.0  & 0.07 & 21.26 & 1.0  & 0.07 & 20.99  & 1.14 \\
1506 & \nod & \nod & \nod  & 1.2  & 0.14 & 23.75 & 0.9  & 0.09 & 21.66 & 1.1  & 0.12 & 21.59  & 1.42 \\
1577 & 0.7  & 0.09 & 23.52 & 0.8  & 0.08 & 21.07 & 0.9  & 0.07 & 19.79 & 1.0  & 0.08 & 19.88  & 1.26 \\
1032 & 2.3  & 0.49 & 23.20 & 2.3  & 0.51 & 21.36 & 2.3  & 0.52 & 20.66 & 2.0  & 0.45 & 20.10  & 0.72 \\
1021 & \nod & \nod & \nod  & 0.9  & 0.11 & 23.58 & 1.1  & 0.14 & 22.52 & 1.3  & 0.21 & 22.27  & 1.48 \\
1088 & \nod & \nod & \nod  & \nod & \nod & \nod  & 1.5  & 0.18 & 21.95 & 1.7  & 0.25 & 21.65  & \nod \\
 811 & \nod & \nod & \nod  & 2.0  & 0.39 & 24.03 & 2.1  & 0.40 & 22.55 & 2.0  & 0.39 & 21.80  & 1.52 \\
1354 & 0.7  & 0.08 & 23.92 & 0.9  & 0.11 & 21.68 & 1.0  & 0.12 & 20.50 & 1.4  & 0.25 & 21.01  & 1.27 \\
1364 & 1.5  & 0.24 & 23.68 & 1.9  & 0.27 & 22.04 & 1.9  & 0.27 & 21.00 & 1.8  & 0.28 & 20.69  & 0.98 \\
1633 & \nod & \nod & \nod  & 1.5  & 0.20 & 22.82 & 1.6  & 0.18 & 21.24 & 1.5  & 0.18 & 20.65  & 1.47 \\
 655 & 0.2  & 0.13 & 23.17 & 0.6  & 0.14 & 22.08 & 0.9  & 0.15 & 21.35 & 0.9  & 0.16 & 21.22  & 0.97 \\
 681 & \nod & \nod & \nod  & 0.9  & 0.16 & 22.70 & 1.2  & 0.20 & 21.43 & 1.2  & 0.20 & 20.88  & 1.62 \\
 664 & \nod & \nod & \nod  & 0.9  & 0.08 & 21.81 & 1.2  & 0.10 & 20.73 & 1.4  & 0.13 & 20.85  & 1.54 \\
 752 & \nod & \nod & \nod  & 2.3  & 0.44 & 23.33 & 2.3  & 0.42 & 22.31 & 2.2  & 0.43 & 21.97  & 0.94 \\
 253 & \nod & \nod & \nod  & 1.6  & 0.12 & 21.94 & 2.2  & 0.21 & 22.29 & 2.1  & 0.25 & 22.18  & 0.67 \\
 256 & \nod & \nod & \nod  & 1.0  & 0.19 & 23.08 & 1.1  & 0.20 & 21.46 & 1.0  & 0.18 & 20.46  & 1.69 \\
1261 & \nod & \nod & \nod  & 1.8  & 0.57 & 25.12 & 2.0  & 0.61 & 23.74 & 2.0  & 0.64 & 22.82  & 1.51 \\
1285 & \nod & \nod & \nod  & 1.0  & 0.16 & 23.96 & 1.3  & 0.17 & 22.49 & 1.1  & 0.15 & 21.37  & 1.71 \\
1336 & \nod & \nod & \nod  & 0.9  & 0.18 & 21.26 & 0.7  & 0.14 & 20.63 & 0.7  & 0.15 & 20.84  & 0.27 \\
1328 & \nod & \nod & \nod  & 1.4  & 0.18 & 22.84 & 1.8  & 0.29 & 22.63 & 1.9  & 0.37 & 22.40  & 1.01 \\
1297 & \nod & \nod & \nod  & 1.7  & 0.33 & 24.75 & 1.5  & 0.26 & 22.88 & 1.6  & 0.31 & 22.23  & 1.53 \\
 876 & \nod & \nod & \nod  & 2.2  & 0.43 & 23.25 & 2.3  & 0.43 & 22.22 & 2.3  & 0.48 & 22.00  & 1.04 \\
 815 & \nod & \nod & \nod  & 1.0  & 0.10 & 22.27 & 1.8  & 0.19 & 21.86 & 1.4  & 0.16 & 21.03  & 1.48 \\
 930 & \nod & \nod & \nod  & \nod & \nod & \nod  & 2.0  & 0.39 & 23.36 & 2.4  & 0.63 & 23.25  & \nod \\
1706 & \nod & \nod & \nod  & 1.4  & 0.24 & 23.77 & 1.7  & 0.27 & 22.47 & 1.8  & 0.33 & 21.90  & 1.57 \\
1741 & \nod & \nod & \nod  & 0.9  & 0.19 & 22.29 & 0.9  & 0.19 & 21.73 & 1.0  & 0.23 & 21.70  & 0.52 \\
1503 & \nod & \nod & \nod  & 1.0  & 0.15 & 22.75 & 1.3  & 0.19 & 21.27 & 1.5  & 0.24 & 21.01  & 1.90 \\
1464 & \nod & \nod & \nod  & 0.9  & 0.09 & 21.78 & 1.1  & 0.10 & 20.65 & 1.1  & 0.11 & 20.47  & 1.32 \\
 959 & \nod & \nod & \nod  & 1.1  & 0.17 & 23.62 & 1.2  & 0.19 & 22.09 & 1.1  & 0.20 & 21.38  & 1.68 \\
1559 & \nod & \nod & \nod  & 1.3  & 0.11 & 23.02 & 1.4  & 0.12 & 21.71 & 1.3  & 0.11 & 21.16  & 1.46 \\
1046 & 2.0  & 0.44 & 24.12 & 2.5  & 0.56 & 23.01 & 2.6  & 0.58 & 22.39 & 2.4  & 0.56 & 21.97  & 0.69 \\
1020 & \nod & \nod & \nod  & 0.6  & 0.10 & 22.43 & 0.9  & 0.11 & 21.60 & 0.9  & 0.12 & 21.51  & 1.03 \\
1084 & \nod & \nod & \nod  & 1.6  & 0.64 & 24.23 & 2.0  & 0.84 & 23.14 & 2.2  & 1.02 & 22.88  & 1.57 \\
1076 & \nod & \nod & \nod  & 1.1  & 0.09 & 22.62 & 1.2  & 0.12 & 21.12 & 1.2  & 0.12 & 20.75  & 1.80 \\
1066 & \nod & \nod & \nod  & 0.6  & 0.12 & 21.71 & 0.8  & 0.14 & 20.87 & 0.8  & 0.13 & 20.80  & 1.05 \\
 818 & \nod & \nod & \nod  & 1.8  & 0.28 & 24.11 & 2.1  & 0.35 & 23.14 & 2.2  & 0.40 & 22.76  & 1.36 \\
1399 & \nod & \nod & \nod  & 1.4  & 0.17 & 22.09 & 1.8  & 0.21 & 21.68 & 1.8  & 0.25 & 21.57  & 0.82 \\
1430 & \nod & \nod & \nod  & 0.8  & 0.08 & 20.93 & 0.7  & 0.07 & 19.60 & 0.9  & 0.08 & 19.70  & 1.18 \\
1287 & \nod & \nod & \nod  & 2.0  & 0.58 & 24.00 & 2.0  & 0.51 & 22.19 & 2.0  & 0.56 & 21.58  & 1.63 \\
 648 & 2.2  & 0.64 & 24.68 & 2.9  & 1.05 & 24.08 & 2.8  & 0.76 & 22.81 & 2.8  & 0.83 & 22.66  & 0.78 \\
 609 & \nod & \nod & \nod  & 0.9  & 0.16 & 22.96 & 1.2  & 0.18 & 21.18 & 1.3  & 0.21 & 20.82  & 1.97 \\
1236 & \nod & \nod & \nod  & 1.7  & 0.38 & 24.05 & 1.9  & 0.38 & 22.54 & 2.0  & 0.43 & 22.04  & 1.56 \\
 748 & \nod & \nod & \nod  & 0.8  & 0.08 & 22.96 & 1.1  & 0.11 & 21.58 & 1.1  & 0.12 & 20.91  & 1.85 \\
 206 & \nod & \nod & \nod  & 1.4  & 0.19 & 23.32 & 2.1  & 0.28 & 22.66 & 1.8  & 0.28 & 21.80  & 1.41 \\
 793 & \nod & \nod & \nod  & 1.7  & 0.34 & 24.79 & 1.8  & 0.34 & 23.25 & 1.8  & 0.37 & 22.56  & 1.54 \\
 329 & 1.0  & 0.20 & 22.80 & 1.9  & 0.44 & 21.99 & 1.9  & 0.45 & 21.03 & 1.8  & 0.45 & 20.69  & 0.96 \\
 264 & 1.7  & 0.46 & 24.27 & 2.3  & 0.56 & 22.64 & 2.4  & 0.57 & 21.63 & 2.4  & 0.63 & 21.41  & 1.04 \\
 241 & 1.8  & 0.67 & 25.40 & 2.5  & 0.99 & 24.08 & 2.6  & 0.99 & 22.99 & 2.6  & 1.06 & 22.68  & 1.10 \\
 244 & \nod & \nod & \nod  & 1.3  & 0.18 & 21.25 & 1.5  & 0.22 & 20.46 & 1.4  & 0.23 & 20.19  & 1.04 \\
 291 & 1.4  & 0.22 & 23.56 & 2.0  & 0.29 & 22.23 & 2.1  & 0.30 & 21.23 & 2.0  & 0.31 & 20.95  & 1.02 \\
 302 & \nod & \nod & \nod  & 2.1  & 0.34 & 23.73 & 2.5  & 0.45 & 22.89 & 2.5  & 0.51 & 22.61  & 1.33 \\
  43 & 1.9  & 0.45 & 24.64 & 3.0  & 1.33 & 24.31 & 2.9  & 1.07 & 22.90 & 2.8  & 1.08 & 22.47  & 1.07 \\
1346 & 1.6  & 0.26 & 23.78 & 2.1  & 0.36 & 22.98 & 2.1  & 0.33 & 21.57 & 2.2  & 0.37 & 21.39  & 1.26 \\
1292 & \nod & \nod & \nod  & 2.1  & 0.35 & 23.11 & 2.3  & 0.39 & 22.25 & 2.2  & 0.40 & 21.84  & 1.05 \\
1303 & \nod & \nod & \nod  & 1.4  & 0.34 & 23.57 & 1.7  & 0.37 & 22.29 & 1.8  & 0.41 & 21.93  & 1.46 \\
1342 & \nod & \nod & \nod  & 1.8  & 0.31 & 23.07 & 1.9  & 0.33 & 22.01 & 2.0  & 0.40 & 21.91  & 1.18 \\
1353 & \nod & \nod & \nod  & 1.9  & 0.28 & 23.45 & 2.1  & 0.31 & 22.13 & 2.0  & 0.32 & 21.63  & 1.48 \\
1329 & \nod & \nod & \nod  & 2.1  & 0.44 & 24.16 & 2.3  & 0.46 & 22.94 & 2.2  & 0.47 & 22.49  & 1.32 \\
 884 & 2.2  & 0.80 & 24.26 & 2.5  & 0.88 & 22.58 & 2.6  & 0.87 & 21.72 & 2.6  & 0.92 & 21.50  & 0.84 \\
 841 & 0.3  & 0.06 & 23.03 & 1.2  & 0.14 & 21.00 & 1.2  & 0.13 & 19.85 & 1.1  & 0.14 & 19.60  & 1.07 \\
 885 & 0.6  & 0.06 & 22.69 & 0.8  & 0.07 & 20.62 & 0.9  & 0.07 & 19.63 & 1.0  & 0.08 & 19.66  & 1.00 \\
 872 & \nod & \nod & \nod  & 1.1  & 0.06 & 20.45 & 1.6  & 0.11 & 21.01 & 1.3  & 0.11 & 20.35  & 0.46 \\
 912 & \nod & \nod & \nod  & 1.7  & 0.31 & 24.50 & 1.8  & 0.33 & 23.09 & 1.9  & 0.38 & 22.43  & 1.54 \\
 804 & \nod & \nod & \nod  & 1.7  & 0.37 & 23.24 & 1.8  & 0.36 & 22.14 & 1.8  & 0.39 & 21.90  & 1.07 \\
 854 & \nod & \nod & \nod  & 1.7  & 0.30 & 23.85 & 1.5  & 0.23 & 22.03 & 1.5  & 0.26 & 21.75  & 1.45 \\
 353 & \nod & \nod & \nod  & 1.8  & 0.35 & 24.65 & 2.3  & 0.54 & 23.90 & 2.2  & 0.56 & 23.12  & 1.47 \\
 468 & 2.3  & 0.91 & 24.05 & 2.5  & 0.87 & 22.25 & 2.5  & 0.83 & 21.40 & 2.4  & 0.83 & 21.07  & 0.77 \\
1718 & 1.7  & 0.34 & 24.31 & 2.2  & 0.45 & 22.57 & 2.3  & 0.44 & 21.35 & 2.3  & 0.49 & 21.13  & 1.20 \\
1362 & 1.0  & 0.10 & 20.63 & 1.1  & 0.11 & 19.63 & 1.2  & 0.11 & 18.84 & 1.3  & 0.13 & 19.01  & 0.86 \\
1709 & 1.0  & 0.23 & 23.81 & 1.1  & 0.16 & 21.68 & 1.2  & 0.18 & 20.25 & 1.1  & 0.18 & 19.81  & 1.60 \\
1740 & \nod & \nod & \nod  & 0.8  & 0.07 & 20.69 & 0.8  & 0.06 & 19.60 & 1.1  & 0.08 & 20.02  & 0.99 \\
1454 & 1.7  & 0.30 & 24.66 & 1.8  & 0.32 & 22.55 & 2.0  & 0.34 & 21.54 & 1.9  & 0.34 & 21.14  & 1.14 \\
1491 & \nod & \nod & \nod  & 0.8  & 0.10 & 21.86 & 1.0  & 0.12 & 20.56 & 1.0  & 0.13 & 20.34  & 1.54 \\
1440 & 0.6  & 0.11 & 24.37 & 1.3  & 0.28 & 22.56 & 1.3  & 0.27 & 21.33 & 1.3  & 0.29 & 21.10  & 1.16 \\
 957 & \nod & \nod & \nod  & 2.2  & 0.46 & 23.24 & 2.4  & 0.49 & 22.28 & 2.3  & 0.50 & 21.87  & 1.09 \\
 933 & \nod & \nod & \nod  & 2.1  & 0.45 & 24.43 & 2.3  & 0.49 & 22.97 & 2.4  & 0.61 & 22.52  & 1.62 \\
1558 & \nod & \nod & \nod  & 2.0  & 0.31 & 23.78 & 2.3  & 0.38 & 22.66 & 2.2  & 0.41 & 22.05  & 1.45 \\
1543 & \nod & \nod & \nod  & 0.7  & 0.08 & 21.59 & 1.2  & 0.14 & 20.65 & 1.0  & 0.12 & 20.20  & 1.78 \\
 951 & \nod & \nod & \nod  & 1.1  & 0.15 & 23.28 & 1.2  & 0.17 & 21.60 & 1.2  & 0.18 & 20.94  & 1.87 \\
 982 & \nod & \nod & \nod  & 0.9  & 0.09 & 21.73 & 0.8  & 0.08 & 20.30 & 1.0  & 0.11 & 20.11  & 1.24 \\
1061 & \nod & \nod & \nod  & 1.9  & 0.36 & 23.58 & 2.4  & 0.53 & 22.72 & 2.4  & 0.58 & 22.26  & 1.55 \\
 454 & 1.5  & 0.22 & 24.51 & 1.7  & 0.27 & 23.23 & 1.9  & 0.32 & 22.60 & 1.9  & 0.36 & 22.40  & 0.91 \\
 505 & 2.0  & 0.24 & 21.97 & 2.1  & 0.30 & 20.54 & 1.9  & 0.27 & 19.66 & 1.9  & 0.31 & 19.60  & 0.70 \\
1107 & 2.0  & 0.57 & 24.20 & 2.4  & 0.69 & 22.70 & 2.4  & 0.68 & 21.82 & 2.3  & 0.68 & 21.49  & 0.85 \\
 506 & 0.7  & 0.10 & 21.22 & 1.1  & 0.13 & 19.75 & 1.2  & 0.13 & 18.86 & 1.2  & 0.14 & 18.83  & 0.94 \\
 783 & \nod & \nod & \nod  & 1.6  & 0.21 & 22.43 & 1.6  & 0.20 & 21.38 & 1.5  & 0.22 & 21.13  & 1.01 \\
 761 & \nod & \nod & \nod  & 1.5  & 0.37 & 24.03 & 1.8  & 0.47 & 22.86 & 1.9  & 0.54 & 22.06  & 1.56 \\
 774 & \nod & \nod & \nod  & 2.0  & 0.42 & 23.54 & 2.3  & 0.51 & 22.40 & 2.2  & 0.52 & 21.90  & 1.48 \\
 827 & 0.9  & 0.09 & 23.35 & 1.8  & 0.24 & 22.10 & 1.6  & 0.19 & 20.63 & 1.4  & 0.18 & 20.20  & 1.11 \\
1436 & \nod & \nod & \nod  & 2.3  & 0.57 & 23.95 & 2.4  & 0.55 & 22.90 & 2.4  & 0.62 & 22.69  & 1.02 \\
 738 & \nod & \nod & \nod  & \nod & \nod & \nod  & 2.3  & 0.18 & 22.30 & 2.3  & 0.23 & 22.04  & \nod \\
1323 & \nod & \nod & \nod  & \nod & \nod & \nod  & 1.4  & 0.13 & 21.07 & 1.5  & 0.17 & 21.04  & \nod 

\enddata

\tablecomments{
%Table \ref{tab:photpars} will be published in its entirety 
%in the electronic edition of the {\it Astrophysical Journal}. A portion is 
%shown here for guidance regarding its form and content.
Col.\@ (1): internal galaxy ID.  
Cols.\@ (2)--(4): \sersic\ $n$ shape parameter, effective radius, and 
 effective surface brightness (uncorrected), respectively, from the light 
 profile modeling (see \S\ref{sec:BDdecomps}) measured in the 
 GOODS-ACS $B$-band (F435W) filter.  
Cols.\@ (5)--(7): same as (2)--(4) but for the GOODS-ACS $V$-band (F606W).
Cols.\@ (8)--(10): same as (2)--(4) but for  the GOODS-ACS $i$-band (F775W).
Cols.\@ (11)--(13): same as (2)--(4) but for  the GOODS-ACS $z$-band (F850LP).
Col.\@ (14): Observed $V-i$ color measured out to $1\,r_e$ 
 (see \S\ref{sec:colors}).
}

\end{deluxetable*}
\clearpage

\begin{deluxetable*}{rr@{\mpm}rr@{\mpm}rr@{\mpm}rr@{\mpm}rcr@{\mpm}lr@{\mpm}r}
%\tabletypesize{\scriptsize}
\tablecaption{Fundamental Plane Parameters. \label{tab:FPpars}}

\tablehead{\colhead{ID} & \multicolumn{2}{c}{$n$} & 
\multicolumn{2}{c}{$R_e$} & \multicolumn{2}{c}{SBe} & 
\multicolumn{2}{c}{$\sigma_{ap}$} & \colhead{$B_{rest}$} & 
\multicolumn{2}{c}{$\log_{10}(M/L_{B})$} & 
\multicolumn{2}{c}{$\log_{10}(M)$} \\
\colhead{} & \multicolumn{2}{c}{} & \multicolumn{2}{c}{(kpc)} & 
\multicolumn{2}{c}{(\magsqarc)} & \multicolumn{2}{c}{(km s$^{-1}$)} & 
\colhead{(AB mag)} & \multicolumn{2}{c}{($M/L_{\odot,B}$)} & 
\multicolumn{2}{c}{($M_{\odot}^{-1}$)} \\
\colhead{(1)} & \multicolumn{2}{c}{(2)} & \multicolumn{2}{c}{(3)} & 
\multicolumn{2}{c}{(4)} & \multicolumn{2}{c}{(5)} & \colhead{(6)} & 
\multicolumn{2}{c}{(7)} & \multicolumn{2}{c}{(8)} }

\startdata
% get the following table from /scr/lam/tommaso/FP.quantities
%Gal   Sn    dSn     Re     dRe     SBe    dSBe    sigma    dsig    Brest  l(M/L_B)dl(M/L_B)  l(M) dl(M)  
  62 & 2.0 & 0.1 &  1.21 &  0.03 & 18.25 & 0.40 &  143.8 &    7.3 & -20.60 & 10.62 & 0.05 &  0.19 &  0.17 \\ 
 198 & 1.4 & 0.3 &  2.31 &  0.58 & 19.40 & 0.40 &  140.0 &    8.7 & -20.78 & 10.91 & 0.12 &  0.38 &  0.30 \\ 
 280 & 1.1 & 0.3 &  1.13 &  0.39 & 19.54 & 0.60 &   70.3 &    9.2 & -19.18 &  9.99 & 0.19 &  0.15 &  0.43 \\ 
 462 & 1.5 & 0.1 &  1.39 &  0.03 & 19.34 & 0.17 &  230.5 &    7.7 & -19.80 & 11.12 & 0.03 &  1.01 &  0.08 \\ 
 519 & 1.0 & 0.2 &  2.12 &  0.53 & 20.72 & 0.40 &   75.4 &   11.5 & -19.27 & 10.32 & 0.17 &  0.40 &  0.32 \\ 
 873 & 1.1 & 0.1 &  0.51 &  0.05 & 18.67 & 0.17 &  142.7 &   12.7 & -18.31 & 10.27 & 0.09 &  0.76 &  0.14 \\ 
 928 & 1.7 & 0.1 &  0.78 &  0.07 & 17.66 & 0.24 &  164.2 &   10.6 & -20.39 & 10.56 & 0.07 &  0.28 &  0.14 \\ 
 975 & 1.2 & 0.1 &  1.79 &  0.29 & 20.46 & 0.14 &   81.9 &   11.4 & -18.87 & 10.33 & 0.14 &  0.45 &  0.21 \\ 
1050 & 1.9 & 0.1 &  2.20 &  0.27 & 19.55 & 0.17 &  240.7 &    8.7 & -20.87 & 11.33 & 0.06 &  0.91 &  0.14 \\ 
1377 & 1.7 & 0.1 &  2.28 &  1.40 & 20.10 & 0.17 &  170.3 &    8.5 & -20.05 & 11.06 & 0.27 &  0.82 &  0.60 \\ 
 211 & 1.2 & 0.1 &  2.35 &  0.31 & 21.19 & 0.13 &   42.7 &   12.9 & -19.06 &  9.88 & 0.27 &  0.06 &  0.30 \\ 
 254 & 1.2 & 0.1 &  1.85 &  0.22 & 20.33 & 0.12 &  136.1 &   11.3 & -19.50 & 10.79 & 0.09 &  0.83 &  0.14 \\ 
 289 & 0.8 & 0.2 &  1.13 &  0.00 & 20.13 & 0.60 &   46.5 &   13.2 & -18.71 &  9.61 & 0.25 & -0.01 &  0.34 \\ 
 548 & 1.6 & 0.1 &  1.36 &  0.13 & 19.24 & 0.15 &  240.4 &    5.9 & -19.86 & 11.14 & 0.05 &  1.01 &  0.11 \\ 
 698 & 1.8 & 0.4 &  2.34 &  0.58 & 21.09 & 0.40 &   48.2 &    5.6 & -19.08 &  9.97 & 0.15 &  0.11 &  0.31 \\ 
 891 & 1.0 & 0.1 &  0.62 &  0.02 & 18.07 & 0.14 &  151.9 &   24.9 & -19.27 & 10.39 & 0.14 &  0.49 &  0.16 \\ 
1008 & 0.8 & 0.0 &  1.89 &  0.46 & 21.51 & 0.00 &   48.3 &    8.0 & -18.30 &  9.86 & 0.18 &  0.36 &  0.28 \\ 
1037 & 0.9 & 0.1 &  1.11 &  0.14 & 19.22 & 0.14 &   58.6 &    5.8 & -19.53 &  9.81 & 0.10 & -0.15 &  0.16 \\ 
  27 & 1.5 & 0.1 &  0.68 &  0.03 & 17.84 & 0.14 &  219.9 &    3.6 & -19.67 & 10.76 & 0.03 &  0.68 &  0.07 \\ 
 181 & 1.2 & 0.1 &  1.69 &  0.18 & 18.37 & 0.15 &  332.7 &   78.3 & -21.28 & 11.52 & 0.21 &  0.86 &  0.24 \\ 
 272 & 1.0 & 0.1 &  1.82 &  0.19 & 18.48 & 0.14 &  233.6 &   16.6 & -21.21 & 11.24 & 0.08 &  0.55 &  0.13 \\ 
 971 & 1.2 & 0.1 &  0.88 &  0.08 & 18.61 & 0.14 &  260.9 &   12.8 & -19.44 & 11.03 & 0.06 &  1.03 &  0.11 \\ 
 973 & 1.3 & 0.1 &  1.44 &  0.22 & 17.98 & 0.16 &  264.7 &   16.3 & -21.28 & 11.26 & 0.09 &  0.57 &  0.17 \\ 
 463 & 1.4 & 0.1 &  1.80 &  0.28 & 18.46 & 0.16 &  274.0 &   18.6 & -21.44 & 11.38 & 0.09 &  0.70 &  0.17 \\ 
 509 & 1.2 & 0.1 &  1.45 &  0.21 & 18.16 & 0.16 &  324.9 &   41.8 & -21.12 & 11.44 & 0.13 &  0.82 &  0.19 \\ 
 661 & 0.6 & 0.1 &  0.98 &  0.04 & 17.25 & 0.10 &  113.8 &   11.9 & -21.14 & 10.28 & 0.09 & -0.36 &  0.11 \\ 
1478 & 1.1 & 0.2 &  0.92 &  0.01 & 18.12 & 0.24 &  119.1 &    8.3 & -20.20 & 10.36 & 0.06 &  0.13 &  0.11 \\ 
 635 & 1.6 & 0.1 &  2.49 &  0.24 & 19.29 & 0.14 &  212.0 &   15.6 & -21.29 & 11.29 & 0.08 &  0.66 &  0.13 \\ 
 683 & 1.5 & 0.1 &  0.85 &  0.18 & 18.71 & 0.17 &  139.6 &   21.2 & -19.61 & 10.47 & 0.16 &  0.54 &  0.25 \\ 
 199 & 1.0 & 0.1 &  2.14 &  0.26 & 18.09 & 0.11 &  242.4 &   15.9 & -22.00 & 11.34 & 0.08 &  0.36 &  0.14 \\ 
 273 & 1.8 & 0.3 &  2.03 &  0.53 & 19.93 & 0.39 &  215.0 &   12.5 & -19.96 & 11.20 & 0.13 &  1.00 &  0.30 \\ 
 237 & 2.1 & 0.2 &  3.64 &  0.36 & 18.90 & 0.24 &  290.8 &   20.0 & -22.41 & 11.70 & 0.07 &  0.58 &  0.15 \\ 
 292 & 1.0 & 0.1 &  0.74 &  0.06 & 18.87 & 0.15 &  217.5 &   12.4 & -18.88 & 10.79 & 0.06 &  1.04 &  0.11 \\ 
 377 & 1.3 & 0.2 &  1.77 &  0.13 & 19.93 & 0.16 &  234.6 &   26.4 & -19.76 & 11.24 & 0.10 &  1.16 &  0.14 \\ 
  18 & 1.3 & 0.1 &  1.86 &  0.15 & 18.89 & 0.14 &  159.8 &    6.1 & -20.84 & 10.93 & 0.05 &  0.39 &  0.10 \\ 
  17 & 1.1 & 0.1 &  1.10 &  0.08 & 18.42 & 0.09 &  203.6 &   11.1 & -20.11 & 10.91 & 0.06 &  0.63 &  0.10 \\ 
  47 & 1.2 & 0.1 &  1.32 &  0.05 & 19.29 & 0.11 &  281.8 &    8.5 & -19.69 & 11.27 & 0.03 &  1.19 &  0.06 \\ 
 402 & 0.9 & 0.1 &  0.69 &  0.05 & 18.90 & 0.14 &  165.6 &    6.1 & -18.75 & 10.51 & 0.05 &  0.84 &  0.10 \\ 
  38 & 1.4 & 0.3 &  1.54 &  0.38 & 19.09 & 0.40 &  147.2 &    6.1 & -20.21 & 10.77 & 0.11 &  0.48 &  0.29 \\ 
  83 & 1.2 & 0.1 &  0.83 &  0.07 & 18.89 & 0.16 &  226.9 &   16.6 & -19.14 & 10.88 & 0.07 &  1.04 &  0.12 \\ 
  63 & 1.2 & 0.1 &  0.64 &  0.05 & 18.67 & 0.30 &  161.8 &   12.0 & -18.74 & 10.47 & 0.07 &  0.78 &  0.15 \\ 
 369 & 1.5 & 0.1 &  1.56 &  0.65 & 19.64 & 0.36 &  109.6 &    8.9 & -19.40 & 10.52 & 0.19 &  0.43 &  0.43 \\ 
 115 & 1.2 & 0.2 &  0.98 &  0.14 & 18.11 & 0.24 &  278.9 &   13.3 & -20.37 & 11.13 & 0.07 &  0.83 &  0.17 \\ 
 400 & 1.2 & 0.4 &  0.66 &  0.12 & 18.39 & 0.57 &  226.6 &   17.1 & -19.32 & 10.78 & 0.10 &  0.94 &  0.29 \\ 
1267 & 1.1 & 0.0 &  0.97 &  0.04 & 18.34 & 0.09 &  281.7 &   10.6 & -19.98 & 11.13 & 0.04 &  0.94 &  0.07 \\ 
1373 & 2.0 & 0.2 &  5.43 &  0.41 & 18.90 & 0.23 &  218.0 &    8.1 & -23.14 & 11.63 & 0.05 &  0.16 &  0.12 \\ 
1308 & 1.0 & 0.1 &  0.74 &  0.01 & 18.09 & 0.14 &  123.7 &    7.5 & -19.96 & 10.30 & 0.05 &  0.23 &  0.08 \\ 
 871 & 2.2 & 0.3 &  3.19 &  0.55 & 20.00 & 0.19 &  136.8 &    8.6 & -20.87 & 10.98 & 0.09 &  0.41 &  0.19 \\ 
 879 & 1.2 & 0.1 &  2.48 &  0.25 & 19.97 & 0.11 &  164.7 &   10.1 & -19.92 & 11.08 & 0.07 &  0.72 &  0.12 \\ 
 870 & 1.4 & 0.1 &  1.37 &  0.19 & 18.71 & 0.26 &  169.6 &    8.2 & -20.49 & 10.85 & 0.07 &  0.50 &  0.17 \\ 
 895 & 1.4 & 0.1 &  1.22 &  0.20 & 19.01 & 0.16 &  130.7 &    5.7 & -19.75 & 10.57 & 0.08 &  0.44 &  0.18 \\ 
 405 & 2.1 & 0.2 &  2.84 &  0.44 & 20.05 & 0.34 &  108.9 &   10.3 & -20.64 & 10.74 & 0.11 &  0.29 &  0.22 \\ 
 399 & 1.0 & 0.1 &  0.53 &  0.04 & 18.50 & 0.15 &  171.8 &    8.2 & -18.58 & 10.44 & 0.05 &  0.83 &  0.10 \\ 
 345 & 1.2 & 0.2 &  0.85 &  0.21 & 17.95 & 0.32 &  292.6 &   18.8 & -20.27 & 11.11 & 0.12 &  0.88 &  0.28 \\ 
 346 & 1.5 & 0.1 &  3.08 &  0.27 & 19.36 & 0.15 &  284.0 &   26.8 & -21.61 & 11.64 & 0.09 &  0.85 &  0.13 \\ 
1743 & 1.3 & 0.2 &  1.29 &  0.21 & 19.74 & 0.16 &  193.1 &   13.5 & -19.21 & 10.93 & 0.09 &  1.05 &  0.18 \\ 
1403 & 1.4 & 0.3 &  0.73 &  0.27 & 17.94 & 0.40 &  149.4 &   14.1 & -20.15 & 10.47 & 0.18 &  0.35 &  0.40 \\ 
1385 & 1.1 & 0.1 &  0.82 &  0.09 & 18.30 & 0.15 &  197.6 &   11.8 & -19.69 & 10.75 & 0.07 &  0.69 &  0.14 \\ 
1463 & 1.3 & 0.1 &  0.67 &  0.08 & 19.59 & 0.16 &   75.0 &    8.6 & -17.89 &  9.83 & 0.11 &  0.45 &  0.16 \\ 
1506 & 1.1 & 0.2 &  0.67 &  0.06 & 19.07 & 0.23 &  114.1 &    7.6 & -18.99 & 10.19 & 0.07 &  0.61 &  0.14 \\ 
1032 & 2.3 & 0.1 &  1.79 &  0.44 & 20.03 & 0.40 &  186.1 &    9.8 & -19.67 & 10.98 & 0.12 &  0.93 &  0.29 \\ 
1021 & 1.3 & 0.2 &  1.28 &  0.40 & 19.55 & 0.26 &  290.1 &   28.1 & -19.78 & 11.28 & 0.16 &  1.33 &  0.33 \\ 
1088 & 1.7 & 0.2 &  1.88 &  0.32 & 18.56 & 0.27 &  202.6 &   16.2 & -21.39 & 11.13 & 0.10 &  0.44 &  0.21 \\ 
 811 & 2.0 & 0.2 &  3.29 &  0.50 & 19.12 & 0.29 &  227.2 &   13.2 & -21.85 & 11.45 & 0.08 &  0.50 &  0.19 \\ 
1364 & 1.8 & 0.2 &  1.61 &  0.14 & 19.61 & 0.16 &  163.0 &    7.9 & -19.80 & 10.86 & 0.06 &  0.74 &  0.12 \\ 
1633 & 1.5 & 0.2 &  1.45 &  0.16 & 18.05 & 0.25 &  363.4 &   12.6 & -21.16 & 11.53 & 0.06 &  0.87 &  0.15 \\ 
 655 & 0.9 & 0.1 &  1.10 &  0.06 & 19.21 & 0.10 &  117.5 &    7.3 & -19.47 & 10.41 & 0.06 &  0.45 &  0.09 \\ 
 681 & 1.2 & 0.1 &  1.61 &  0.26 & 18.37 & 0.13 &  373.0 &   28.1 & -21.10 & 11.60 & 0.10 &  0.98 &  0.18 \\ 
 664 & 1.4 & 0.2 &  0.71 &  1.11 & 18.37 & 1.29 &  172.7 &   11.0 & -19.41 & 10.58 & 0.68 &  0.66 &  1.60 \\ 
 752 & 2.2 & 0.1 &  2.78 &  0.58 & 20.48 & 0.36 &  117.0 &   10.2 & -20.11 & 10.78 & 0.12 &  0.52 &  0.26 \\ 
 253 & 2.1 & 0.2 &  1.87 &  0.34 & 19.00 & 0.37 &   89.0 &   11.9 & -20.91 & 10.38 & 0.14 & -0.12 &  0.26 \\ 
 256 & 1.0 & 0.1 &  1.55 &  0.16 & 17.66 & 0.13 &  336.7 &   13.2 & -21.66 & 11.49 & 0.06 &  0.61 &  0.12 \\ 
1261 & 2.0 & 0.2 &  5.53 &  0.61 & 19.50 & 0.24 &  263.4 &   20.2 & -22.63 & 11.80 & 0.08 &  0.56 &  0.16 \\ 
1285 & 1.1 & 0.0 &  1.34 &  0.05 & 18.37 & 0.00 &  182.3 &   17.1 & -20.65 & 10.90 & 0.08 &  0.43 &  0.09 \\ 
1336 & 0.7 & 0.1 &  1.07 &  0.09 & 19.08 & 0.07 &   71.3 &   15.9 & -19.18 &  9.94 & 0.20 & -0.05 &  0.21 \\ 
1328 & 1.9 & 0.2 &  2.47 &  0.28 & 19.49 & 0.31 &  263.3 &   31.9 & -21.15 & 11.46 & 0.12 &  0.91 &  0.20 \\ 
1297 & 1.6 & 0.1 &  2.65 &  0.28 & 19.01 & 0.16 &  191.7 &   21.5 & -21.53 & 11.23 & 0.11 &  0.43 &  0.15 \\ 
 876 & 2.3 & 0.1 &  2.87 &  0.52 & 20.21 & 0.25 &  144.1 &   10.4 & -20.49 & 10.97 & 0.10 &  0.57 &  0.21 \\ 
 815 & 1.4 & 0.1 &  1.22 &  0.34 & 19.22 & 0.55 &  190.2 &    5.7 & -19.83 & 10.90 & 0.12 &  0.85 &  0.35 \\ 
 930 & 2.4 & 0.3 &  4.39 &  0.77 & 19.95 & 0.44 &  245.0 &   16.4 & -21.89 & 11.60 & 0.10 &  0.74 &  0.25 \\ 
1706 & 1.8 & 0.1 &  2.50 &  0.38 & 18.95 & 0.17 &  227.3 &   11.1 & -21.59 & 11.34 & 0.08 &  0.57 &  0.17 \\ 
1741 & 1.0 & 0.4 &  1.20 &  0.18 & 20.09 & 0.08 &   69.7 &   10.2 & -18.71 & 10.01 & 0.14 &  0.32 &  0.20 \\ 
1503 & 1.5 & 0.1 &  1.45 &  0.29 & 18.43 & 0.15 &  230.0 &   14.0 & -21.18 & 11.13 & 0.10 &  0.63 &  0.21 \\ 
1464 & 1.1 & 0.1 &  0.69 &  0.07 & 18.50 & 0.15 &  157.6 &    9.5 & -19.16 & 10.48 & 0.07 &  0.64 &  0.12 \\ 
 959 & 1.1 & 0.1 &  1.57 &  0.19 & 18.94 & 0.11 &  213.5 &   15.8 & -20.52 & 11.10 & 0.08 &  0.73 &  0.14 \\ 
1046 & 2.4 & 0.2 &  2.40 &  0.34 & 21.07 & 0.31 &   57.2 &   16.2 & -19.30 & 10.08 & 0.25 &  0.18 &  0.31 \\ 
1020 & 0.9 & 0.1 &  0.67 &  0.10 & 20.09 & 0.08 &   44.0 &    7.7 & -17.51 &  9.34 & 0.17 &  0.17 &  0.22 \\ 
1084 & 2.2 & 0.2 &  6.67 &  1.30 & 20.08 & 0.27 &  365.6 &   32.8 & -22.72 & 12.15 & 0.12 &  0.97 &  0.23 \\ 
1076 & 1.2 & 0.1 &  0.94 &  0.12 & 18.29 & 0.15 &  238.8 &   23.8 & -20.05 & 10.98 & 0.10 &  0.79 &  0.16 \\ 
1066 & 0.8 & 2.8 &  0.96 &  0.10 & 18.82 & 0.00 &   83.6 &   18.1 & -19.57 & 10.05 & 0.19 &  0.05 &  0.21 \\ 
 818 & 2.2 & 0.2 &  3.11 &  0.62 & 19.67 & 0.34 &  167.5 &   16.6 & -21.31 & 11.14 & 0.12 &  0.46 &  0.25 \\ 
1399 & 1.8 & 0.2 &  1.32 &  0.15 & 19.35 & 0.25 &   78.9 &    5.8 & -19.79 & 10.15 & 0.08 &  0.09 &  0.16 \\ 
1430 & 0.9 & 0.1 &  0.52 &  0.03 & 18.25 & 0.06 &  148.7 &    8.5 & -18.67 & 10.30 & 0.06 &  0.60 &  0.08 \\ 
1287 & 2.0 & 0.1 &  4.26 &  0.50 & 18.84 & 0.16 &  350.8 &   21.7 & -22.83 & 11.94 & 0.07 &  0.65 &  0.14 \\ 
 648 & 2.8 & 0.2 &  3.81 &  0.31 & 22.10 & 0.28 &   69.4 &   13.2 & -19.37 & 10.41 & 0.17 &  0.53 &  0.21 \\ 
 609 & 1.3 & 0.2 &  1.41 &  0.19 & 18.35 & 0.22 &  287.8 &   16.4 & -21.09 & 11.32 & 0.08 &  0.80 &  0.16 \\ 
1236 & 2.0 & 0.1 &  3.21 &  0.31 & 19.23 & 0.15 &  226.4 &   10.9 & -21.83 & 11.43 & 0.06 &  0.56 &  0.12 \\ 
 748 & 1.1 & 0.2 &  1.02 &  0.12 & 18.14 & 0.24 &  256.9 &   20.5 & -20.36 & 11.07 & 0.09 &  0.76 &  0.17 \\ 
 206 & 1.8 & 0.2 &  2.40 &  0.43 & 18.82 & 0.32 &  212.5 &   16.3 & -21.48 & 11.27 & 0.10 &  0.48 &  0.23 \\ 
 793 & 1.8 & 0.2 &  3.06 &  0.61 & 19.49 & 0.29 &  115.1 &   12.8 & -21.39 & 10.84 & 0.13 &  0.10 &  0.24 \\ 
 329 & 1.8 & 0.1 &  2.48 &  0.40 & 19.76 & 0.16 &  134.6 &    2.7 & -20.63 & 10.88 & 0.07 &  0.44 &  0.17 \\ 
 264 & 2.4 & 0.2 &  3.60 &  0.40 & 19.76 & 0.25 &  236.5 &   16.1 & -21.44 & 11.49 & 0.08 &  0.72 &  0.16 \\ 
 241 & 2.6 & 0.2 &  6.56 &  1.02 & 20.94 & 0.21 &  308.7 &   16.5 & -21.55 & 11.96 & 0.08 &  1.14 &  0.18 \\ 
 244 & 1.4 & 0.1 &  1.19 &  5.50 & 18.80 & 0.27 &  210.9 &    6.7 & -20.11 & 10.98 & 2.00 &  0.78 &  4.47 \\ 
 291 & 2.0 & 0.2 &  1.97 &  0.16 & 19.24 & 0.23 &  155.9 &   10.6 & -20.63 & 10.90 & 0.07 &  0.45 &  0.14 \\ 
 302 & 2.5 & 0.2 &  3.20 &  0.72 & 20.08 & 0.36 &  136.5 &   12.7 & -20.95 & 10.95 & 0.13 &  0.41 &  0.27 \\ 
  43 & 2.8 & 0.2 &  8.00 &  1.81 & 21.29 & 0.31 &  242.5 &   10.3 & -21.48 & 11.82 & 0.11 &  0.97 &  0.26 \\ 
1346 & 2.2 & 0.2 &  2.59 &  0.23 & 18.96 & 0.17 &  128.0 &    7.1 & -21.46 & 10.83 & 0.06 &  0.03 &  0.12 \\ 
1292 & 2.2 & 0.2 &  2.42 &  0.33 & 20.13 & 0.32 &   93.2 &    6.8 & -20.26 & 10.52 & 0.09 &  0.25 &  0.20 \\ 
1303 & 1.8 & 0.2 &  2.87 &  0.22 & 19.50 & 0.20 &  160.8 &    8.1 & -21.22 & 11.10 & 0.06 &  0.43 &  0.12 \\ 
1342 & 2.0 & 0.1 &  2.39 &  0.32 & 19.56 & 0.23 &  115.0 &    6.9 & -20.78 & 10.72 & 0.08 &  0.22 &  0.17 \\ 
1353 & 2.0 & 0.2 &  2.58 &  0.34 & 18.79 & 0.21 &  133.5 &    7.1 & -21.72 & 10.88 & 0.07 &  0.01 &  0.16 \\ 
1329 & 2.2 & 0.2 &  3.32 &  0.54 & 20.45 & 0.33 &  132.4 &    7.2 & -20.59 & 10.97 & 0.08 &  0.55 &  0.21 \\ 
 884 & 2.6 & 0.2 &  4.38 &  0.61 & 20.55 & 0.15 &  192.7 &    3.5 & -21.08 & 11.37 & 0.06 &  0.75 &  0.15 \\ 
 841 & 1.1 & 1.0 &  0.89 &  0.03 & 18.41 & 0.09 &  176.1 &    7.6 & -19.68 & 10.69 & 0.04 &  0.60 &  0.06 \\ 
 885 & 1.0 & 0.2 &  0.47 &  0.03 & 18.25 & 0.19 &  179.5 &    7.7 & -18.50 & 10.42 & 0.05 &  0.82 &  0.10 \\ 
 872 & 1.3 & 0.1 &  0.97 &  0.10 & 17.39 & 0.16 &  200.8 &   24.8 & -20.97 & 10.84 & 0.12 &  0.27 &  0.16 \\ 
 912 & 1.9 & 0.2 &  3.05 &  0.41 & 19.39 & 0.23 &  155.0 &   10.3 & -21.51 & 11.09 & 0.08 &  0.32 &  0.17 \\ 
 804 & 1.8 & 0.1 &  2.58 &  0.22 & 20.05 & 0.15 &   93.7 &    7.3 & -20.39 & 10.59 & 0.08 &  0.23 &  0.12 \\ 
 854 & 1.5 & 0.2 &  1.93 &  0.39 & 19.72 & 0.38 &  122.3 &   10.2 & -19.97 & 10.71 & 0.11 &  0.47 &  0.26 \\ 
 353 & 2.2 & 0.1 &  4.66 &  0.77 & 20.02 & 0.20 &   99.9 &   12.4 & -21.74 & 10.87 & 0.13 & -0.02 &  0.21 \\ 
 468 & 2.4 & 0.2 &  3.95 &  0.43 & 20.44 & 0.14 &  159.5 &    8.2 & -20.93 & 11.18 & 0.07 &  0.61 &  0.13 \\ 
1718 & 2.3 & 0.2 &  3.11 &  0.39 & 19.19 & 0.25 &  175.5 &    6.3 & -21.66 & 11.17 & 0.06 &  0.31 &  0.16 \\ 
1362 & 1.3 & 0.3 &  0.80 &  0.08 & 16.56 & 0.17 &  228.4 &   15.3 & -21.38 & 10.87 & 0.07 &  0.13 &  0.13 \\ 
1709 & 1.1 & 0.1 &  1.44 &  0.10 & 17.26 & 0.11 &  240.0 &   10.5 & -21.95 & 11.17 & 0.05 &  0.19 &  0.09 \\ 
1740 & 1.1 & 0.1 &  0.49 &  0.04 & 17.57 & 0.13 &  111.1 &    5.9 & -19.25 & 10.03 & 0.06 &  0.11 &  0.11 \\ 
1454 & 1.9 & 0.1 &  2.22 &  0.23 & 19.50 & 0.19 &  183.6 &    5.4 & -20.68 & 11.10 & 0.05 &  0.65 &  0.13 \\ 
1491 & 1.0 & 0.1 &  0.81 &  0.03 & 18.49 & 0.08 &  173.0 &   15.2 & -19.55 & 10.63 & 0.08 &  0.65 &  0.09 \\ 
1440 & 1.3 & 0.1 &  1.87 &  0.35 & 19.67 & 0.15 &  112.5 &    5.8 & -20.06 & 10.63 & 0.09 &  0.39 &  0.20 \\ 
 957 & 2.3 & 0.1 &  3.08 &  0.57 & 20.26 & 0.23 &  257.8 &    9.4 & -20.63 & 11.50 & 0.09 &  1.07 &  0.20 \\ 
 933 & 2.4 & 0.2 &  4.17 &  0.84 & 19.61 & 0.26 &  311.7 &   34.7 & -22.13 & 11.79 & 0.13 &  0.83 &  0.24 \\ 
1558 & 2.2 & 0.2 &  3.19 &  0.41 & 19.23 & 0.32 &  109.1 &    5.8 & -21.78 & 10.78 & 0.07 & -0.10 &  0.18 \\ 
1543 & 1.0 & 0.2 &  1.12 &  0.17 & 17.65 & 0.22 &  314.4 &   11.9 & -20.88 & 11.29 & 0.07 &  0.69 &  0.17 \\ 
 951 & 1.2 & 0.1 &  1.44 &  0.26 & 18.44 & 0.16 &  258.2 &   16.2 & -20.84 & 11.23 & 0.10 &  0.74 &  0.19 \\ 
 982 & 1.0 & 0.1 &  0.82 &  0.10 & 17.26 & 0.13 &  164.6 &   11.0 & -20.85 & 10.59 & 0.08 &  0.11 &  0.14 \\ 
1061 & 2.4 & 0.2 &  3.99 &  0.66 & 19.77 & 0.26 &  249.8 &   16.0 & -21.76 & 11.58 & 0.09 &  0.72 &  0.20 \\ 
 454 & 1.9 & 0.2 &  1.89 &  0.33 & 20.53 & 0.27 &   53.8 &    6.1 & -19.38 &  9.96 & 0.12 &  0.06 &  0.22 \\ 
 505 & 1.9 & 0.2 &  1.19 &  0.15 & 18.84 & 0.27 &  148.8 &    8.6 & -20.02 & 10.64 & 0.08 &  0.47 &  0.17 \\ 
1107 & 2.3 & 0.2 &  3.47 &  0.67 & 20.64 & 0.24 &  113.8 &    4.5 & -20.48 & 10.84 & 0.09 &  0.46 &  0.21 \\ 
 506 & 1.2 & 0.1 &  0.88 &  0.05 & 17.01 & 0.15 &  228.3 &   10.2 & -21.14 & 10.91 & 0.04 &  0.27 &  0.09 \\ 
 783 & 1.5 & 0.1 &  1.36 &  0.13 & 19.58 & 0.16 &  114.9 &    8.7 & -19.47 & 10.50 & 0.08 &  0.51 &  0.13 \\ 
 761 & 1.9 & 0.1 &  4.57 &  0.49 & 18.78 & 0.16 &  384.3 &   36.0 & -22.96 & 12.05 & 0.09 &  0.69 &  0.15 \\ 
 774 & 2.2 & 0.2 &  4.03 &  0.47 & 19.35 & 0.26 &  171.6 &    8.6 & -22.12 & 11.27 & 0.07 &  0.25 &  0.16 \\ 
 827 & 1.4 & 0.2 &  1.45 &  0.22 & 19.16 & 0.25 &  169.0 &    7.1 & -19.84 & 10.87 & 0.08 &  0.65 &  0.18 \\ 
1436 & 2.4 & 0.2 &  3.63 &  0.58 & 21.01 & 0.32 &  116.2 &    9.7 & -20.18 & 10.87 & 0.10 &  0.60 &  0.21 \\ 
 738 & 2.3 & 0.2 &  1.76 &  0.33 & 18.78 & 0.36 &  186.5 &   11.8 & -20.98 & 10.98 & 0.10 &  0.44 &  0.24 \\ 
1323 & 1.5 & 0.3 &  1.31 &  0.29 & 17.94 & 0.44 &  229.7 &   18.1 & -21.19 & 11.09 & 0.12 &  0.47 &  0.29 

\enddata

\tablecomments{
%Table \ref{tab:FPpars} will be published in its entirety 
%in the electronic edition of the {\it Astrophysical Journal}. A portion is 
%shown here for guidance regarding its form and content.
Table \ref{tab:FPpars} lists the derived fundamental plane parameters of 
the spheroidal component of the 147 galaxies in our spectroscopic 
sample (see \S\ref{sec:sample} for sample description) as follows:
Col.\@ (1): internal galaxy ID.  
Col.\@ (2): the \sersic\ $n$ shape parameter.
Col.\@ (3): effective radius in kpc.
Col.\@ (4): average restframe $B$-band SB within $R_e$. 
Col.\@ (5): velocity dispersion in \kms\ as measured in our observed 
 aperture (1\,$\times$\,0.1185\arcsec).
Col.\@ (6): total restframe $B$-band AB magnitude of the spheroid.
Col.\@ (7): logarithm of the total $B$-band mass-to-light ratio, 
 in solar units, of the spheroid.
Col.\@ (8): logarithm of the dynamical stellar mass, in solar masses, of 
 the spheroid.
}
\end{deluxetable*}

\end{document}